\documentclass{aa} 

\usepackage{graphicx}
\usepackage{txfonts}
\usepackage{xfrac}
\usepackage{footmisc}
\usepackage[colorlinks=true,urlcolor=blue,citecolor=blue,
            linkcolor=blue]{hyperref}
\usepackage{mhchem}

\def\aap{A\&A}

\def\apjl{ApJL}
\def\apjs{ApJS}

\def\la{\mathrel{\mathchoice {\vcenter{\offinterlineskip\halign{\hfil
$\displaystyle##$\hfil\cr<\cr\sim\cr}}}
{\vcenter{\offinterlineskip\halign{\hfil$\textstyle##$\hfil\cr
<\cr\sim\cr}}}
{\vcenter{\offinterlineskip\halign{\hfil$\scriptstyle##$\hfil\cr
<\cr\sim\cr}}}
{\vcenter{\offinterlineskip\halign{\hfil$\scriptscriptstyle##$\hfil\cr
<\cr\sim\cr}}}}}
\def\ga{\mathrel{\mathchoice {\vcenter{\offinterlineskip\halign{\hfil
$\displaystyle##$\hfil\cr>\cr\sim\cr}}}
{\vcenter{\offinterlineskip\halign{\hfil$\textstyle##$\hfil\cr
>\cr\sim\cr}}}
{\vcenter{\offinterlineskip\halign{\hfil$\scriptstyle##$\hfil\cr
>\cr\sim\cr}}}
{\vcenter{\offinterlineskip\halign{\hfil$\scriptscriptstyle##$\hfil\cr
>\cr\sim\cr}}}}}

\def\kabs{\kappa_\nu^{\rm abs}}
\def\ksca{\kappa_\nu^{\rm sca}}
\def\kapC{\kappa_{\rm C}}
\def\kapL{\kappa_{\rm L}}
\def\tauC{\tau_{\!\rm C}}
\def\tauL{\tau_{\rm L}}
\def\tauCup{\tau_{\!\rm C}^{\uparrow}}
\def\tauLup{\tau_{\rm L}^{\uparrow}}
\def\tauCdown{\tau_{\!\rm C}^{\downarrow}}
\def\tauLdown{\tau_{\rm L}^{\downarrow}}
\def\tauCrad{\tau_{\!\rm C}^{\rm rad}}
\def\tauLrad{\tau_{\rm L}^{\rm rad}}
\def\Nirad{N_i^{\rm rad}}
\def\Niver{N_i^{\uparrow}}
\def\SC{S_{\!\rm C}}
\def\SL{S_{\!\rm L}}
\def\Jcont{J_\nu^{\rm cont}}
\def\Jbar{\bar{J}_{ul}}
\def\nuD{\nu_{\rm D}}
\def\smax{s_{\rm max}}
\def\Ppump{P_{ul}^{\rm\,pump}}
\def\Pesc{P_{ul}^{\rm\,esc}}
\def\Ppumpold{P_{ul}^{\rm\,pump, old}}
\def\Pescold{P_{ul}^{\rm\,esc, old}}
\def\Lcell{L^{\rm cell}}
\def\Fline{F_{\rm line}}
\def\Tg{T_{\rm g}}
\def\Td{T_{\rm d}}
\def\nH{n_{\langle\rm H\rangle}}

\def\rhom{\rho_{\rm m}}
\def\cs{c_{\rm S}}
\def\St{S\hspace*{-0.3ex}t\hspace*{0.1ex}}
\def\pabl#1#2{\frac{\partial #1}{\partial #2}}
\def\vth{v_{\rm th}}

\def\vdreq{{v^{\hspace{-0.8ex}^{\circ}}_{\rm dr}}}
\def\Dd{D_{\rm\hspace{0.05ex}d}}
\def\Dgas{D_{\rm gas}}
\def\vz{\langle v_z\rangle}
\def\tstop{\tau_{\rm stop}}
\def\tedd{\tau_{\rm eddy}}
\def\Mdot{\dot{M}_{\rm acc}}

\begin{document} 

\title{2D disc modelling of the JWST line spectrum of EX\,Lupi}

\author{P.~Woitke\inst{1}
       \and
       W.-F.~Thi\inst{2}
       \and
       A.~M.~Arabhavi\inst{3}
       \and
       I.~Kamp\inst{3}
       \and
       \'A.~K\'osp\'al\inst{4,5}
       \and
       P.~\'Abrah\'am\inst{4,5}
}

\institute{Space Research Institute, Austrian Academy of Sciences,
           Schmiedlstr.~6, A-8042, Graz, Austria
         \and
           Max-Planck-Institut f\"ur extraterrestrische Physik,
           Giessenbachstrasse 1, 85748, Garching, Germany
         \and
           Kapteyn Astronomical Institute, University of Groningen,
           PO Box 800, 9700 AV Groningen, The Netherlands
         \and
           Konkoly Observatory, Research Centre for Astronomy and
           Earth Sciences, E\"ov\"os Lor\'and Research Network (ELKH),
           Konkoly-Thege Mikl\'os \'ut 15–17, 1121 Budapest, Hungary
         \and  
           CSFK, MTA Centre of Excellence, Konkoly Thege Mikl\'os \'ut
           15–17, 1121, Budapest, Hungary
}

\date{Received 16/08/2023; accepted 30/11/2023}

\abstract{We introduce a number of new theoretical approaches and
  improvements to the thermo-chemical disc modelling code ProDiMo to
  better predict and analyse the JWST line spectra of protoplanetary
  discs. We develop a new line escape probability method for disc
  geometries, a new scheme for dust settling, and discuss how to apply
  UV molecular shielding factors to photorates in 2D disc geometry.
  We show that these assumptions are crucial for the determination of
  the gas heating/cooling rates and discuss how they affect the
  predicted molecular concentrations and line emissions.  We apply our
  revised 2D models to the protoplanetary disc around the T\,Tauri
  star EX\,Lupi in quiescent state.  We calculate infrared line
  emission spectra between 5 and 20\,$\mu$m by CO, H$_2$O, OH, CO$_2$,
  HCN, C$_2$H$_2$ and H$_2$, including lines of atoms and ions, using
  our full 2D predictions of molecular abundances, dust opacities, gas
  and dust temperatures.  We develop a disc model with a slowly
  increasing surface density structure around the inner rim that can
  simultaneously fit the spectral energy distribution, the overall
  shape of the JWST spectrum of EX\,Lupi, and the main observed
  molecular characteristics in terms of column densities, emitting
  areas and molecular emission temperatures, which all result from one
  consistent disc model. The spatial structure of the line emitting
  regions of the different molecules is discussed. High abundances of
  HCN and \ce{C2H2} are caused in the model by stellar X-ray
  irradiation of the gas around the inner rim.}


\maketitle


\section{Introduction}

Medium-resolution spectroscopy of protoplanetary discs with the
Spitzer Space Telescope has revealed a dense forest of \ce{H2O}
emission lines \citep{Carr2008,Salyk2008}, in addition to overlapping
line emission features from OH, \ce{C2H2}, HCN and \ce{CO2}. From the
emission temperatures of these molecules, a few 100\,K to about
1000\,K, and the derived sizes of the line emitting areas, it was
immediately clear that these molecular spectra probe the chemistry in
the terrestrial planet-forming regions, see e.g.\ review by
\citet{Pontoppidan2014}. Over the past 15 years, more than 100
Spitzer/IRS disc spectra have been analysed by simple slab models to
reveal certain properties of these line emissions in terms of
molecular column densities, emission temperatures, and line emission
areas, see \citet{Pontoppidan2010, Carr2011, Salyk2011}.
Some statistics are available as well, and certain trends have been
identified concerning the influence of the accretion
luminosity, the disc mass, the depletion of molecules in discs with large inner dust
cavities, and chemical differences as function of stellar type, see,
for example, \citet{Pascucci2013, Najita2013, Walsh2015, Banzatti2017}.

However, until the present day it remains somewhat unclear what these
measured molecular properties actually mean, where in the disc these
molecules are present, and how this all fits together. The slab models
used for the analysis of the spectra do not provide answers to these
basic questions, they also have certain principle weaknesses, such as
their inability to predict absorption lines, and an uncertainty how to
include the effects of Keplerian broadening, continuous dust emission
and absorption under a given inclination angle.

Ultimately, we are seeking a higher level of understanding, where the
predictions of molecular emission features would be in accordance with
disc chemistry and heating/cooling balance, based on a model for the
physical structure of the gas and dust in the disc that can predict
the emission features of all molecules simultaneously.  However,
thermo-chemical disc models, such as \citet{Akimkin2013, Zhang2013,
  Walsh2015, Bruderer2015, Woitke2018, Anderson2021}, are still to
demonstrate that this is possible.

The puzzle of revealing the physical structure of the disc, which
produces these emission lines, is complicated by the fact that we
actually do not know the element abundances of the gas in the inner
disc very well. Indeed, there are very good reasons to assume that the
transport of icy grains in the disc can profoundly change these
element abundances during disc evolution
\citep[e.g.][]{Ciesla2006}. For example, pebble drift through the
water snowline has been demonstrated in hydrodynamical disc models to
increase the oxygen abundance in the inner disc \citep{Bosman2017,
  Booth2019}.  {However, results from the VLT/MATISSE
  interferometer \citep{Kokoulina2021} suggest that the dust in the
  inner 10\,au of the disc around HD\,179218 is carbon-enriched in
  form of nano-carbon grains. These conclusions are based on the
  spectral shape of the solid opacities.  Studying new VLT/MATISSE and
  VLT/GRAVITY data of HD\,144432, Varga et\,al.\,(accepted by A\&A)
  concluded that most of the carbon in the inner disk is in the gas
  phase, whereas solid iron and FeS are the main souces of dust
  opacity at wavelengths in the $L$-, $M$- and $N$-bands}.  Additional
clues about element transport processes have been obtained by linking
the mid-IR line spectra to ALMA continuum images. Expanding on earlier
results from \citet{Najita2013}, \citet{Banzatti2020} reported on an
anti-correlation between the infrared water line luminosity tracing
the gas within a few au, and the distribution of solid pebbles at
10-200\,au in discs. That is, radially compact discs show stronger
water lines than large discs. {Based on new JWST data,
  \citet{Banzatti2023} see more evidence for this anti-correlation.
  \cite{vantHoff2020} and \cite{Nazari2023} have proposed that the
  inward transport and subsequent sublimation of carboneceous dust
  grains and PAHs might enrich the inner disc with carbon.}

{New evidence for carbon-rich/oxygen-poor inner discs} have recently
been found in the first published disc spectra obtained with the James
Webb Space Telescope (JWST), see \citet{Kospal2023, Grant2023,
  Tabone2023, Banzatti2023, Perotti2023}.  A rich hydrocarbon
chemistry has been identified by \citet{Tabone2023} in the inner disc
around the M5-star 2MASS-J16053215-1933159, with exceptionally
optically thick \ce{C2H2} emission features around 7\,$\mu$m and
14\,$\mu$m, which are so strong that they form a wide quasi-continuum
by tens of thousands of overlapping weak lines, without detecting any
water lines. This is very clear evidence for a strong oxygen depletion
in the inner disc of this star. \citet{Tabone2023} also mention a
carbon-enrichment of the inner disc by the destruction of PAHs or
carbonaceous grains as possible cause for the large column densities
of hydro-carbon molecules. {However this seems inconsistent with the
  very weak CO fundamental ro-vibrational lines as measured with
  JWST. To fit these lines, it seems to be required to remove the
  oxygen rather than to enrich the carbon (Kanwar et\,al.\,2024 in
  prep.)}.

In this paper, we take the challenge to fit the first published JWST
spectrum of a T\,Tauri star \citep{Kospal2023} by the 2D
thermo-chemical disc model {\sc ProDiMo} \citep{Woitke2009, Woitke2016}.  In
Sect.~\ref{sec:CodeChanges}, we report on a number of important
{\sc ProDiMo} code changes that resulted in considerably improved
predictions of the various mid-IR molecular emission features in
general.  In Sect.~\ref{sec:EXLup}, we explain how we set up our disc
model and how we fitted that model to the Spectral Energy Distribution
(SED) of EX\,Lupi and its JWST/MIRI spectrum.
Section~\ref{sec:analysis} contains a thorough analysis of the best
fitting disc model in terms of the physical and chemical disc
structure, including the calculated dust and gas temperatures. We also
derive molecular column densities and emission temperatures from this
model that can be compared to the values derived from slab models, and
analyse the chemical pathways in the line forming regions of the
various molecules in this section. Our conclusions are presented in
Sect.~\ref{sec:conclusions}.

\section{Code improvements}
\label{sec:CodeChanges}

\citet{Woitke2018} have demonstrated the problems that current
standard thermo-chemical models for T\,Tauri discs have to explain the
about equally strong emission features of \ce{H2O}, \ce{CO2}, \ce{HCN}
and \ce{C2H2}. First, a regular gas/dust mass ratio of 100 generally
results in too weak mid-IR molecular emissions. Second, while the
molecular emission features of \ce{H2O} and OH can be pushed to the
observed flux levels by using a larger gas/dust ratio of 1000
\citep{Bruderer2015,Woitke2018,Greenwood2019}, the \ce{CO2} emission
feature at 15\,$\mu$m becomes too strong. And third, the emission
lines of \ce{HCN} and \ce{C2H2} mostly come from a radially extended
optically thin photodissociation region high above the disc in these
models, but the emission lines created in these layers are generally
optically thin, and their emission characteristics are in conflict
with observations, in particular the molecular column densities are
too low in these models, e.g.\ \citet{Bruderer2015} for HCN.
\cite{Kanwar2023} have thoroughly analysed these dilute
photodissociation regions using an extended hydro-carbon chemical
network.

Increasing the C/O ratio can indeed boost the \ce{HCN} and \ce{C2H2}
line fluxes, but at the same time, one looses the \ce{CO2} emission
feature \citep{Woitke2018}, and a very fine-tuned C/O value would be
required to explain the ratios the emission line features of different
molecules. In contrast, what the observations tell us is that the
emission features of \ce{H2O}, \ce{CO2}, \ce{HCN} and \ce{C2H2} are
optically thick in most cases, with column densities in excess of
$10^{15}\rm\,cm^{-2}$, which causes the flux levels of these features
to be on a similar level in a more robust way. We have therefore
carefully analysed our {\sc ProDiMo} models to see how we can possibly
create larger amounts of optically thick \ce{HCN} and \ce{C2H2}. These
efforts resulted in a number of important code changes that are
reported in this section. They are essential for fitting the first
JWST spectrum as described in Sect.~\ref{sec:EXLup}.

\subsection{A new escape probability theory for discs}
\label{sec:newProDiMo}

Many published line fluxes from models of protoplanetary discs rely on
the concept of escape probability, for example \cite{Nomura2007}.
Different variants of the escape probability formalism have been
developed \cite{Woods2009,Woitke2009,Du2014}.  In {\sc ProDiMo}, while
we normally perform formal solutions of the 3D line transfer problem
to calculate the line fluxes, an escape probability method is still
required to compute the non-LTE populations and the heating/cooling
rates.  The previously used escape probability method in {\sc ProDiMo}
is based on Eq.\,(73) in \citet{Woitke2009}, which is a rather crude
approximation as it only considers radiative pumping by continuum
photons in the radial direction, and line photon escape into the
upward vertical direction.  In particular, the IR-pumping from the
disc below is incorrectly attenuated by the radial line optical depth
$\tau_{ul}^{\rm rad}$, which can be huge, i.e.\ we are likely
underestimating the true IR pumping. In addition, in the outer
optically thin disc regions, the escape probability is currently
limited to $\Pesc\!\leq\!\frac{1}{2}$, whereas it should be
$\Pesc\!\to\!1$ in the optically thin limit.

This paper introduces a new idea how to calculate the line averaged
mean intensity $\Jbar$ by taking into account continuum radiative
transfer effects in the line resonance region
\begin{equation}
  \Jbar = \frac{1}{4\pi}
  \int\!\!\!\int \phi_\nu\,I_\nu(\vec{n})\;d\Omega\,d\nu \ ,
  \label{eq:Jbar}
\end{equation}  
where $\phi_\nu$ is the line profile function associated with the
spectral line $u\to l$ and $I_\nu$ is the spectral intensity. Finding
$\Jbar$ is the key to formulate the effective non-LTE rates and
heating/cooling functions associated with line absorption and
emission, see Eqs.\,(80),\,(81),\,(85)\,and\,(86) in
\citet{Woitke2009}. Here we will show that, by making some assumptions
about disc geometry and locality of radiative transfer quantities
explained below, it is possible to arrive at a final expression that
reads again
\begin{equation}
  \Jbar = \Ppump \Jcont + (1-\Pesc) \SL
  \label{eq:escpro}
\end{equation}  
but now with new pumping and escape probabilities, $\Ppump$ and
$\Pesc$, that depend not only on line optical depths, but also on
continuum optical depths, and take into account the pumping from and
the escape into all directions. $\SL$ is the line source function and
$\Jcont$ the continuous mean intensity that would result if the line
has zero opacity. We write the continuum and line radiative transfer
equation as
\begin{equation}
  \frac{dI_\nu}{ds} = -(\kapC + \kapL\phi_\nu)\,I_\nu
                 \,+\, \kapC \SC \,+\, \kapL\phi_\nu \SL \ ,
  \label{eq:RT}
\end{equation}
where the continuum, absorption and scattering opacities $\rm[cm^{-1}]$,
and the continuum source function, assuming isotropic scattering, are
given by
\begin{eqnarray}
       \kapC &\!\!\!=\!\!\!& \kabs + \ksca \ ,\\
  \kapC\,\SC &\!\!\!=\!\!\!& \kabs B_\nu(\Td) + \ksca \Jcont \ .
\end{eqnarray}
The line opacity $\rm[cm^{-1}\,Hz]$, Gaussian line profile function
$\rm[Hz^{-1}]$, and line source function
$\rm[erg\,cm^{-2}\,s^{-1}\,Hz^{-1}\,sr^{-1}]$ are
\begin{eqnarray}
  \kapL &\!\!\!=\!\!\!& \frac{h\nu_{ul}}{4\pi}
            \big(n_l B_{lu} - n_u B_{ul}\big) \ ,\\
  \phi_\nu &\!\!\!=\!\!\!& \frac{1}{\sqrt{\pi}\Delta\nuD}
     \exp\left(-\left(\frac{\nu-\nu_{ul}}{\Delta\nuD}\right)^2\right) \ ,\\
  \SL &\!\!\!=\!\!\!& \frac{2h\nu_{ul}^3}{c^2}
     \left(\frac{g_u\,n_l}{g_l\,n_u}-1\right)^{-1} \ ,
\end{eqnarray}
where $\nu_{ul}$ is the line centre frequency, $\nuD=\nu_{ul}\,\Delta
v/c$ the Doppler width and $\Delta v=\big(v_{\rm th}^2+v_{\rm
    turb}^2\big)^{1/2}$ the line width in velocity space. Introducing the
optical depth and maximum optical depth as
\begin{eqnarray}
  t_\nu(s) &\!\!\!=\!\!\!& \int_0^s \kapC + \kapL\phi_\nu \,ds \ ,\\
  \tau_\nu &\!\!\!=\!\!\!& t_\nu(\smax) \ ,
\end{eqnarray}
the formal solution of Eq.\,(\ref{eq:RT}) is
\begin{equation}
  I_\nu(\vec{n}) = I_\nu^{\,0}(\vec{n})\;{\rm e}^{-\tau_\nu}
      + \int_0^{\smax}\!\!\!\left(\kapC \SC + \kapL \phi_\nu \SL\right)
      \,{\rm e}^{-t_\nu} \,ds  \ ,
      \label{eq:formSol}
\end{equation}
where $\vec{n}$ is the direction of the ray (unit vector) and $\smax$
is the distance backward along the ray up to the point where the ray
enters the model volume. $I_\nu^{\,0}(\vec{n})$ is the incident intensity
at that point in direction $\vec{n}$.

\paragraph{Assumption 1 (locality):} We assume that all
radiative transfer quantities, that is $\kapC$, $\kapL$, $\SL$, $\SC$
and $\phi_\nu$, are independent of location and direction, and given by
their local values.  Assuming $\SL\!=\rm\!const$ is standard in escape
probability theory, because the line photons typically interact only
within a small ``resonance volume'' around the point of interest.
Generalising this concept to the continuum transfer properties is a new
idea (however, see \citealt{Hummer1985}).

This way, we can capture the most relevant continuum radiative
transfer effects in that resonance region, where there is a
competition between line and continuum absorption and emission.
Constant $\phi_\nu$ also means that we neglect line velocity shifts,
i.e.\ we assume a static medium. The only quantities that are allowed
to depend on direction are $I_\nu^{\,0}(\vec{n})$ and $\smax$ (and hence
$\tau_\nu$).  The formal solution of the RT equation
(\ref{eq:formSol}) then simplifies to
\begin{equation}
  I_\nu(\vec{n}) = I_\nu^{\,0}(\vec{n})\;{\rm e}^{-\tauC-\tauL\phi_\nu}
  \,+\, \frac{\tauC \SC + \tauL \phi_\nu \SL}{\tauC + \tauL\phi_\nu}
  \left(1-\,{\rm e}^{-\tauC-\tauL\phi_\nu}\right)  \ ,
  \label{eq:formSol1}
\end{equation}
where we have used the continuum and line centre optical depths
$\tauC\!=\!\kapC\smax$ and $\tauL\!=\!\kapL\smax$.

\begin{figure*}
  \centering
  \begin{tabular}{cc}
    {\sf\hspace*{2mm} old implementation} &
    {\sf\hspace*{2mm} new implementation}\\
    \includegraphics[height=62mm,width=75mm,trim=0 0 700 0,clip]
                    {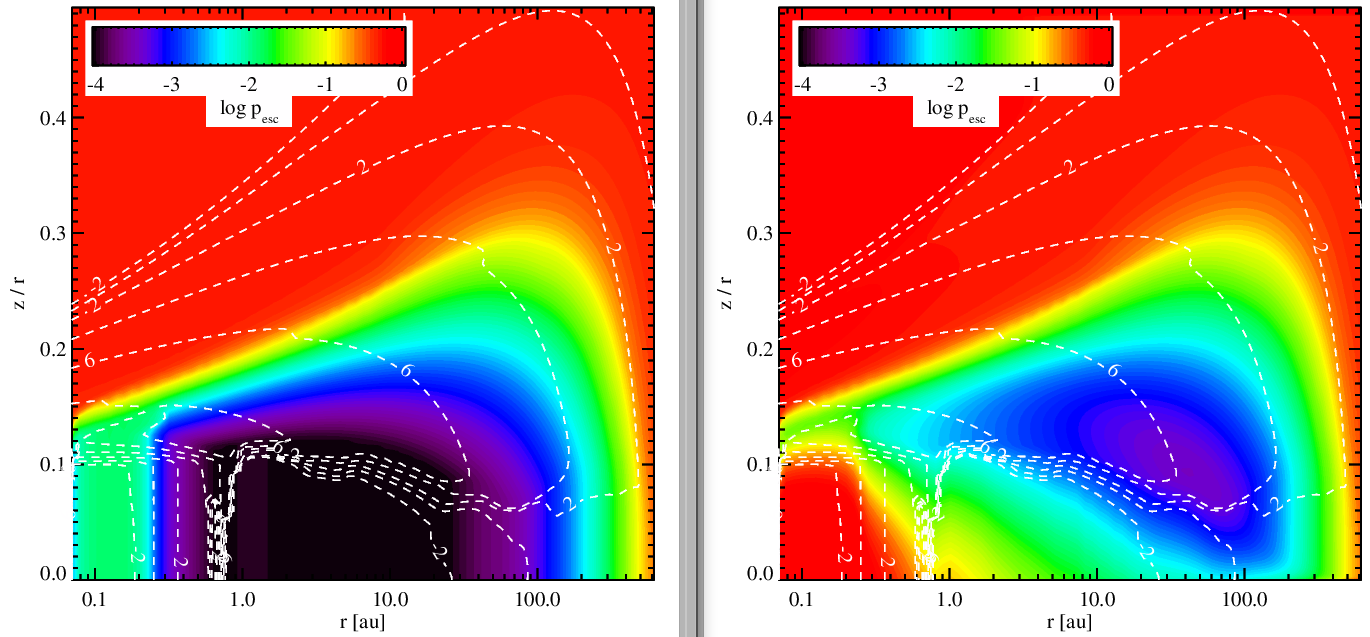} &
    \includegraphics[height=62mm,width=75mm,trim=715 0 0 0,clip]
                    {Figures/OI63_pesc.png}\\
    \includegraphics[height=62mm,width=75mm,trim=0 0 700 0,clip]
                    {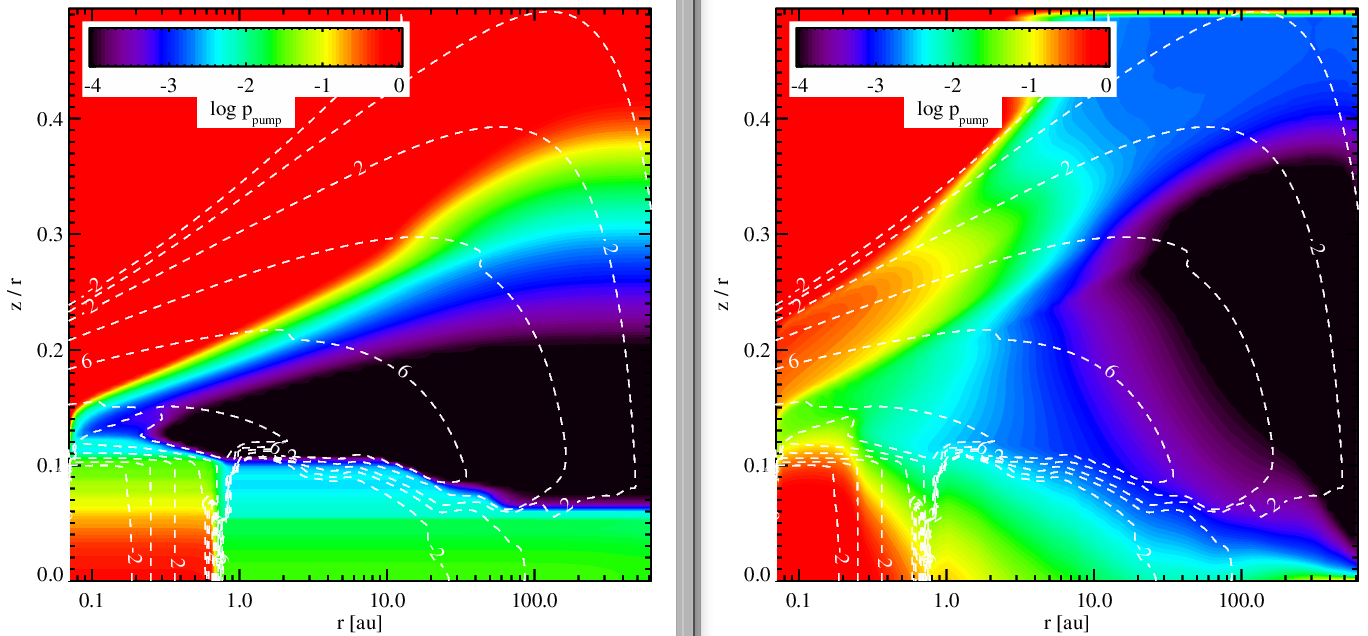} &
    \includegraphics[height=62mm,width=75mm,trim=715 0 0 0,clip]
                    {Figures/OI63_pump.png}
  \end{tabular}
  \caption{Escape probability ($\Pesc$, top) and continuum pumping
    probability ($\Ppump$, bottom) of the [OI]\,63\,$\mu$m line in the
    {\sc ProDiMo} standard T\,Tauri model. Left and right plots show the
    results from the previous and the new escape probability theory. The
    white dashed contours show the particle density of atomic oxygen
    $\log_{10} n_{\rm OI}\rm[\,cm^{-3}]=\{8,6,4,2,0,-2\}$.}
  \label{fig:newescape}
\end{figure*}

\paragraph{Assumption 2 (disc geometry):} We consider
three major directions: radial (the rays coming from the stellar
surface -- often important for pumping), vertically upwards and
vertically downwards. At any given location in the disc, the star occupies
a small solid angle $\Omega_\star$.  The other two principle directions
cover half of $\Omega_{\rm d} = 4\pi-\Omega_\star$ each.  All rays
coming from the stellar surface are represented by a single radial
ray.  For the upward and downward directions, we approximate the disc
as being plane-parallel with $\mu=\cos(\theta)$ where $\theta$ is the
angle with the vertical.  Thus we find from Eq.\,(\ref{eq:Jbar})
\begin{eqnarray}
  \Jbar &\!\!\!=\!\!\!& \frac{\Omega_\star}{4\pi}
                \int \phi_\nu\,I_\nu^{\rm rad}\,d\nu
           \,+\,\frac{\Omega_{\rm d}}{8\pi}
                \int \phi_\nu \int_0^1   I_\nu(\mu)\,d\mu\,d\nu \nonumber\\
        & &\hspace{24mm}\,+\,\frac{\Omega_{\rm d}}{8\pi}
                \int \phi_\nu \int_{-1}^{\,0} \,I_\nu(\mu)\,d\mu\,d\nu \ .
  \label{eq:Jbar1}         
\end{eqnarray}
Introducing the vertically upwards and downwards (across the disc)
line and continuum optical depths, $\tauLup$, $\tauLdown$, $\tauCup$
and $\tauCdown$, respectively, and the radially inward line and
continuum optical depths, $\tauLrad$ and $\tauCrad$, and using
plane-parallel geometry $\tau(\mu)=\tau(\mu\!=\!0)/\mu$ we obtain from
Eqs.\,(\ref{eq:formSol1}) and (\ref{eq:Jbar1})
\begin{eqnarray}
  \Jbar &\!\!\!=\!\!\!& \frac{\Omega_\star}{4\pi} \int \phi_\nu\,
  \bigg( I_\nu^\star\;{\rm e}^{-\tauCrad-\tauLrad\phi_\nu}
  \\
  &\!\!\!\!\!\!& \hspace*{16mm}
  + \frac{\tauCrad \SC + \tauLrad \phi_\nu \SL}{\tauCrad + \tauLrad\phi_\nu}
  \left(1-\,{\rm e}^{-\tauCrad-\tauLrad\phi_\nu}\right)\bigg)\,d\nu
  \nonumber\\
  &\!\!\!+\!\!\!&
  \frac{\Omega_{\rm d}}{8\pi} \int \phi_\nu
  \int_0^1 I_\nu^{\rm ISM}\;{\rm e}^{-\frac{\tauCup}{\mu}-\frac{\tauLup}{\mu}\phi_\nu}
  \nonumber\\
  &\!\!\!\!\!\!& \hspace*{19mm}
  + \frac{\tauCup \SC + \tauLup \phi_\nu \SL}{\tauCup + \tauLup\phi_\nu}
  \left(1-\,{\rm e}^{-\frac{\tauCup}{\mu}-\frac{\tauLup}{\mu}\phi_\nu}\right)d\mu\;d\nu 
  \nonumber\\
  &\!\!\!+\!\!\!&
  \frac{\Omega_{\rm d}}{8\pi} \int \phi_\nu
  \int_{-1}^0 I_\nu^{\rm ISM}\;{\rm e}^{-\frac{\tauCdown}{|\mu|}-\frac{\tauLdown}{|\mu|}\phi_\nu}
  \nonumber\\
  &\!\!\!\!\!\!& \hspace*{19mm}
  + \frac{\tauCdown \SC + \tauLdown \phi_\nu \SL}{\tauCdown + \tauLdown\phi_\nu}
  \left(1-\,{\rm
    e}^{-\frac{\tauCdown}{|\mu|}-\frac{\tauLdown}{|\mu|}\phi_\nu}\right)d\mu\;d\nu
  \nonumber\ ,
\end{eqnarray}
where $I_\nu^\star$ and $I_\nu^{\rm ISM}$ are the stellar and
interstellar incident intensities, respectively. Introducing the
second exponential integral function $E_2(\tau)=\int_0^1
\exp\big(-\frac{\tau}{\mu}\big)\,d\mu$ we find
\begin{eqnarray}
  \Jbar &\!\!\!=\!\!\!& \frac{\Omega_\star}{4\pi} \int \phi_\nu\,
  \bigg( I_\nu^\star\;{\rm e}^{-\tauCrad-\tauLrad\phi_\nu}
  \\
  &\!\!\!\!\!\!& \hspace*{15mm}
  + \frac{\tauCrad \SC + \tauLrad \phi_\nu \SL}{\tauCrad + \tauLrad\phi_\nu}
  \left(1-\,{\rm e}^{-\tauCrad-\tauLrad\phi_\nu}\right)\bigg)\,d\nu
  \nonumber\\
  &\!\!\!+\!\!\!& \frac{\Omega_{\rm d}}{8\pi} \int \phi_\nu\,
  \bigg(I_\nu^{\rm ISM}\,E_2(\tauCup+\tauLup\phi_\nu)
  \nonumber\\
  &\!\!\!\!\!\!& \hspace*{15mm}  
  + \frac{\tauCup \SC + \tauLup \phi_\nu \SL}{\tauCup + \tauLup\phi_\nu}
  \left(1-E_2(\tauCup+\tauLup\phi_\nu)\right)\bigg)\,d\nu
  \nonumber\\
  &\!\!\!+\!\!\!& \frac{\Omega_{\rm d}}{8\pi} \int \phi_\nu\,
  \bigg(I_\nu^{\rm ISM}\,E_2(\tauCdown+\tauLdown\phi_\nu)
  \nonumber\\
  &\!\!\!\!\!\!& \hspace*{15mm}    
  + \frac{\tauCdown \SC + \tauLdown \phi_\nu \SL}{\tauCdown + \tauLdown\phi_\nu}
  \left(1-E_2(\tauCdown+\tauLdown\phi_\nu)\right)\bigg)\,d\nu
  \nonumber\ .
\end{eqnarray}
We now introduce the dimensionless profile function
$\phi(x)=\exp(-x^2)/\sqrt{\pi}$ with $x=(\nu-\nu_{ul})/\Delta\nuD$ and
six basic 2D-functions, which all produce values between 0 and 1, to
further simplify the result
\begin{eqnarray}
  \alpha_0(\tauL,\tauC) &\!\!\!=\!\!\!\!\!&
  \int\! \phi(x)\,{\rm e}^{-\tauC-\tauL\phi(x)}\,dx \ , \\
  \alpha_1(\tauL,\tauC) &\!\!\!=\!\!\!\!\!&
  \int\! \phi(x)\,\frac{\tauC}{\tauC+\tauL\phi(x)}
              \,\big(1-\,{\rm e}^{-\tauC-\tauL\phi(x)}\big)\,dx \ , \\
  \alpha_2(\tauL,\tauC) &\!\!\!=\!\!\!\!\!&
  \int\! \phi(x)\,\frac{\tauL\phi(x)}{\tauC + \tauL\phi(x)}
              \,\big(1-\,{\rm e}^{-\tauC-\tauL\phi(x)}\big)\,dx \ , \\
  \beta_0(\tauL,\tauC) &\!\!\!=\!\!\!\!\!&
  \int\! \phi(x)\,E_2\big(\tauC+\tauL\phi(x)\big)\,dx \ , \\
  \beta_1(\tauL,\tauC) &\!\!\!=\!\!\!\!\!&
  \int\! \phi(x)\,\frac{\tauC}{\tauC+\tauL\phi(x)}
              \,\Big(1-E_2\big(\tauC+\tauL\phi(x)\big)\Big)\,dx \ , \\
  \beta_2(\tauL,\tauC) &\!\!\!=\!\!\!\!\!&
  \int\! \phi(x)\,\frac{\tauL\phi(x)}{\tauC + \tauL\phi(x)}
              \,\Big(1-E_2\big(\tauC+\tauL\phi(x)\big)\Big)\,dx \ , 
\end{eqnarray}
where, using the Einstein relations,
\begin{eqnarray}
  \tauL &\!\!\!=\!\!\!& \int \frac{\kapL}{\Delta\nuD}\,ds
        = \int \frac{h\nu_{ul}}{4\pi\,\Delta\nuD}
               \big(n_l B_{lu} - n_u B_{ul}\big)\,ds \nonumber\\
        &\!\!\!=\!\!\!& \int \frac{A_{ul}}{8\pi\,\Delta v}
               \Big(\frac{c}{\nu_{ul}}\Big)^3
               \big(n_l \frac{g_u}{g_l} - n_u\big)\,ds \ .
\end{eqnarray}  
The results are
\begin{eqnarray}
  \Jbar
  &\!\!\!=\!\!\!& \frac{\Omega_\star}{4\pi} 
  \Big( I_\nu^\star\,\alpha_0(\tauLrad,\tauCrad)
  + \SC\,\alpha_1(\tauLrad,\tauCrad)
  + \SL\,\alpha_2(\tauLrad,\tauCrad)\Big)
  \nonumber\\
  &\!\!\!+\!\!\!& \frac{\Omega_{\rm d}}{8\pi} 
  \Big( I_\nu^{\rm ISM}\,\beta_0(\tauLup,\tauCup)
  + \SC\,\beta_1(\tauLup,\tauCup)
  + \SL\,\beta_2(\tauLup,\tauCup)\Big)
  \nonumber\\
  &\!\!\!+\!\!\!& \frac{\Omega_{\rm d}}{8\pi} 
  \Big( I_\nu^{\rm ISM}\,\beta_0(\tauLdown,\tauCdown)
  + \SC\,\beta_1(\tauLdown,\tauCdown)
  + \SL\,\beta_2(\tauLdown,\tauCdown)\Big) \ .
  \label{eq:solution}
\end{eqnarray}  
From this geometric model we also get a prediction of the continuum,
i.e.\ the mean intensity at line centre frequency as would be present
if the line opacity was zero 
\begin{eqnarray}
  \Jcont
  &\!\!\!=\!\!\!& \frac{\Omega_\star}{4\pi} 
  \Big( I_\nu^\star\,\alpha_0(0,\tauCrad)
  + \SC\,\alpha_1(0,\tauCrad)\Big)
  \label{eq:Jcont}\\
  &\!\!\!+\!\!\!& \frac{\Omega_{\rm d}}{8\pi}\, 
  \Big( I_\nu^{\rm ISM}\,\beta_0(0,\tauCup)
  + \SC\,\beta_1(0,\tauCup)\Big)
  \nonumber\\
  &\!\!\!+\!\!\!& \frac{\Omega_{\rm d}}{8\pi}\, 
  \Big( I_\nu^{\rm ISM}\,\beta_0(0,\tauCdown)
  + \SC\,\beta_1(0,\tauCdown)\Big) \ .
  \nonumber 
\end{eqnarray}  
Equation~(\ref{eq:Jcont}) is used to determine $\SC$. This means that
we can calibrate $\SC$ in such a way that our simple 3-way RT model
with constant continuum quantities results in the correct $\Jcont$ as
known from the proper solution of the continuum radiative transfer
problem.  This is a great advantage as it allows us to eliminate most
the principle problems arising from that simplified 3-way RT model in
the continuum.

Comparing Eq.\,(\ref{eq:solution}) to Eq.\,(\ref{eq:escpro}) we find
the definitions of our new pumping and escape probabilities
\begin{eqnarray}
  \Ppump
  &\!\!\!\!\!=\!\!\!\!& \frac{1}{\Jcont}
  \Bigg(\frac{\Omega_\star}{4\pi}
  \bigg[I_\nu^\star\,\alpha_0(\tauLrad,\tauCrad)
    +\SC\,\alpha_1(\tauLrad,\tauCrad)\bigg]
  \\
  &\!\!\!\!\!\!\!\!&\hspace*{7mm}
  + \frac{\Omega_{\rm d}}{8\pi} \bigg[ I_\nu^{\rm ISM}
  \left(\beta_0(\tauLup,\tauCup)+\beta_0(\tauLdown,\tauCdown)\right)
  \nonumber\\
  &\!\!\!\!\!\!\!\!&\hspace*{15mm}
  +\,\SC
  \left(\beta_1(\tauLup,\tauCup)+\beta_1(\tauLdown,\tauCdown)\right)\bigg]
  \Bigg)\ ,
  \nonumber\\
  \Pesc
  &\!\!\!\!\!=\!\!\!\!& 1 - \frac{\Omega_\star}{4\pi} \alpha_2(\tauLrad,\tauCrad)
  - \frac{\Omega_{\rm d}}{8\pi}
  \Big(\beta_2(\tauLup,\tauCup)+\beta_2(\tauLdown,\tauCdown)\Big) \ .
\end{eqnarray}
The interpretation of these results is as follows. $\Ppump$ is the
{\it probability that continuum photons make it to the considered
  point}, given the presence of line opacity.  The average is taken
over all continuum photons that would contribute to $\Jcont$ when the
line was not existing, whether they have been emitted by the star or
by any point in the disc, averaged over the local line profile
function. $\Pesc$ is the {\it probability that line photons emitted
  anywhere in the disc, in the direction of the considered point, do
  not make it to the considered point}, averaged over the local line
profile function. $\Pesc<1$ means that these line photons are either
re-absorbed by line opacity, or re-absorbed/scattered by continuum
opacity along the path to the considered point.

\begin{figure*}
  \centering
  \begin{tabular}{cc}
    {\sf\hspace*{10mm} old implementation} &
    {\sf\hspace*{10mm} new implementation}\\
    \includegraphics[height=94mm,width=85mm,trim=20 0 30 0,clip]
                    {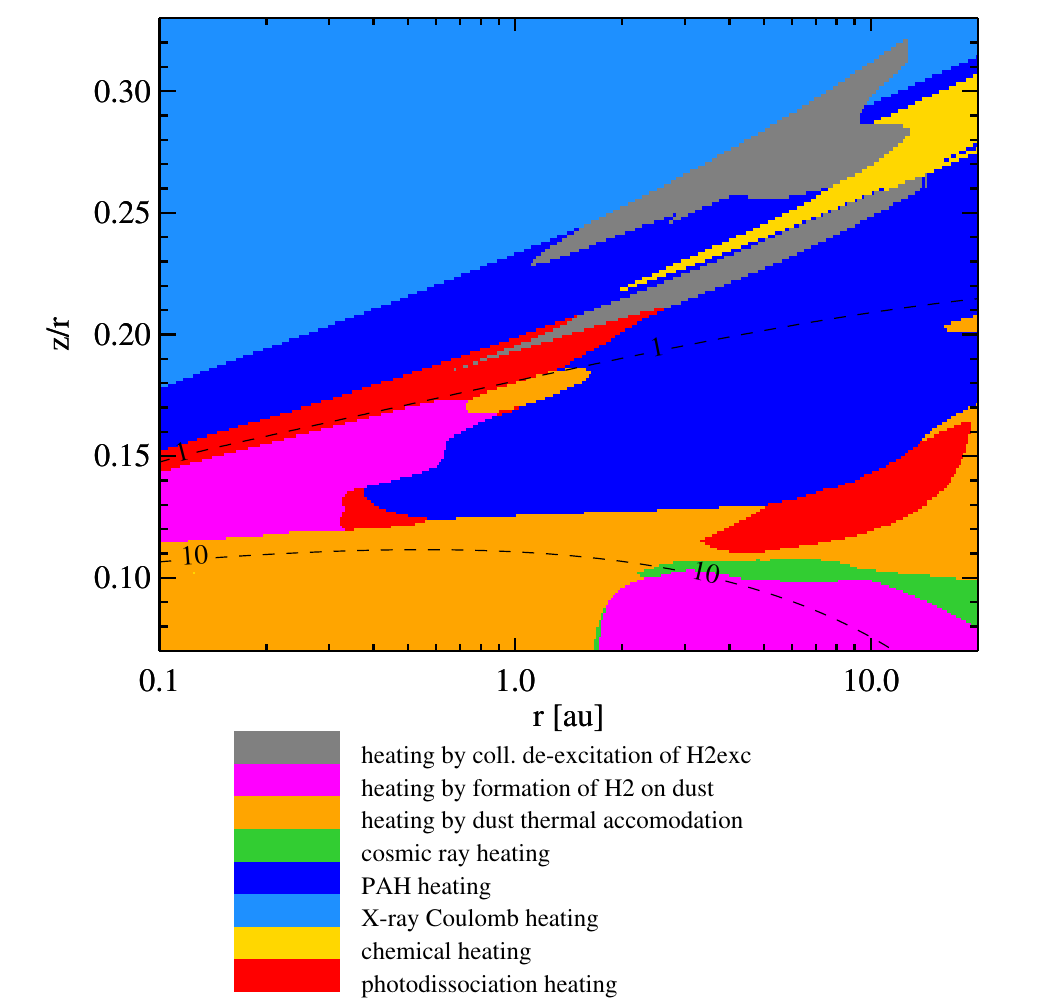} &
    \includegraphics[height=94mm,width=85mm,trim=20 0 30 0,clip]
                    {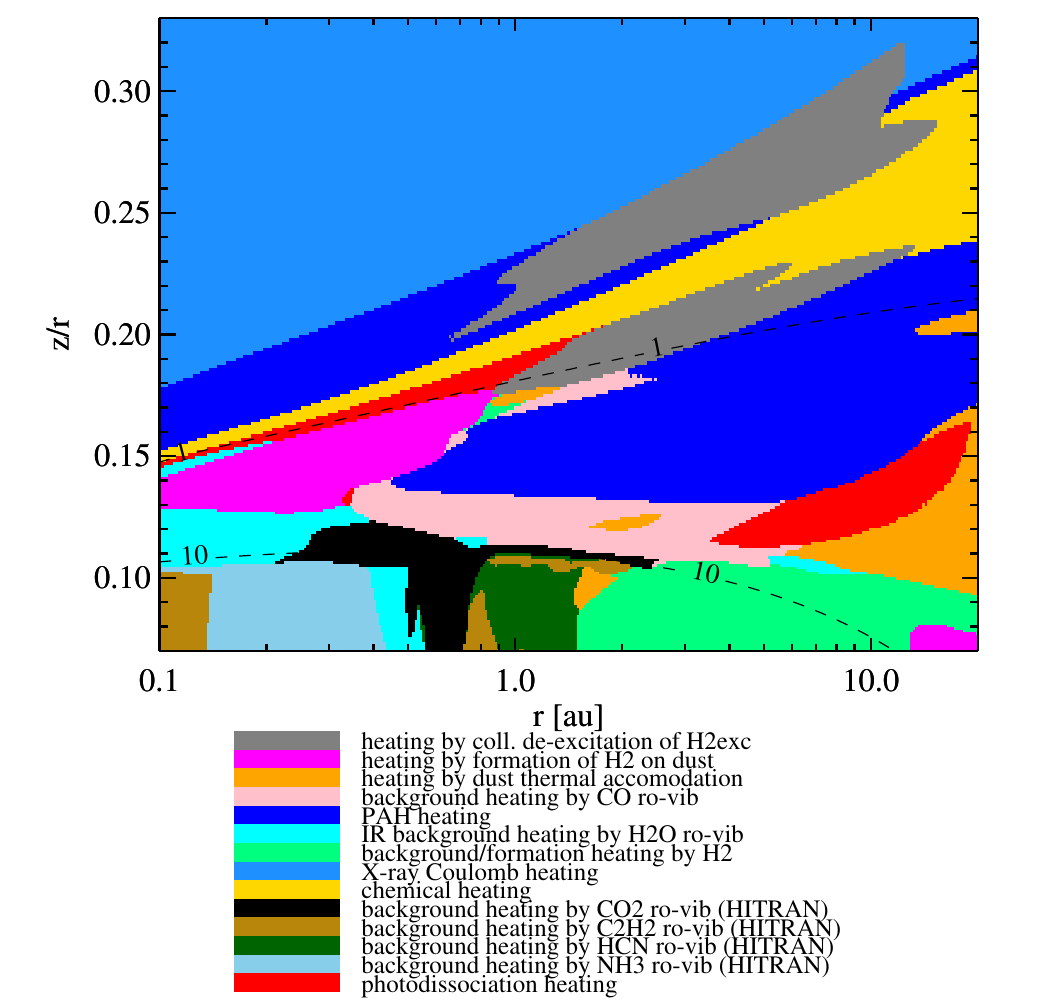} \\
    \includegraphics[height=94mm,width=85mm,trim=20 0 30 0,clip]
                    {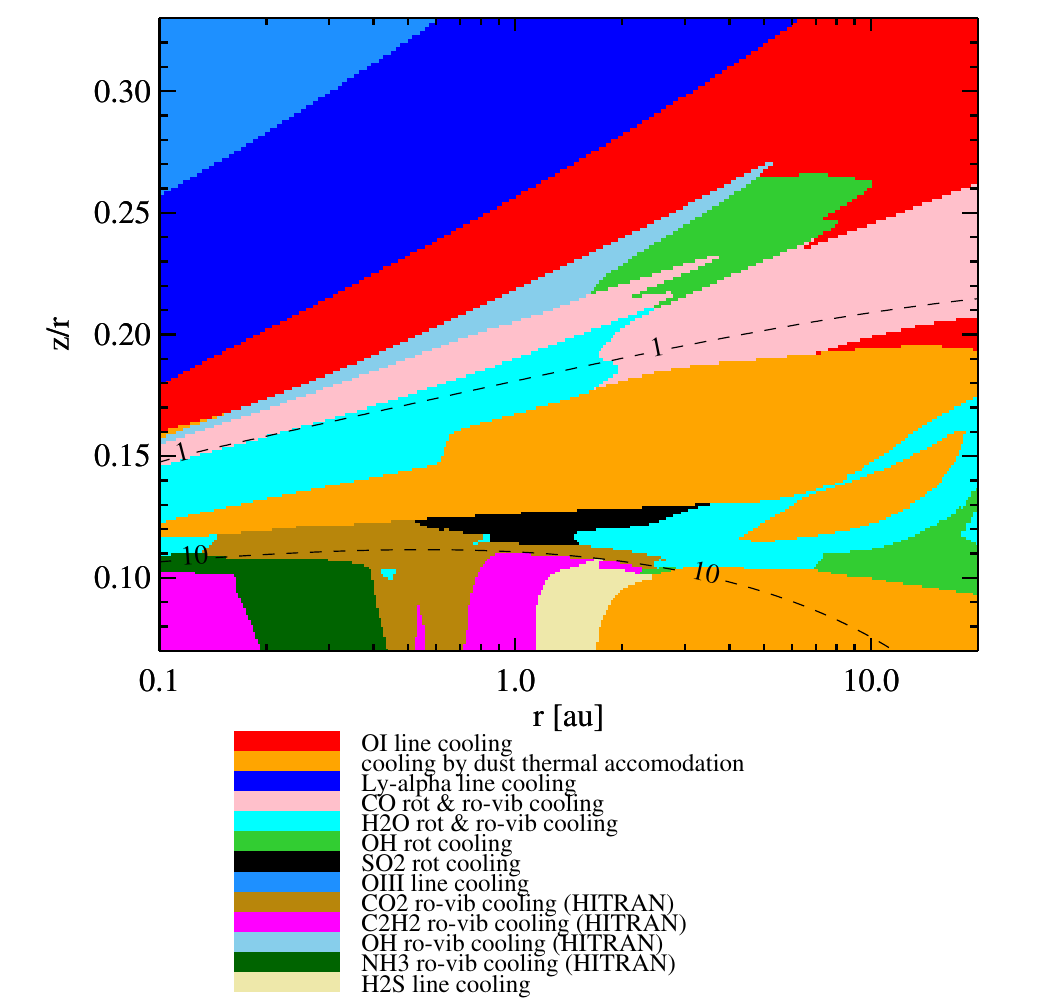} &
    \includegraphics[height=94mm,width=85mm,trim=20 0 30 0,clip]
                    {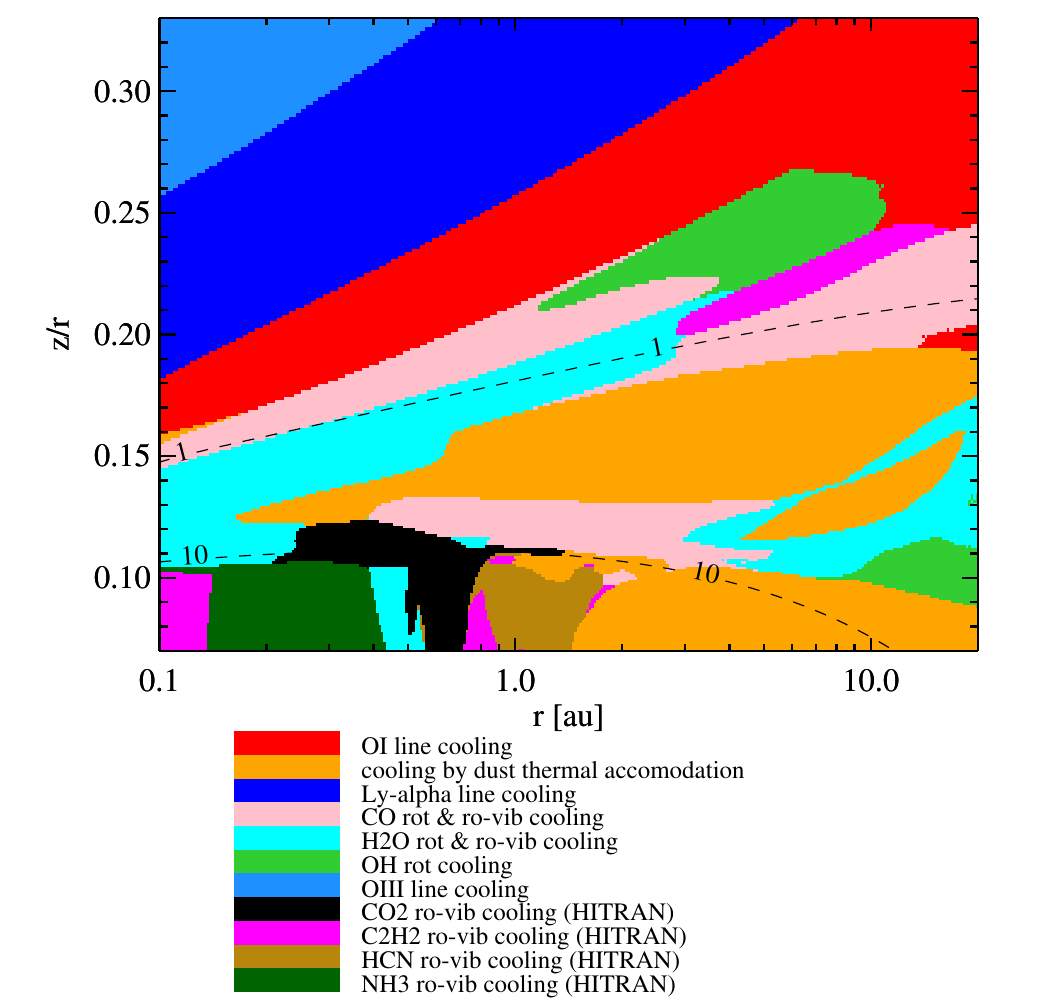} 
  \end{tabular}
  \caption{Change of heating/cooling balance with the new escape
    probability treatment and the new 2D concept to apply UV molecular
    shielding factors. The plots show the leading heating process
    (top) and the leading cooling process (bottom) in a zoomed-in disc
    region responsible for the disc IR line emissions. Processes are
    only plotted when they fill at least 1\% of the area in these
    plots. The dashed lines mark the visual disc surface $A_{V,\rm
      rad}\!=\!1$ and the vertical optical extinction $A_V\!=\!10$.
    The model includes 103 heating and 95 cooling processes with
    altogether 64910 spectral lines.}
  \label{fig:heatcool}
\end{figure*}

Since the order of photon emission \& re-absorption and the optical
depths effects along the photon paths can be reversed, $\Pesc$ can
also be interpreted as {\it the probability of a line photon emitted
  from the considered point not to have another line interaction in
  the disc}. It is important to realise the emphasis on further line
interaction here.  $\Pesc$ is {\bf not} the probability of line
photons to escape the disc! That interpretation is only valid when
there are no continuum opacities.  Continuum opacity will prevent
further line interaction, so it increases $\Pesc$.  The latter could
be interpreted as ``intrinsic escape'', meaning that, once a line
photon is absorbed and re-emitted by the dust, that continuum photon
will escape the disc somehow, but this is actually not what the
equations tell us.  In particular, in the optically thick limiting
case with $\tauL\ll\tauC$, the result is $\Pesc\!\to\!1$ and
$\Ppump\!\to\!1$.

The six basic 2D-functions $\alpha_0(\tauL,\tauC)\,...\,
\beta_2(\tauL,\tauC)$ are pre-calculated in form of 2D numerical
tables.  $I_\nu^\star$ and $I_\nu^{\rm ISM}$ are known functions,
and $\tauCrad$, $\tauCup$ and $\tauCdown$ are available after the
initialisation of the disc structure and the dust opacities in
{\sc ProDiMo}. Since $\Jcont$ is available from the proper solution of the
continuum transfer problem, we can use Eq.\,(\ref{eq:Jcont}) to
determine and pre-calculate $\SC$ for every line at any point in the
disc.  The centre line optical depths $\tauLrad$ and $\tauLup$ are
integrated during the downward/outward sweep of the chemistry and
heating/cooling balance in {\sc ProDiMo}, but we also need $\tauLdown$ here,
which requires another assumption.

\paragraph{Assumption 3 (downward line optical depths):}
We calculate $\tauLdown$ by assuming that the local line/continuum
opacity ratio remains the same on the way down and across the disc
\begin{equation}
  \tauLdown ~=~ \tauCdown\;\frac{\kapL}{\kapC} \ ,
\end{equation}
where $\kapL$ and $\kapC$ are the local line and continuum opacities.
The old escape probability method in {\sc ProDiMo} was based on
\begin{align}
  \Ppumpold &~=~ \int \phi(x)\,{\rm e}^{-\tauLrad\phi(x)}\,dx \ , \\
  \Pescold  &~=~ \frac{1}{2} \int \phi(x)\,E_2\big(\tauLup\phi(x)\big)\,dx \ .
\end{align}
As can be easily shown, $\Ppump=\Ppumpold$ is only valid when the
direct stellar illumination dominates, i.e.\ $\Jcont =
\frac{\Omega_\star}{4\pi} I_\nu^\star\,{\rm e}^{-\tauCrad}$. In all
other cases, the new $\Ppump$ probabilities are larger.
$\Pesc=\Pescold$ is only valid when $\Omega_\star\to 0$,
$\tauCup\!=\!0$ and $\tauCdown\to\infty$. Under any other, more realistic
circumstances, the new $\Pesc$ probabilities are also generally
larger. Both effects are caused by an inclusion of line-continuum
interactions, namely (i) locally produced continuum photons can be
absorbed in the line, and (ii) line photons can locally be absorbed in
the continuum. We have used both limiting cases discussed above to
check the numerical implementation of the new escape probability
method in {\sc ProDiMo}.

Figure \ref{fig:newescape} shows a comparison between old and new
escape and pumping probabilities {based on our T\,Tauri standard
  disc model. This model was introduced by \citet{Woitke2016} and has
  stellar, dust and disc parameters as listed in
  Table~\ref{tab:parameter_standard}.}  The results of the new
treatment show larger $\Ppump$ and $\Pesc$ in general. The deviations
are particularly obvious in the disc midplane close to the star, where
the continuum is optically thick. We see large and about equal
$\Ppump$ and $\Pesc$ here, in contrast to the old treatment.  More
significant for line observations are the changes in the more
transparent upper disc regions, where $\Pesc\!\sim\!\Pescold$ and
$\Ppump\!>\!\Ppumpold$ in the inner disc. But in the outer disc, this
relation turns around and we find $\Ppump\!<\!\Ppumpold$. The reasons
for these changes are more complex as $\Jcont$ is not given by the
attenuated stellar irradiation, but is created mostly by continuum
emission from the disc below.

Figure \ref{fig:heatcool} shows the resulting leading heating and cooling
processes in the innermost 10\,AU of the disc (T\,Tauri standard model).
{Figure \ref{fig:Qheatcool} shows additional vertical cuts at
  selected radii with more details.}
There are two significant changes:

(i) In the disc surface regions inside of 1\,au, which are borderline
optically thin and relevant for the IR line emission, we find increased
H$_2$O heating and cooling rates, whereas previously there was a
balance between UV driven heating by H$_2$ formation versus H$_2$O
cooling. This is due to the increased pumping and escape probabilities
in these regions.

(ii) Thermal accommodation in the midplane is now quite unimportant for
both heating and cooling, unless the midplane becomes virtually
molecule-free due to ice formation. In the optically thick midplane,
dust and gas now exchange energy very efficiently via line photons,
more efficiently than by inelastic collisions. And since the
formulation of the line heating/cooling in {\sc ProDiMo} is robust including
all reverse processes and stimulated emission, $\Jcont=B_\nu(\Td)$
implies that a balance between line heating and cooling can only be
achieved when $\Tg=\Td$.

\subsection{New escape probabilities for line flux estimations}
\label{sec:escprofluxes}

\begin{figure}[!t]
  \includegraphics[width=90mm]{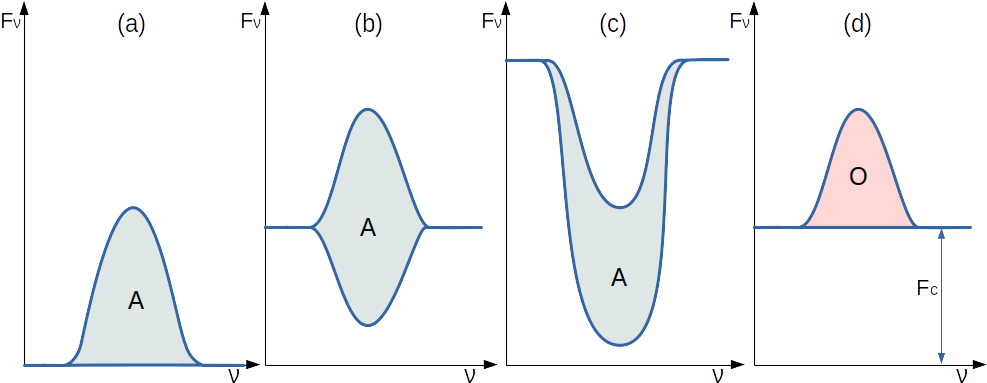}\\[-3mm]
  \caption{Effect of continuum emission on line flux. Case (a)
    shows an emission line on a cold continuum $\SC=0$. Case (b) is an
    emission line on a warm continuum $\SL>\SC$, and case (c) is an
    absorption line where the continuum is hot $\SL<\SC$. The area
    {\bf A} is the flux of photon energy [$\rm erg/s/cm^2$] carried
    away by line photons escaping from the object (after continuum
    extinction and line self-absorption) and is the same in cases (a),
    (b) and (c).  The task, however, is to derive the area {\bf O},
    which is the line flux as measured by observers.  For a warm
    continuum, clearly ${\bf O<A}$, and for an absorption line ${\bf
    O}<0$ (although ${\bf A}>0$).}
  \label{fig:lineflux}
  \vspace*{-1mm}
\end{figure}

In {\sc ProDiMo}, the observable line fluxes are estimated by using the escape
probabilities, summing up the contributions from all spatial cells.
The following expression is used to calculate the luminosity [$\rm
  erg/s$] of photon energy escaping in form of line photons from a
single cell in the old model
\begin{equation}
  \Lcell = \Delta V\,n_{\rm u} A_{\rm ul}\,h\nu_{\rm ul}
           \int \phi(x)\,{\rm e}^{-\tauL\phi(x)-\tauC}\,dx \ ,
\end{equation}
where $\Delta V\rm\,[cm^{-3}]$ is the volume of a cell $k$ centred
around point grid point $(ix,iz)$.  The line flux $\rm[erg/s/cm^2]$ is
then calculated as
\begin{equation}
  \Fline = \frac{\sum_k \Lcell_k}{4\pi d^2} \ ,
\end{equation}
where $d$ is the distance to the object. We consider the vertical
escape here (face-on disc), so the line and continuum optical depths
are $\tauL\!=\!\tauLup$ and $\tauC\!=\!\tauCup$.  Using the definition of the
line emission coefficient, we have $n_{\rm u} A_{\rm ul} h\nu_{\rm ul}
= 4\pi \kapL \SL$, and we can express the line luminosity of a cell,
according to the old implementation, by
\begin{equation}
  \Lcell = 4\pi\,\Delta V\,\kapL \SL\,{\rm e}^{-\tauC}
  \int \phi(x)\,{\rm e}^{-\tauL\phi(x)}\,dx
  \label{eq:Lcellold}\ .
\end{equation}
This result corresponds to the case of a cold background $\SC=0$,
where the result is always positive, see area {\bf A} in
Fig.\,\ref{fig:lineflux}.  However, line and continuum 
interfere in more complicated ways than that. It is not a simple
superposition, as the sketch (d) in Fig~\ref{fig:lineflux} suggests,
because the escaping continuum photons are partly blocked by line
opacity.  This effect is visualised by the lower graph in case (b).

\begin{figure}[!t]
  \vspace*{-3mm}
  \centering
  \includegraphics[width=80mm]{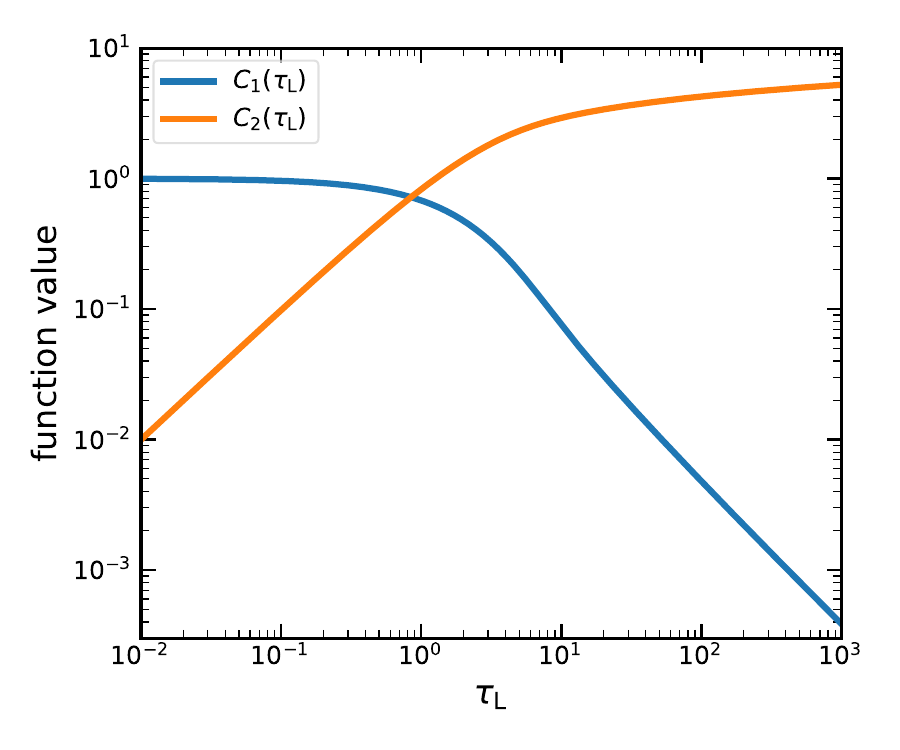}\\[-6mm]
  \caption{The functions $C_1(\tauL)$ and $C_2(\tauL)$ as introduced in
    Eq.\,(\ref{eq:Lcell}).}
  \label{fig:C1C2}
  \vspace*{-2mm}
\end{figure}

In a more detailed model, we need to compute the upper graph in case
(b) and the pure continuum, and then subtract the two to simulate the
way that line fluxes are measured from observations. Without continuum
we have area {\bf A} in Fig~\ref{fig:lineflux}
\begin{equation}
  \Lcell = 4\pi\,\Delta V \int \kapL \SL \phi_\nu\,
         {\rm e}^{-\tauL\phi(x)-\tauC}\,d\nu \ .
\end{equation}
With continuum, the (infinite) result is
\begin{equation}
  \Lcell = 4\pi\,\Delta V \int (\kapL \SL \phi_\nu+\kapC \SC)\,
         {\rm e}^{-\tauL\phi(x)-\tauC}\,d\nu \ .
  \label{eq:tmp1}
\end{equation}
Considering only the continuum, the (infinite) result is
\begin{equation}
  \Lcell = 4\pi\,\Delta V \int \kapC \SC\,
         {\rm e}^{-\tauC}\,d\nu \ .
  \label{eq:tmp2}\\
\end{equation}
The continuum-subtracted line flux (area {\bf O} in
Fig~\ref{fig:lineflux}) is
\begin{eqnarray}
  \Lcell &\!\!\!=\!\!\!& 4\pi\,\Delta V\,{\rm e}^{-\tauC}
  \label{eq:Lcell}\\
  &&\hspace*{-10mm} \times\left(\kapL \SL
  \underbrace{\int \phi(x)\,{\rm e}^{-\tauL\phi(x)}\,dx}_{C_1(\tauL)} 
  ~-~\kapC \SC \Delta\nu_{\rm D}
  \underbrace{\int \big(1-{\rm e}^{-\tauL\phi(x)}\big)\,dx}_{C_2(\tauL)}\right)
  \nonumber \ .
\end{eqnarray}
Equation~(\ref{eq:Lcell}) shows that there is an additional correction
term with respect to the old way of calculating the escaping line
luminosity from a cell (Eq.~\ref{eq:Lcellold}), which is always
negative.  This new term describes the loss of continuum flux due to
line absorption. The integration boundaries in these equations are
$-\infty$ and $+\infty$, so both intermediate results
(Eqs.\,\ref{eq:tmp1} and \ref{eq:tmp2}) are infinite, but the
subtraction of the two (Eq.~\ref{eq:Lcell}) is finite. The function
$C_1(\tauL)$ describes the probability of an outgoing line photon not
to be re-absorbed in the line again, and $C_2(\tauL)$ describes the
reduction of continuum photons escaping the object due to line
opacity. The two functions $C_1(\tauL)$ and $C_2(\tauL)$ are plotted
in Fig.~\ref{fig:C1C2}.

According to this new formulation of escape probability, the line flux
is {\it always smaller}\, when continuum emission effects are
included, and can become negative when the continuum source function
$\SC$ is as large as the line source function $\SL$. We note, however,
that the continuum is usually emitted from a different disc region than the
line.

\begin{figure*}
  \centering
  \vspace*{-4mm}
  \hspace*{-2mm}
  \includegraphics[width=175mm,height=70mm]{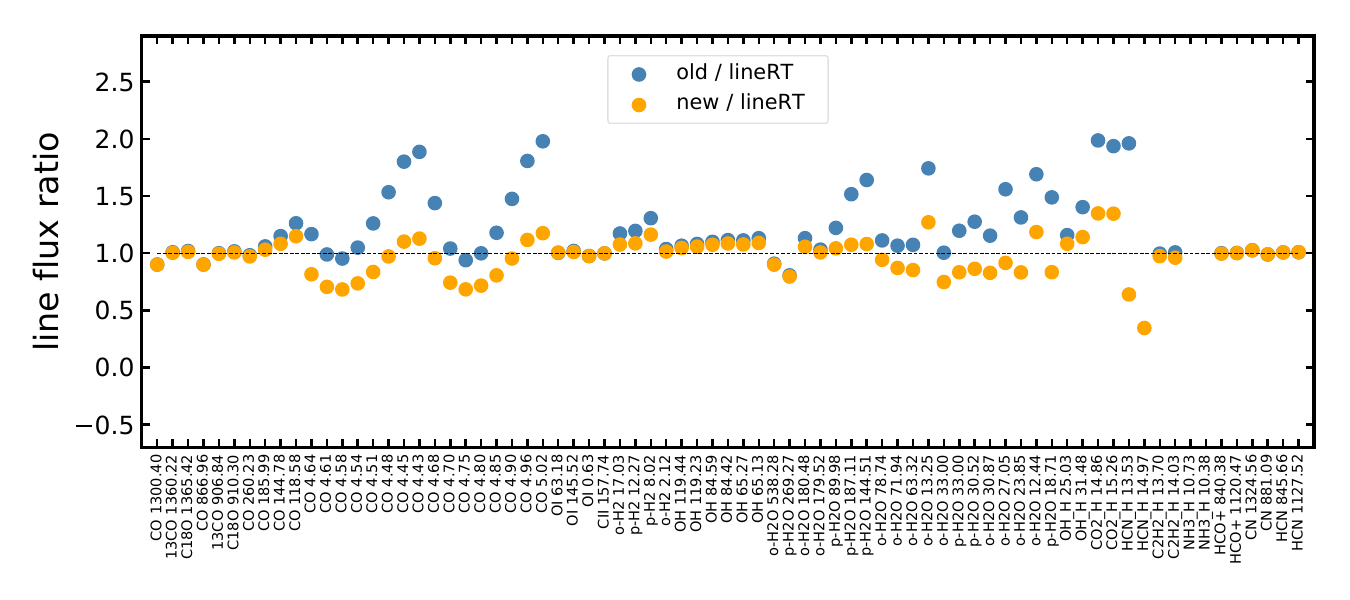}\\[-4mm]
  \caption{Line fluxes predicted by the old escape probability
    treatment (blue) and new escape probability treatment
    (orange). All line fluxes are divided by the results from the full
    line RT for zero inclination (face-on).}
  \label{fig:compare_lflux}
\end{figure*}

Figure~\ref{fig:compare_lflux} shows a comparison between the line
flux predictions obtained with the old and the new escape probability
treatment.  All results are shown with respect to the proper full {\sc
  ProDiMo} line RT results for face-on inclination.  We generally see
a good agreement between the results of all three methods of order
$10-30\%$. The differences become more noticeable at shorter
wavelengths where the lines are mostly emitted from the inner
disc. The new escape probability treatment is superior and more
balanced around the $y\!=\!1$ line, whereas the old escape probability
treatment has the tendency to over-predict the line fluxes. Two
\ce{NH3} at around 10.5\,$\mu$m turn out to be absorption lines
according to the proper line RT, and are not plotted in
Fig.~\ref{fig:compare_lflux}.  Both escape probability methods fail to
predict these lines correctly. The old escape probability method
predicts (by design) emission lines.  The new escape probability
method does produce absorption lines as it should, but the negative
fluxes are too large.  We note that the both escape probability
methods can predict the sub-mm lines very well (e.g.\ low-$J$ CO
lines, CN, HCN, HCO$^+$ in Fig.~\ref{fig:compare_lflux}) and also the far-IR lines
(e.g.\ Herschel/HIFI water lines, fine-structure cooling lines of
oxygen); their fluxes are predicted better than 10\%.

\subsection{Photo cross sections}

{\sc ProDiMo} now uses the new high-resolution photoionisation and
photodissociation cross sections $\sigma^{\rm mol}_{\rm ph}$ of
\citet{Heays2017} obtained from the Leiden
database\footnote{\resizebox{84mm}{!}{\url{https://home.strw.leidenuniv.nl/~ewine/photo/index.html}
  \label{LeidenDatabase}}}.

\subsection{UV molecular shielding and self-shielding}

Shielding factors are used in {\sc ProDiMo} to reduce the molecular
photoionisation and photodissociation rates
\begin{equation}
  R_{\rm ph} = 4\pi \int \sigma_{\rm ph}(\nu) \frac{J_\nu}{h\nu}\,d\nu
  \ ,
  \label{eq:photorate}
\end{equation}
where $\sigma_{\rm ph}(\nu)$ is the photo-dissociation cross section
and $J_\nu$ is the local mean intensity.  Without molecular shielding,
we have $J_\nu\!=\!\Jcont$, where $\Jcont$ is the result from the
continuum radiative transfer (RT).  However, the UV molecular
opacities are not included in the continuum RT.  They generally
cause $J_\nu\!<\!\Jcont$, because there is no molecular UV emission in
the disc. Hence, these opacities reduce the photo rates.

There are various molecular shielding effects included, both
self-shielding $i\!\to\!i$ and shielding of a molecule $j$ by another
molecule $i$, see for example \cite{Bethell2009}.  Water shielding has
been recently discussed by \cite{Bosman2022a,Bosman2022b}. In some
cases, the shielding-factors $s_{i\to j}$ are taken from published 1D
semi-infinite slab models, which resolved the UV lines and have
published efficiency factors as function of column density of the
molecular absorber ``towards the source'' and (local) temperature.
\begin{equation}
  s_{i\to j} = s_{i\to j}(N_i,\Tg) \quad\mbox{with}\quad
  N_i =\!\!\int\! n(i)\,ds
\end{equation}
In general, the shielding factors can be applied as a product, since
$\exp(-\tau_{{\rm all}\to j}) = \exp\big(-\sum_i \sigma_{\rm i}(\nu)
N_i\big) = \Pi_i\exp(-\sigma_{\rm i}(\nu) N_i)$, at least
approximately. Table~\ref{tab:shield} summarises the basic
self-shielding functions used. In {\sc ProDiMo}, all photorates are
calculated by numerical integration of Eq.\,(\ref{eq:photorate}) over
the frequency-dependent photo cross sections and the local UV
radiation field $J_\nu$. The latter is interpolated between the values
known on $\sim\!7$ UV spectral gridpoints used in the continuum
radiative transfer, see \citet{Woitke2009} for details.  In cases where
we do not have detailed photo cross sections, but only the $\alpha$,
$\beta$, $\gamma$ coefficients from a chemical database, we fit a
generic cross-section $\sigma_{\rm ph}(\lambda)=\sigma_0
\exp\big(-(\lambda_0-\lambda)^2/\Delta\lambda^2\big)$, with two free
parameters $\sigma_0$ and $\lambda_0$, until the photorates are
correct for optical extinctions $A_V=0$ and $A_V=0.5$, using standard
interstellar dust extinction properties.

Therefore, it is relatively straightforward to use any
frequency-dependent recipes to account for the effect of molecular
shielding.  In particular, and this is new in {\sc ProDiMo}, we use
the following idea to account for generic self-shielding
\begin{equation}
  s_{i\to i} = \int \sigma_{\rm ph}(\nu) \frac{J_\nu}{h\nu}
                  {\rm e}^{-N_i\sigma_{\rm ph}(\nu)} \,d\nu
    \;\Bigg/ \int \sigma_{\rm ph}(\nu) \frac{J_\nu}{h\nu}\,d\nu \ .
\end{equation}
Other recipes are applied to the shielding of any molecule by \ce{H2}
and C, see Table~\ref{tab:shield}.  For example, Eq.~(8) in
\citet{Kamp2000} describes the $T$-dependent reduction of UV light by
\ce{H2}-lines, when detected with the flat opacity of neutral C, which
is assumed to provide a first approximation of the shielding effect on
any molecule with uncorrelated $\sigma_{\rm ph}(\lambda)$ for
wavelengths shorter than $1100\,$\AA.

\begin{table}[!t]
  \vspace*{1mm}
  \caption{Molecular shielding and self-shielding factors in ProDiMo}
  \label{tab:shield}
  \vspace*{-3mm}\hspace*{-1.5mm}
  \resizebox{90mm}{!}{
  \begin{tabular}{l|c|c}
    \hline
    &&\\[-2.2ex]
    effect$^{(0)}$ & formula / reference & flag \\
    \hline
    &&\\[-2.2ex]
    self-sh. $\rm H_2 \to H_2$ & \!\!\citet[][Eq.\,37]{Draine1996}\!\! & default \\
    self-sh. $\rm CO \to CO$\!\!& \cite{Kamp2000} & default\\
    self-sh. $\rm C \to C$     & $\exp(-\sigma_0^{\rm\,C} N_{\rm C})^{(3)}$ & default \\
    self-sh. $\rm N_2 \to N_2^{(5)}$ & \cite{Li2013}, see footnote \footref{LeidenDatabase} & default\\
    self-sh. $\rm i \to i^{(1)}$ & $\exp(-\sigma^{\,i}_\nu N_i)$ under $\nu$-integral
      & \!\!\resizebox{20mm}{!}{\tt self\_shielding}\!\!\\
    \hline
    &&\\[-2.2ex]
    sh. $\rm H_2 \to CO$ &  \citet[][Eq.\,23]{Kamp2000} & default \\
    sh. $\rm H_2 \to C$  & \citet[][Eq.\,8]{Kamp2000} & default \\
    sh. $\rm H_2 \to all^{(2)}$ & as $\rm H_2 \to C$, but for $\lambda\!<\!1100\,$\AA$^{(4)}$
      & \!\!\resizebox{18mm}{!}{\tt H2\_shielding}\!\! \\    
    sh. $\rm C \to H_2,CO$ & $\exp(-\sigma_0^{\rm\,C} N_{\rm C})$
    for $\lambda\!<\!1100\,$\AA$^{(4)}$ & default \\
    sh. $\rm C \to all^{(2)}$ & $\exp(-\sigma_0^{\rm\,C} N_{\rm
      C})$ for $\lambda\!<\!1100\,$\AA$^{(4)}$
      & \!\!\resizebox{17mm}{!}{\tt C\_shielding}\!\!\\
  \hline    
  \end{tabular}}\\[1mm]
  \small\noindent
  $^{(0)}$: self-shielding is abbreviated by 'self-sh.' and shielding by 'sh.'.\\ 
  $^{(1)}$: for all molecules $i$ without explicit self-shielding
  treatment\\
  $^{(2)}$: for all target molecules without explicit shielding
  treatment\\
  $^{(3)}$: constant $\sigma_0^{\rm\,C}\!=\!1.1\times
  10^{-17}\rm\,cm^2$\\
  $^{(4)}$: shielding factor applied under the $\nu$-integral for
  $\lambda\!<\!1100\,$\AA\\
  $^{(5)}$: contains the $\rm H_2 \to N_2$ shielding, see also \cite{Visser2009}\\
  \vspace*{-3mm}
\end{table}

\subsection{New 2D treatment of molecular shielding}

The question arises how to deal with the problem of molecular
shielding in 2D. Which column density $N_i$ is to be considered?  The
radial one, the vertically upward one, a mixture, or something else?
Equation (\ref{eq:Jcont}) holds an answer to that question. Using the
explicit shapes of $\alpha_0$, $\beta_0$ and $\beta_1$ for
$\tauL\!\to\!0$, and neglecting the $\alpha_1$ term, which is the
continuum emission of the disc in the solid angle of the star that is
usually very small, this equation reads
\begin{eqnarray}
  \Jcont
    &\!\!\!=\!\!\!& \frac{\Omega_\star}{4\pi} I_\nu^\star
        \,{\rm e}^{-\tauCrad} 
    ~+~ \frac{\Omega_{\rm d}}{8\pi} I_\nu^{\rm ISM}
        \Big(E_2(\tauCup) + E_2(\tauCdown)\Big)  \label{eq:Jcont2}\\
    &\!\!\!\!\!\!& \hspace*{17mm}
    ~+~\,\frac{\Omega_{\rm d}}{8\pi} \SC
        \Big(2-E_2(\tauCup)-E_2(\tauCdown)\Big) \nonumber \ .
\end{eqnarray}
Equation (\ref{eq:Jcont2}) states the influence of the three principle
sources (the star, the ISM background, and the disc itself) on the
calculated $\Jcont$. Contrary to the IR pumping of molecules, which
often comes from the warm extended dust below the point of interest,
the UV radiation mainly arrives from two directions, the direct radial
and the vertically downward directions. Clearly, if direct
illumination from the star dominates, then
$\Jcont\approx\frac{\Omega_\star}{4\pi} I_\nu^\star\,{\rm
  e}^{-\tauCrad}$ and one should use the radial column densities
towards the star to compute the shielding factors. If interstellar UV
irradiation dominates, one should take the vertical column densities.
However, in the IR line emitting disc regions, none of these two terms
are usually the important ones, but it is the downward scattering of
star light by the small dust in the disc's upper layers, which is
represented by $\SC$ and expressed by the third term in
Eq.~(\ref{eq:Jcont2}).

It is not obvious which molecular column densities to consider for the
molecular shielding factors applied to the third term in
Eq.~(\ref{eq:Jcont2}).  The proper approach would be to consider the
{\em molecular column density along the average scattered photon path},
but this information is not available. One could argue that taking
the radial column density at a much higher disc level up to the
scattering point, and then adding the downward vertical column density
from that point is correct, when considering single vertical
scattering events, but then the point at which that scattering
event takes place needs to be determined.

Instead we follow a much more simple approach here.  We assume that
the second part of the molecular column density along the scattered
photon path, namely the downward
vertical column density from the scattering site is dominant, and since this
column density increases exponentially on the way down, because
(a) the density increases and (b) the molecular concentration
increases, it is mostly determined locally, and there is no big
difference when simply taking the full vertical column density.

According to this approach, we use the radial column densities for the
1$^{\rm st}$, and the vertical column densities for the 2$^{\rm nd}$
and 3$^{\rm rd}$ terms in Eq.\,(\ref{eq:Jcont2}), which can be
reformulated as
\begin{align}
  J_{\nu,{\rm rad}}^{\rm cont} ~=~& \frac{\Omega_\star}{4\pi} I_\nu^\star
                             \,{\rm e}^{-\tauCrad} \\
  J_{\nu,{\rm ver}}^{\rm cont} ~=~& \Jcont - J_{\nu,{\rm rad}}^{\rm cont} 
  \label{eq:better}\\
  J_\nu ~=~& s_{i\to j}(N_i^{\rm rad},\Tg)\,J_{\nu,\rm rad}^{\rm cont}
        ~+~ s_{i\to j}(N_i^{\rm ver},\Tg)\,J_{\nu,\rm vert}^{\rm cont}
  \ .
  \label{eq:Jcont3}      
\end{align}
In previous {\sc ProDiMo} models, we had used the stellar and interstellar
irradiation terms (parts 1 and 2 in Eq.\,\ref{eq:Jcont2}) to
calculate weighting factors for the application of shielding factors
according to this scheme
\begin{align}
  J_{\nu,{\rm rad}}^{\rm cont} ~=~& \frac{\Omega_\star}{4\pi} I_\nu^\star
                \,{\rm e}^{-\tauCrad} \\
  J_{\nu,{\rm ver}}^{\rm cont} ~=~& I_\nu^{\rm ISM} E_2(\tauCup)
  \label{eq:questionable} \\
  w_{\rm rad} ~=~& J_{\nu,{\rm rad}}^{\rm cont}\Big/
    (J_{\nu,{\rm rad}}^{\rm cont}+J_{\nu,{\rm ver}}^{\rm cont}) \\
  w_{\rm vert} ~=~& J_{\nu,{\rm ver}}^{\rm cont}\Big/
    (J_{\nu,{\rm rad}}^{\rm cont}+J_{\nu,{\rm ver}}^{\rm cont}) \\*[-0.4ex]
  s_{i\to j} ~=~& w_{\rm rad}\,s_{i\to j}(\Nirad,\Tg)
           ~+~ w_{\rm vert}\,s_{i\to j}(\Niver,\Tg) \label{eq:shield_old} \\
  J_\nu ~=~& s_{i\to j}\,\Jcont  \ ,
\end{align}
where $w_{\rm rad}$ and $w_{\rm vert}$ are the radial and vertical
weights. But this approach is obviously a far more questionable
assumption, because the often most important term (the third term in
Eq.~\ref{eq:Jcont2}) is not even considered, and thus $\Jcont\gg
J_\nu^{\rm rad}+J_\nu^{\rm vert}$. Instead, based on the mixing ratio
between $J_\nu^{\rm rad}$ and $J_\nu^{\rm vert}$, an effective
shielding factor was calculated and applied to all of $\Jcont$. Since
Eq.\,(\ref{eq:shield_old}) ignores the scattering altogether, the old
vertical part $J_{\nu,{\rm ver}}^{\rm cont}$ is too small, and hence
we switched from radial to vertical column densities too late.  
Since the radial column densities are generally larger than the
vertical ones, {\it the effects of molecular shielding were
  overestimated in the old models}.

\begin{figure*}[!t]
  \centering
  \begin{tabular}{cc}
    {\sf\hspace*{2mm} old implementation} &
    {\sf\hspace*{2mm} new implementation}\\
    \hspace*{-2mm}
    \includegraphics[height=62mm,width=80mm,trim=0 0 0 0,clip]
                    {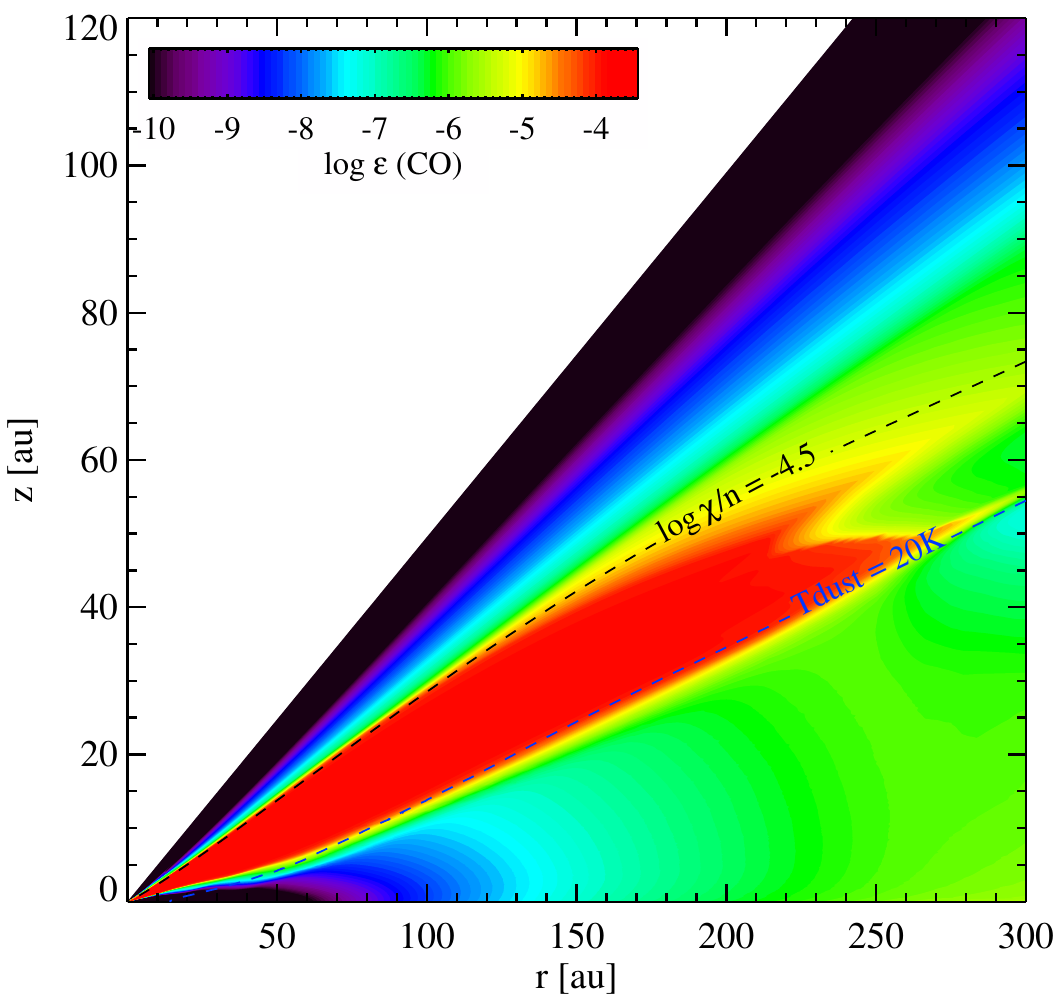} &
    \hspace*{-7mm}
    \includegraphics[height=62mm,width=80mm,trim=0 0 0 0,clip]
                    {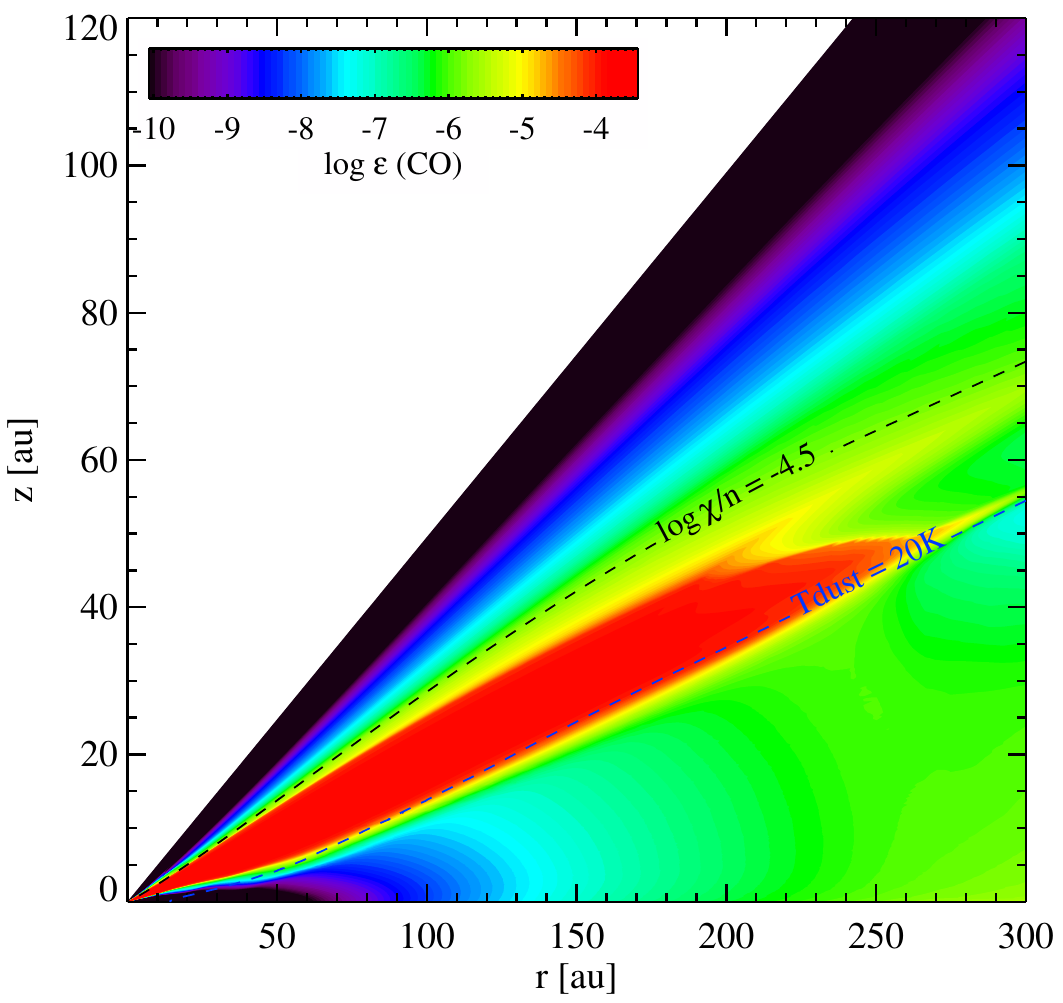}\\[-2mm]
    \includegraphics[height=62mm,width=79mm,trim=0 0 0 0,clip]
                    {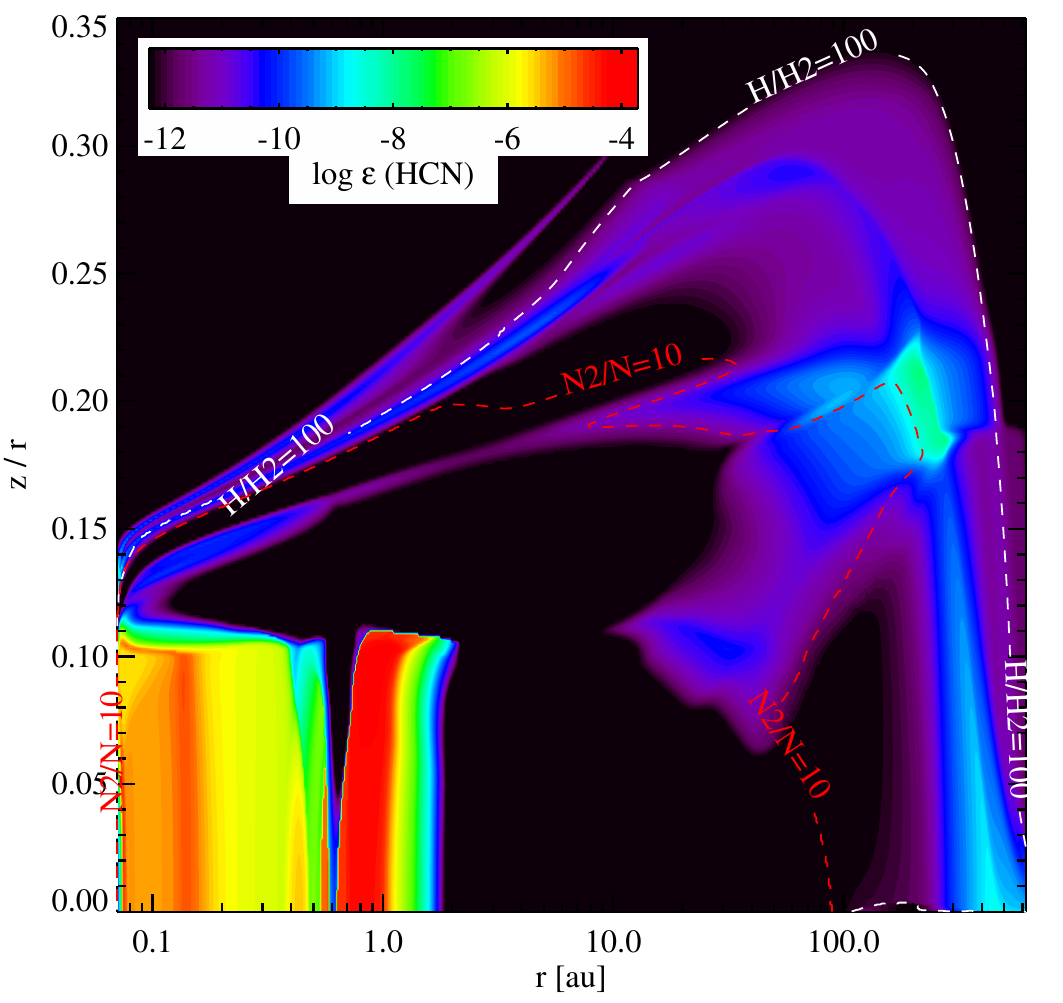} &
    \hspace*{-5mm}
    \includegraphics[height=62mm,width=79mm,trim=0 0 0 0,clip]
                    {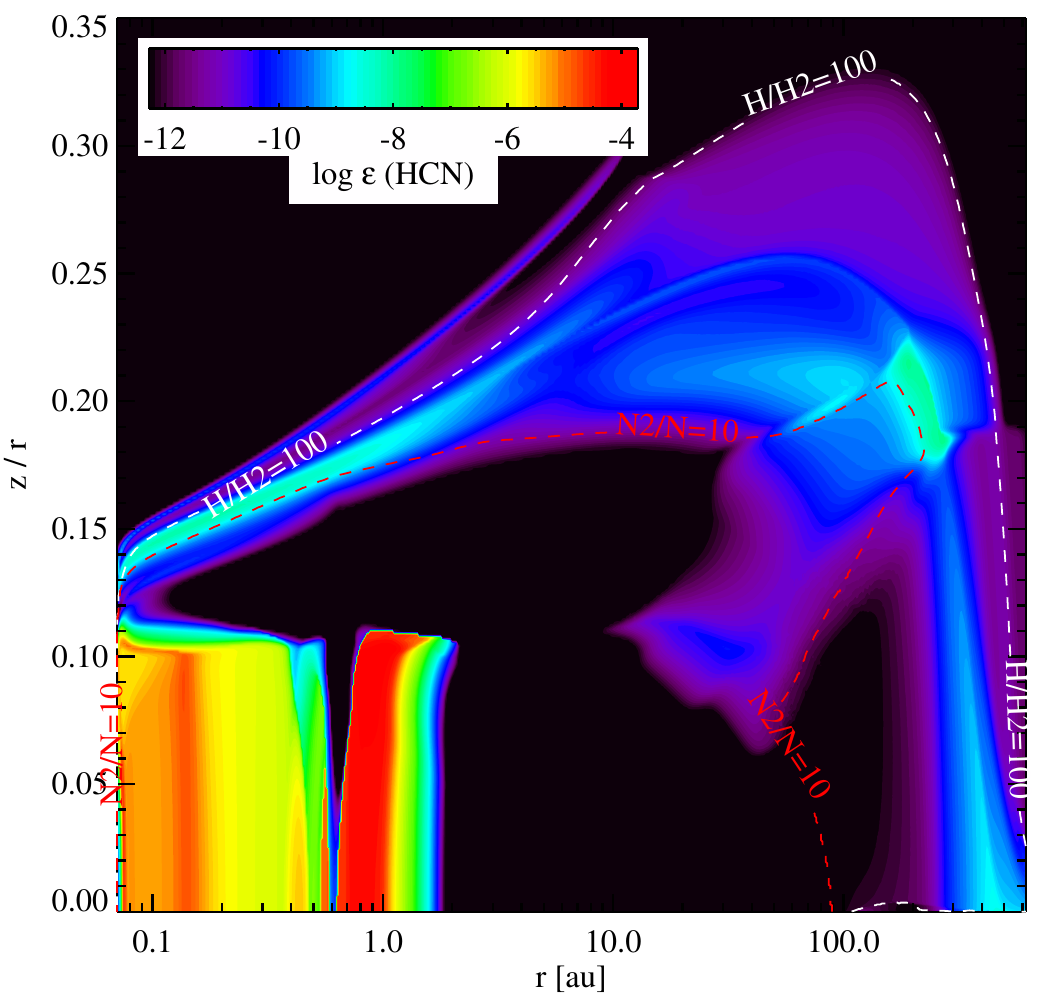}\\[-2mm]
    \includegraphics[height=64mm,width=79mm,trim=0 0 0 0,clip]
                    {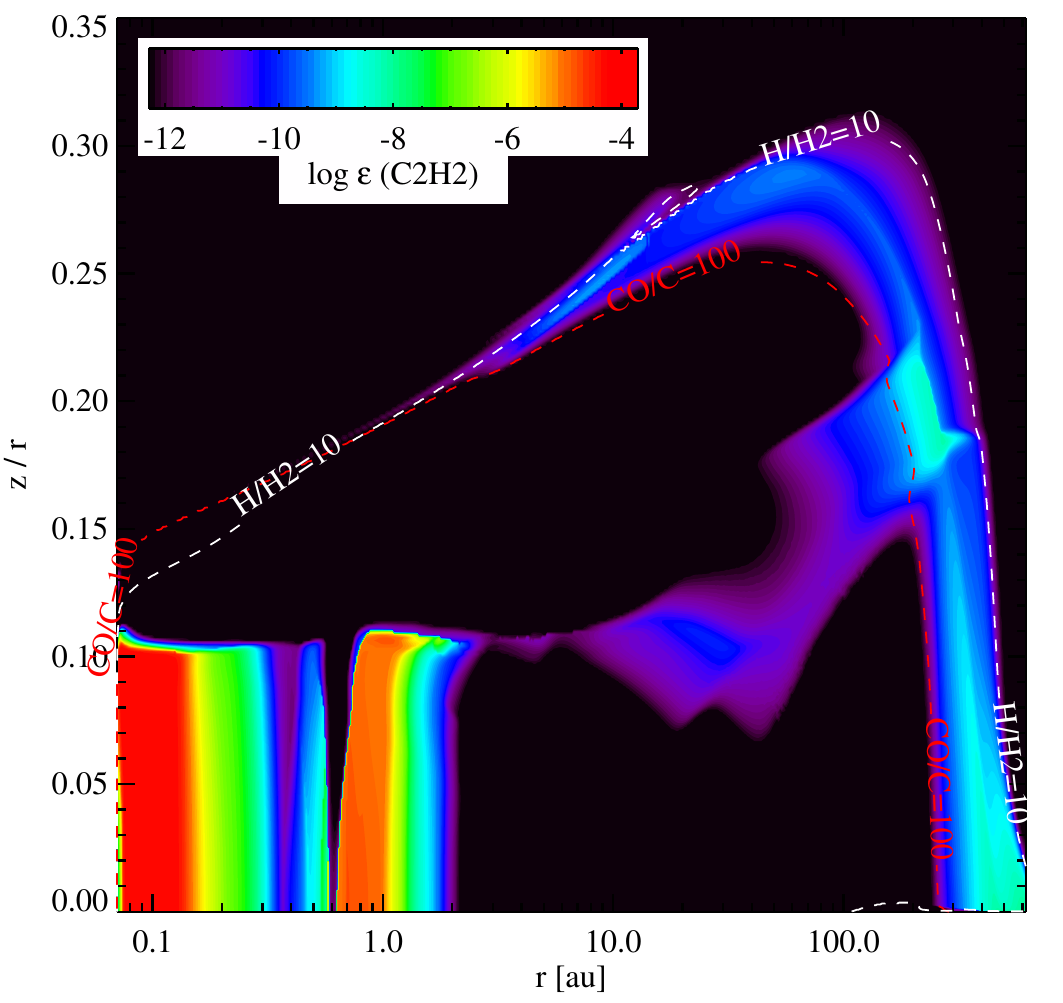} &
    \hspace*{-5mm}
    \includegraphics[height=62mm,width=79mm,trim=0 0 0 0,clip]
                    {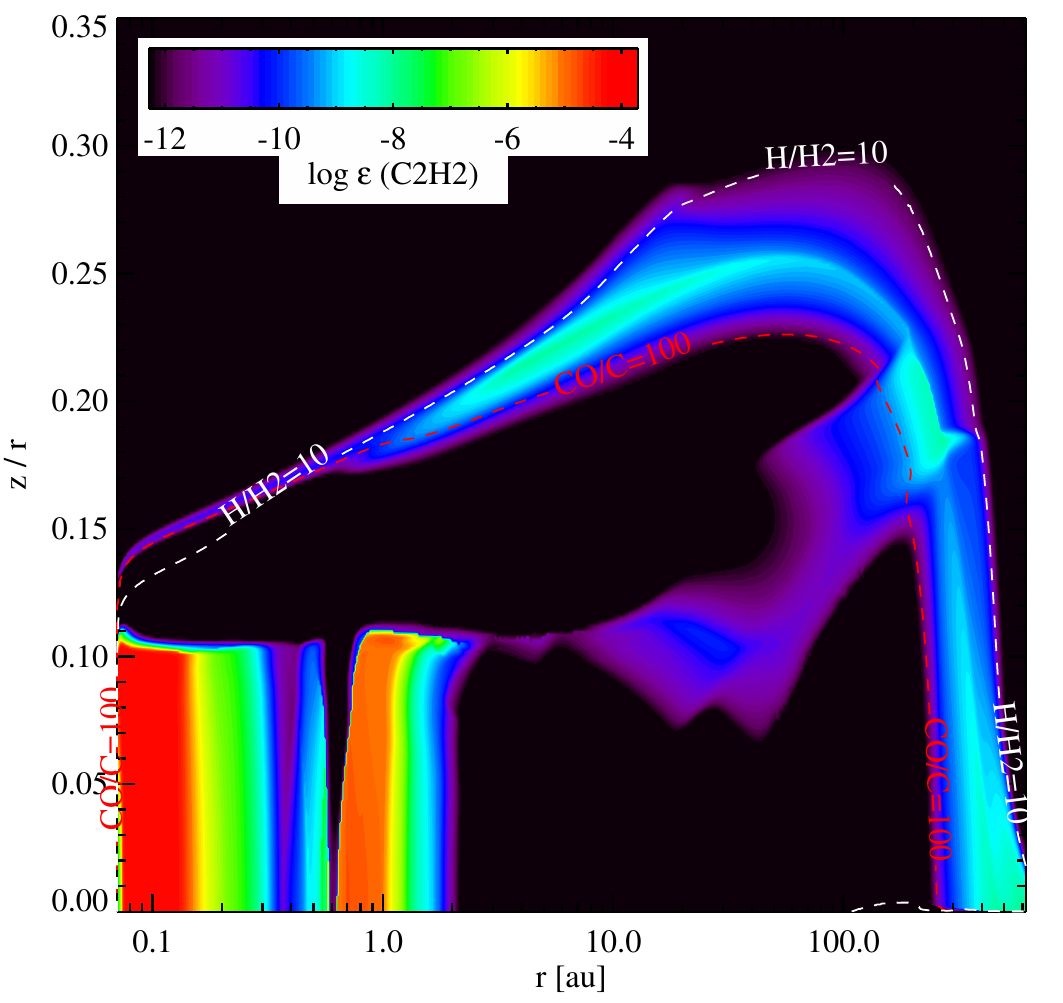}
  \end{tabular}
  \caption{Change of molecular concentrations caused by introducing
    our new 2D concept to apply molecular shielding factors.}
  \label{fig:molconc}
\end{figure*}

Figure~\ref{fig:molconc} shows that, as expected, the formation of CO
happens now slightly deeper in the disc. In both models, CO is found
to be abundant when (i) the ionisation parameter $\chi/n$ is low
enough ($\chi$ is the strength of the UV field, see Eq.\,(42) in
\citet{Woitke2009}, and $n$ is the total hydrogen nuclei density
$\rm[cm^{-3}]$), and when (ii) the dust temperature is high enough
$\Td\,\ga\,20\,$K. These two criteria are related to the CO
photo-dissociation and freeze-out. The critical value for the CO
photodissociation $\log_{10}(\chi/n)$ changes from about -4.5 to -5
in the new models.

The feedback on the more complex molecules like HCN and C$_2$H$_2$ is
remarkable, leading to profoundly different chemical conditions in the
IR line emitting disc regions.  In Fig.~\ref{fig:molconc}, we
concentrate on the upper warm line emitting regions.  There are also
some regions with high concentrations of HCN and C$_2$H$_2$ in the
midplane, but these molecules are covered by optically thick dust and
play no significant role for the IR line emission in this T\,Tauri
standard disc setup. The formation of these IR active molecules in the
upper disc happens in slightly deeper layers according to the new
models, and since the gas densities are larger there, the formation of
more complex molecules is generally favoured in the new models.

In particular, the formation of \ce{C2H2} needs both, \ce{H2} and
neutral carbon atoms, see e.g. \cite{Agundez2008}. In the lower part
of Fig.~\ref{fig:molconc}, we overplot $\rm H/H_2\!=\!10$ and $\rm
CO/C\!=\!100$ contour lines. These two lines nicely encircle the
region of high \ce{C2H2} concentrations in the model. Similar results
were reported by \cite{Walsh2015} and \cite{Anderson2021}.  In the
inner disc, these lines are crossed in our models, and so when \ce{H2}
finally forms, there is no neutral C anymore, and we get no \ce{C2H2},
see also \cite{Kanwar2023}.  In the new model, these two
lines are further apart, there is a wider region to form \ce{C2H2},
and that region now also extends to smaller radii.

A similar effect is found for HCN, where the changes are even more
pronounced. The formation of HCN requires \ce{H2} and neutral N atoms.
In the new models, both transitions $\rm H \to H_2$ and $\rm N\to N_2$
happen deeper in the disc, and they are vertically more separated,
leading to increased HCN column densities in the upper layers of the
inner disc.

\subsection{Heating by photo-dissociation}

One heating process that has so far been partly neglected in {\sc
  ProDiMo} is the ``direct'' heating by photo-dissociation
\begin{equation}
  \Gamma_{\rm pd} = \sum_{\rm mol} n_{\rm mol} R^{\rm ph}_{\rm mol} \Delta E_{\rm mol} \ ,
\end{equation}
where $n_{\rm mol}\rm\,[cm^{-3}]$ is the molecular particle density,
$R^{\rm ph}_{\rm mol}\rm\,[s^{-1}]$ the photo-dissociation rate, and
$\Delta E_{\rm mol}$ the mean energy liberated per photo-dissociation.
Similar to photoionisation, the excess photon energy can create
super-thermal dissociation products, or can lead to vibrationally or
electronically excited molecular fragments, and the key question is
how much of this energy is thermalised by the gas.

\citet{Gu2020} have treated this process with 100\% efficiency, assuming
that the energy released as local heat is set as the incident photon
energy in excess of the respective dissociation threshold.
\begin{equation}
  \Delta E_{\rm mol} = \frac
    {\int_{\nu_0}^\infty \frac{J_\nu}{h\nu} \sigma^{\rm mol}_{\rm ph}(\nu)
    \,h(\nu-\nu_0)\,d\nu}
    {\int_{\nu_0}^\infty \frac{J_\nu}{h\nu} \sigma^{\rm mol}_{\rm
        ph}(\nu)\,d\nu} \ ,
    \label{eq:HeatPD}
\end{equation}
where $\nu_0$ is the photo-dissociation threshold frequency.
However, more detailed studies show that not all of this energy
is thermalised, for example excited dissociation products can decay
radiatively, or can trigger follow-up chemical reactions.

\citet{Glassgold2015} have published detailed energies per
photo-dissociation $\Delta E$ for \ce{H2} (1.2\,eV), \ce{CO}
(1.4\,eV), \ce{H2O} (1.2\,eV), and \ce{OH} (3.6\,eV), and we have now
implemented these data into {\sc ProDiMo}.  For all other molecules we
conservatively assume $\Delta E\!\approx\!2\,$eV. Previously in {\sc
  ProDiMo}, only \ce{H2}, with $\Delta E(\ce{H2})\!=\!0.4\,$eV
\citep{Black1987}, was considered for the heating by
photo-dissociation, besides the heating by photoionisation of \ce{H-},
  C, Fe, Si and Mg. Figure~\ref{fig:heatcool} shows that this additional
heating process can be quite relevant in the IR active molecular
layer. Interestingly, the reduced UV molecular shielding factors cause
larger photorates, which lead to a more active chemistry where the
molecular fragments created by photo-dissociation react further
exothermally, which increases the area where chemical heating is
important.

\subsection{Chemical rate network}
\label{sec:network}

We used the large DIANA standard chemical network \citep{Kamp2017} in
this work, which has 235 chemical species, with reaction rates mostly
taken from the UMIST 2012 database \citep{McElroy2013}, but including
a simple freeze-out and desorption ice chemistry \citep{Woitke2009},
X-ray processes partly based on doubly ionised species
\citep{Aresu2011}, excited molecular hydrogen \citep[see][]{Kamp2017},
and polycyclic aromatic hydrocarbons in five different charging states
\citep{Thi2019}, altogether 3066 reactions. \ce{H2}-formation of grain
surfaces is calculated according to \citet{Cazaux2002,Cazaux2004} with
updates described in \citet{Cazaux2009,Cazaux2010}. {We use our
standard element abundances (Table~\ref{tab:abund}), which
are close to solar ($\rm C/O\!=\!0.46$) except for the depleted metals
Na, Mg, Si, S and Fe.}

PAH molecules are included in the radiative transfer, with opacities
calculated from their overall abundance and an assumed ratio between
charged and neutral PAHs to avoid global iterations, see details in
\citet{Woitke2016} and \citet{Woitke2019}. PAHs are also included in
the chemical network with 5 charging states \citep{Thi2019}, and
included in the calculation of heating/cooling rates, as the
photoelectric effect on PAHs is one of the most important heating
processes in the upper disc layers. The PAH heating rate is calculated
from the calculated abundances of the different charging states, their
individual photo-cross sections, and the calculated local radiation
field.

In recent updates of our chemical network, we added a few hand-picked
gas phase reactions not included in UMIST~2012. One of these reactions
turned out to be critical for the fitting of the JWST spectrum
of EX\,Lupi (Sect.~\ref{sec:EXLup}):
\begin{equation}
  \ce{C} ~+~ \ce{H2O} ~~\longrightarrow~~ \ce{HCO} ~+~ \ce{H}
\end{equation}
from \citet{Hickson2016}, who measured surprisingly large rates for
this reaction at low temperatures, enhanced by quantum mechanical
tunnelling.  Although the Kinetic Database For Astrochemistry
(KIDA\footnote{\url{https://kida.astrochem-tools.org/}}) does not
include this reaction in their standard releases, they list it with
Arrhenius parameters\linebreak $\{\,\alpha\!=\!1.07\times 10^{-8}{\rm
  cm^3s^{-1}}~,~\beta\!=\!-1.59~,~\gamma\!=\!0\,\}$ with reference to
\citet{Hickson2016}. In our models, whenever a CO-bond is broken, for
example by photodissociation or X-ray induced reactions, this reaction
would be very efficient in immediately re-forming that bond, which
would prohibit the subsequent formation of hydrocarbon molecules and
HCN as discussed in Sect.~\ref{sec:chempaths}.  Tests show that if we
would include this reaction with the Arrhenius coefficients from the
KIDA database, the \ce{C2H2} column density in the line emitting
regions of our EX\,Lupi model would decrease by a factor of about 300,
and peak line flux at 13.7\,$\mu$m by a factor of about 5.  Closer
inspection of Fig.~3 in \citet{Hickson2016} shows, however, that the
Arrhenius coefficients $\{\,\alpha\!=\!4.2\times 10^{-13}{\rm
  cm^3s^{-1}}~,~\beta\!=\!-2.5~,~\gamma\!=\!0\,\}$ fit the original
data points at 50\,K, 75\,K and 100\,K much better, and are consistent
with the measurements of \citet{Husain1971} that the rate coefficient
of this reaction is very small at room temperature ($<\!10^{-12}\,{\rm
  cm^3s^{-1}}$).  We therefore decided to use our own Arrhenius fit of
this reaction in this paper.

\subsection{New HITRAN line selection rules}

\begin{table}
  \vspace*{2mm}
  \caption{Line selection criteria for the HITRAN~2020 database.}
  \label{tab:LineSelection}
  \vspace*{-5.5mm}
  \begin{center}
  \begin{tabular}{c|c|cc|ccc}
    \hline
    & $\lambda\,[\mu$m] & $E_u\,$[K]
    & $\rm strength\,[s^{-1}]$ & \!\!\#\,lines\!\! \\
    \hline
    &&&&\\[-2.2ex]
    \ce{OH}   & 4.5-50 & $900-30000^{(1)}$ & $>10^{-5}$ & 1391 \\
    \ce{CO2}  & 4.5-40 & $<5000$    & $>10^{-4}$ &  4809 \\
    \ce{C2H2} & 4.5-40 & $<5000$    & $>10^{-5}$ & 16960 \\
    \ce{HCN}  & 4.5-40 & $<5000$    & $>10^{-5}$ &  3728 \\
    \ce{CH4}  & 4.5-40 & $<4000$    & $>10^{-3}$ &  3414 \\
    \ce{NH3}  & 4.5-40 & $<4000$    & $>10^{-3}$ &  6938 \\
    \hline
  \end{tabular}
  \end{center}
  \vspace*{-3mm}  
  \small $^{(1)}$: The $E_u$ interval for OH is to avoid
  double-accounting of pure rotational lines included
  as another line species treated in non-LTE.
\end{table}

The expansion from the Spitzer/IRS to the JWST/MIRI wavelength range
made it necessary to revise our line selection strategy, by which we
import only a limited number of molecular lines from the HITRAN
database.  We also updated {\sc ProDiMo} to use the HITRAN~2020 catalogue
\citep{HITRAN2020} instead of the HITRAN~2012 catalogue.
The imported data are the upper level energies $E_u$, line
wavelengths $\lambda_{ul}$, Einstein coefficients $A_{ul}$ and level
degeneracies $g_u$.  The new HITRAN~2020 database has much more lines,
which, if we would just import all lines, would blow up {\sc ProDiMo}'s
memory consumption. As explained in \citet{Woitke2018}, we therefore
have to limit our line selection to lines with a minimum line strength
defined by
\begin{equation}
  {\rm line\ strength} ~=~
  A_{ul}\,g_u\,\exp\left(-\frac{E_u}{k\,(1500\rm\,K)}\right) \ .
\end{equation}
In combination with further restrictions on wavelength range and upper
level energy $E_u$, see Table~\ref{tab:LineSelection}, we aim at
selecting only the relevant, strong, observable lines. Together with
the ro-vibrational CO lines of the first four vibrationally excited
states \citep[][]{Thi2013}, ro-vibrational \ce{H2O} lines, rotational
lines of various molecules, and lines of atoms and ions
\citep{Woitke2018}, we are now tracing altogether 64890 lines in this
model. All lines are automatically considered for heating and cooling.

\subsection{Dust settling}

Considering a turbulent one-dimensional column of gas and dust,
\citet{Mulders2012} write down the continuity equation for passive dust
particles with a size distribution function $f(a,z,t)$ as
\begin{equation}
  \pabl{f(a,z)}{t} = \pabl{}{z}\bigg(\underbrace{\rho\,\Dd(a)\, 
            \pabl{}{z}\!\left(\frac{f(a,z)}{\rho}\right)}_{j_{\rm diff}}\!\bigg)
      \,-\, \pabl{}{z}\Big(\underbrace{f(a,z)\,\vdreq(a,z)}_{j_{\rm set}}\Big)
\end{equation}
where $\rho$ is the gas mass density, $t$ the time, $z$ the vertical
coordinate, and $a$ the dust particle radius. Assuming that there is
no bulk gas motion, the grains are transported downwards by
gravitational settling with their size-dependent equilibrium settling
velocity $\vdreq(a,z)$, and transported upwards by eddy diffusion
according to the local dust diffusion coefficient $\Dd$, driven by
dust particle concentration gradients
$\pabl{}{z}\big(\frac{f(a,z)}{\rho}\big)$. Assuming that the grains
have relaxed towards a steady state, we have for each size
$\partial/\partial t\,f(a,z)\to 0$ and zero net flux $j_{\rm
  diff}-j_{\rm set}\to 0$, hence
\begin{equation}
  \rho\,\Dd\,\pabl{}{z}\bigg(\frac{f(a,z)}{\rho}\bigg)
  ~=~ f(a,z)\,\vdreq(a,z) \ ,
\end{equation}
from which we find
\begin{equation}
  \pabl{}{z}\bigg(\ln\frac{f(a,z)}{\rho}\bigg) ~=~ \frac{\vdreq(a,z)}{\Dd}
  \label{basic1} \ .
\end{equation}
For large Knudsen numbers and subsonic particle velocities (Epstein regime),
the equilibrium settling velocity, also called terminal fall speed, is
given by \citet{Schaaf1963}
\begin{equation}
  \vdreq(a,z) = -g_z\,\tstop(a,z) = -z\;\Omega^2\,\tstop(a,z) \ ,
  \label{vdr}
\end{equation}
where $g_z$ is the vertical gravitational acceleration, $\Omega$ the
Keplerian circular frequency.  The stopping timescale, or frictional
coupling timescale, is given by
\begin{equation}
  \tstop(a,z) = \frac{\rhom\,a}{\rho(z)\,\vth} \ ,
  \label{tstop}
\end{equation}
where $\rhom$ the dust material density, and the thermal velocity is
\begin{equation}
  \vth = \sqrt{\frac{8\,k\Tg}{\pi\,\mu}} = \sqrt{\frac{8}{\pi}}\,\cs \ ,
  \label{vth}
\end{equation}
where $k$ the Boltzmann constant, $\Tg$ the local gas temperature, $\mu$ the
mean molecular weight, $\cs$ the local isothermal speed of sound, and
$p=\cs^2\,\rho$ the local gas pressure. Equation~(\ref{tstop}) seems
widely incorrect in the literature, where often $\cs$ is sloppily used
instead of the correct mean of the magnitude of the thermal velocities
of gas particles $\vth$, as required in Schaaf's formula, see
\citet{Woitke2003}.

Eddy diffusion (``turbulent mixing'') is assumed to be dominant over
gas-kinetic diffusion. The eddy diffusion coefficient is
given by a characteristic velocity times a characteristic length
\begin{equation}
  \Dgas = \vz\,L
\end{equation}
where $L$ is the size of the largest eddies (``mixing length'') and
$\vz$ the root-mean square average of the fluctuating part of the
vertical gas velocities. In a Kolmogorov type of power spectrum
$P(k)\!\propto\!k^{-5/3}$, there are different turbulent modes
associated with different wave-numbers $k$ or different spatial eddy
sizes $l$. Any given particle of size $a$ tends to co-move with the
sufficiently large and slow turbulent eddies, whereas its inertia
prevents following the short-term, small turbulence modes. In order to
arrive at an effective particle diffusion coefficient, the advective
effect of all individual turbulent eddies has to be averaged, and
thereby transformed into a collective particle diffusion coefficient.
This procedure has been carried out, with different methods and
approximations, by \citet{Dubrulle1995}, \citet{Schrapler2004},
\citet{Youdin2007}, and \citet{Mulders2012}. The result of
\citet{Mulders2012} is
\begin{equation}
  \Dd = \frac{\Dgas}{1+\St^2}
\end{equation}
where $\St$ is the Stokes number given by
\begin{equation}
  \St = \frac{\tstop}{\tedd}  \ .
\end{equation}
$\tedd$ is the eddy turnover or turbulence correlation
timescale in consideration of the largest eddy, which is assumed to
equal the mixing length $L$. 
\begin{equation}
  \tedd = \frac{L}{\vz} \ .
\end{equation}
According to an $\alpha$ disc model \citep{Shakura1973},
the typical vertical length scale is $L=\sqrt{\alpha}\,H_p$ and 
$\vz=\sqrt{\alpha}\,c_s$ \citep{Ormel2007}, hence
\begin{align}
  \Dgas  &= \alpha\,\cs\,H_p \\
  \tedd  &= \frac{H_p}{\cs} = \frac{1}{\Omega} \ .
\end{align}
Our basic equation (\ref{basic1}) for the balance between
upward mixing and downward gravitational settling is hence
\begin{equation}
  \pabl{}{z}\bigg(\ln\frac{f(a,z)}{\rho}\bigg) 
         ~=~ -\frac{z\,\Omega\,\tstop}{\alpha\,H_p^2}\,(1+\St^2)  \ ,
  \label{basic2}
\end{equation}
All quantities on the right
hand side of Eq.\,(\ref{basic2}), $\Omega$, $\alpha$, as well as
$D_{\rm gas}$, $T$, $\cs$, $\mu$, $\rho$ and $\rhom$, can in general
depend on height $z$. In addition, $\St$ and $\tstop$ depend on 
$z$ and on grain size $a$.

\citet{Dubrulle1995} ignore all $z$-dependencies on the r.h.s.\ of
Eq.\,(\ref{basic2}) and replace all variables by their mid-plane
values. They also drop the $(1+\St^2)$ term. In that case, we 
can carry out the $z$-integration to find
\begin{equation}
  \ln\frac{f(a,z)/f(a,0)}{\rho(z)/\rho(0)}
  = -\frac{\Omega\,\tstop(0)}{\alpha\,\,H_p^2}\,\int_0^z z'\,dz' 
  ~=~ -\frac{z^2}{2\,H_p^2}\,\frac{\Omega\,\tstop(0)}{\alpha}
  \nonumber\ .
\end{equation}
Introducing the size-dependent dust scale height $H_a$ via
$f(a,z)=f(a,0)\exp(-\frac{z^2}{2\,H_a^2})$, and using
$\rho(z)=\rho(0)\exp(-\frac{z^2}{2\,H_p^2})$, the result is
\begin{equation}
  \frac{H_p^2}{H_a^2} ~=~ 1 + \frac{\Omega\,\tstop(0)}{\alpha}
  \label{Dubrulle}
\end{equation}

\citet{Riols2018} also drop the $(1+\St^2)$ term in Eq.\,(\ref{basic2}),
but take into account that $\tstop$ depends on density, and is hence
$z$-dependent. They furthermore explicitly assume a Gaussian for the
vertical gas density distribution.  In that case the stopping
timescale can be expressed as
\begin{equation}
  \tstop = \tstop(0)\,\exp\Big(\frac{z^2}{2\,H_p^2}\Big)  \ .
\end{equation}  
and integration of Eq.\,(\ref{basic2}), assuming that $\alpha$,
$\cs$, $H_p$ and $\Omega$ are height-independent, yields
\begin{equation}
  \ln\frac{f(a,z)/f(a,0)}{\rho(z)/\rho(0)}
  ~=~ -\frac{\Omega\,\tstop(0)}{\alpha\,\,H_p^2}\,\int_0^z z'
              \,\exp\Big(\frac{z'^2}{2\,H_p^2}\Big)\,dz'  \ ,
  \label{eq:RiolsIntegral}
\end{equation}
which reproduces Eq.\,(33) in \citet{Riols2018}. The analytical
solution of Eq.\,(\ref{eq:RiolsIntegral}) is
\begin{equation}
  \frac{H_p^2}{H_a^2(z)}
  ~=~ 1 + \frac{\Omega\,\tstop(0)}{\alpha}\;
      \left(\frac{2\,H_p^2}{z^2}\right)
      \bigg(\exp\Big(\frac{z^2}{2\,H_p^2}\Big)-1\bigg) \ .
  \label{Riols} 
\end{equation}
The decrease of $H_a$ with respect to Eq.\,(\ref{Dubrulle}) is
remarkable, the function $(2/x^2)\big[\exp(x^2/2)-1\big]$ is 3.2 at
$x\!=\!2$, 20 at $x\!=\!3$, and 370 at $x\!=\!4$.  Infrared emission
lines from discs are quite typically emitted from layers located at
about $3-4$ scale-heights \citep{Woitke2018b}.
The Riols \& Lesur settling description reduces the local dust/gas
ratios by orders of magnitude in these regions (see
Fig.\,\ref{fig:discmodel}), even the abundance of the smallest grains
at high altitudes.  We suggest that Eq.~(\ref{Riols}) should become
the standard in disc modelling, have implemented it in {\sc ProDiMo},
and used it in this paper.

However, in order to keep consistency with previous publications, in
{\sc ProDiMo}, we multiply the second terms on the right side of
Eqs.~(\ref{Dubrulle}) and (\ref{Riols}) by a factor
$\sqrt{1+\gamma_0}$, with $\gamma_0\!=\!2$ for compressible
turbulence, and incorrectly use $\cs$ instead of $\vth$ in
Eq.\,(\ref{vdr}). These two changes amount to a factor of
$\sqrt{8\times 3/\pi}\approx 2.8$, which can be compared to the
uncertainty in the settling parameter $\alpha$. Equation (\ref{Riols})
can be generalised to a numerical integration of Eq.~(\ref{basic1})
for cases where $\rho(z)=\rho(0)\exp(-\frac{z^2}{2\,H_p^2})$ is not
valid, and $\cs = H_p\,\Omega$ is not valid, for example in hydrostatic
disc models where both the gas temperature and the mean molecular
weight depend on $z$.

\subsection{Rounded inner rims}

As previous Spitzer and recent JWST observations show
\citep{Banzatti2017,Kospal2023,Grant2023,Tabone2023,Banzatti2023}, the
mid-IR line spectroscopy of discs first and foremost probes the
physics and chemistry in the innermost disc regions $\la 1\,$au, close
to where the dust reaches its sublimation temperature, with line
emitting radii sometimes as small as 0.1\,au for T\,Tauri
stars. Therefore, the shape of the inner rim becomes a significant
factor for understanding the new JWST spectra.  Our knowledge about
these inner disc regions is still rather limited, because of both the
difficulty to observe these regions with sufficient resolution and the
complexity of physics involved. According to radiation hydrodynamics
models, the direct stellar irradiation causes the dust around the
inner rim to evaporate and form a puffed-up inner rim
\citep{Natta2001, Dullemond2001, Dullemond2010}. The radial transition
in terms of gas surface density across the inner rim is not an
infinitely sharp jump, but a relatively diffuse region, so that the
gas pressure stabilises the rim \citep{Isella2005, Kama2009,
  Woitke2009, Flock2017}.  In addition, evolutionary MHD models
suggest that the switch from magnetically active inner disc to the
dead zone may cause a local pressure maximum and formation of gas-dust
rings \citep{Masset2006, Dzyurkevich2010, Kadam2019, Kadam2022}, which
causes a more gradual buildup of the column density towards the first
pressure bump. Inspired by the inner disc structures suggested by
these models, here we introduce a parametric approach, assuming
\begin{align}
  \Sigma(r) &~\propto~
       \exp\bigg[-\bigg(\frac{R_{\rm soft}}{r}\bigg)^s\bigg]\;\; 
       r^{\,-\epsilon}\;
       \exp\bigg[-\bigg(\frac{r}{R_{\rm tap}}\bigg)^{2-\gamma}\bigg]
       \ ,\label{eq:coldens}\\
  R_{\rm soft} &~=~ {\rm raduc} \times R_{\rm in}
  \ ,\nonumber\\
  s &~=~ \frac{\ln\big[\ln(1/{\rm reduc})\big]}{\ln({\rm raduc})}
  \ ,\nonumber
\end{align}
where $\Sigma(r)$ is the mass column density at radius $r$, ${\rm
  raduc}$ is the radius up to which the column density initially
increases, in units of the inner disc radius $R_{\rm in}$, and ${\rm
  reduc}$ is the factor by which the column density is reduced in
Eq.\,(\ref{eq:coldens}) at $R_{\rm in}$. As in previous publications,
$R_{\rm tap}$ is the tapering off radius, and the self-similar
solution is obtained for $\gamma=\epsilon$, but we rather keep
$\epsilon$ and $\gamma$ as independent parameters.

In hydrostatic models, where the hydrostatic equilibrium is solved in
each vertical column \citep{Natta2001, Kama2009, Dullemond2010}, we
see large scale-heights at the inner rim that is directly heated by
the star, followed by a dip of $H_p(r)$ behind the inner rim, because
the inner rim casts a shadow on the subsequent disc where the
temperatures are lower, see e.g.\ Fig.~9 in \citet{Woitke2009} and
Fig.~6 in \citet{Woitke2016}.  We introduce two more parameters to
describe this ubiquitous feature of hydrostatic models
\begin{align}
  H_p(r) &~=~ H_0 \left(\frac{r}{r_0}\right)^\beta \; f(r)\ ,\\
  f(r) &~=~ 1-{\rm deduc}\times\exp
  \bigg(-\Big(\frac{\Delta r}{w}\Big)^2\bigg) \nonumber\ ,\\
  \Delta r &= \ln(r/R_{\rm in})-\ln({\rm daduc}) \nonumber\ ,\\
  w &~=~ \left\{\begin{array}{rl}
       \frac{1}{2}\ln(\rm deduc) & \mbox{, if $\Delta r<0$}\\
                  \ln(\rm deduc) & \mbox{, otherwise}
       \end{array}\right. \nonumber\ ,
\end{align}
where ${\rm daduc}$ is the dip radius in units of the inner radius,
and ${\rm deduc}$ is the reduction of the scale height at the dip
radius. The effect of the new disc shape parameters reduc, raduc,
deduc and daduc on column density structure and scale heights are
visualised in Fig.~\ref{fig:discmodel}, upper panel.

\section{The ProDiMo disc model for EX\,Lupi}
\label{sec:EXLup}

\subsection{Stellar setup}
\label{sec:star}

Our photometric data collection for the quiescent state of EX\,Lupi is
shown in Table~\ref{tab:photo}, and the assumed stellar properties in
Table~\ref{tab:parameter}.  Stellar luminosity, effective temperature
and optical extinction are taken from \citet{Sipos2009}. Stellar
mass, distance to EX\,Lupi, and disc inclination are taken from the
collection of ALMA observations by \citet{White2020}.  The stellar
excess UV parameters are derived from XMM UV photometric observations
(see Table~\ref{tab:photo}). The quantity $f_{\rm UV}\!=\!L_{\rm
  UV}/L_\star$ is the UV excess between 91.2\,nm and 250\,nm. $p_{\rm
  UV}$ is a powerlaw index explained in Appendix~A of
\citet{Woitke2016}. The X-ray properties are explained in Appendix~A
of \citet{Woitke2016} and are taken from measurements of
\citet{Teets2012}, which indicate a significant increase of the X-ray
luminosity during burst phase, in particular for the soft X-rays. The
accretion rate $\Mdot\!=\!4\times10^{-10}\,M_\odot\rm/yr$ is taken from
\citet{Aguilar2015}, which is used for viscous gas heating
according to Eq.~(2) in \citep{Dalessio1998}, see details in
\citep{Woitke2019}.  

\begin{table}
\caption{Photometric data collection for the quiescent state of EX\,Lupi.}
\begin{center}
\vspace*{-6mm}     
\label{tab:photo}
\resizebox{90mm}{!}{\begin{tabular}{c|cc|cc}
  \hline
  &&&&\\[-2.2ex]
  $\lambda\,[\mu\rm m]$ & $F_\nu\,[\rm Jy]$ & $\sigma\,[\rm Jy]$ & filter & reference\\
  \hline
  \hline
  &&&&\\[-2.2ex]
   0.231  & 1.21(-3)  & 0.01(-3)  & XMM.UVM2   & VIZIER\\
   0.291  & 2.68(-3)  & 0.01(-3)  & XMM.UVW1   & VIZIER\\
   0.344  & 4.62(-3)  & 0.01(-3)  & XMM.U      & VIZIER\\
   0.444  & 1.90(-2)  & 1.3(-2)   & JOHNSON.B  & VIZIER\\
   0.482  & 2.43(-2)  & 1.29(-2)  & SDSS.G     & VIZIER\\ 
   0.554  & 3.55(-2)  & 1.47(-2)  & JOHNSON.V  & VIZIER\\ 
   0.625  & 5.12(-2)  & 1.65(-2)  & SDSS.RP    & VIZIER\\  
   0.763  & 9.04(-2)  & 1.68(-2)  & SDSS.IP    & VIZIER\\  
   0.902  & 0.113     & 0.004     & SDSS.Z     & VIZIER\\ 
   1.24   & 0.203     & 0.004     & 2MASS.J    & VIZIER\\
   1.63   & 0.278     & 0.004     & JOHNSON.H  & VIZIER\\
   1.65   & 0.274     & 0.006     & 2MASS.H    & VIZIER\\
   2.16   & 0.270     & 0.005     & 2MASS.KS   & VIZIER\\
   2.19   & 0.261     & 0.005     & JOHNSON.K  & VIZIER\\
   3.6    & 0.190     & 0.004     & IRAC.36    & \cite{Sipos2009}\\
   4.5    & 0.206     & 0.004     & IRAC.45    & \cite{Sipos2009}\\
   5.8    & 0.236     & 0.005     & IRAC.58    & \cite{Sipos2009}\\
   8.0    & 0.317     & 0.007     & IRAC.80    & \cite{Sipos2009}\\
   60     & 1.25      & 0.25      & IRAS.F60   & VIZIER\\
  66.3    & 1.23      & 0.246     & AKARI.N60  & VIZIER\\   
  87.9    & 1.21      & 0.09      & \!\!AKARI.WIDES\!\!\! & VIZIER\\
    70    & 1.03      & 0.34      & PACS.70    & PACS pointsource\\
   100    & 0.99      & 0.33      & PACS.100   & PACS pointsource\\
   100    & 1.27      & 0.17      & IRAS.100   & \cite{Sipos2009}\\
   160    & 0.891     & 0.30      & PACS.160   & PACS pointsource\\
   250    & 0.554     & 0.02      & SPIRE.250  & SPIRE pointsource\\
   350    & 0.311     & 0.023     & SPIRE.350  & SPIRE pointsource\\
   500    & 0.126     & 0.021     & SPIRE.500  & SPIRE pointsource\\
   850    & 4.1(-2)   & 1(-2)     & LABOCA     & \cite{Juhasz2012}\\
  1300    & 1.9(-2)   & 0.4(-2)   & SMA        & \!\!\cite{Lommen2010}\!\!\!\\
  1300    & \!\!17.37(-3)\!\! & 0.15(-3)          & ALMA & \cite{Hales2018}\\
  3000    & \!\!2.720(-3)\!\! & \!\!0.013(-3)\!\! & ALMA & \cite{White2020}\\
  \hline
\end{tabular}}
\end{center}
\vspace*{-2mm} \small The notation $a(-b)$ means $a\times10^{-b}$. The
Herschel PACS and SPIRE point source catalogues are available via
\url{https://irsa.ipac.caltech.edu/applications/Gator}.  The VIZIER
photometric data collections are available at
\url{https://vizier.cds.unistra.fr}.
\vspace*{-2mm}
\end{table}

Recently, \citet{Cruz2023} have
analysed an {\sc XShooter} spectrum of EX\,Lupi taken in July~2022,
just one month before the JWST observations were carried out. This
$0.3-2.2\,\mu$m spectrum suggests a much higher mass accretion rate of
$\Mdot\!\approx\!(2-3)\times10^{-8}\,M_\odot\rm/yr$. Another value has
been reported by \cite{Wang2023},
$\Mdot\!\approx\!3\times10^{-9}\,M_\odot\rm/yr$. The data analysis is
intertwined with the determination of the optical extinction $A_V$ of
the star.  Cruz-S\'aenz de Miera et al.\ arrive at a large value of
$A_V\!\approx\!1.1$ using $R_V\!=\!3.1$.  In contrast, most published
work on EX\,Lupi have so far assumed $A_V\!\approx\!0$, including
\citep{Sipos2009}.  \citet{Varga2018} have used $A_V\!=\!0.2$,
\cite{Alcala2017} have used $A_V\!=\!1.1$, and \cite{Wang2023} have
used $A_V\!=\!0.1$.  If $A_V$ is indeed as large as 1.1, the XMM
photometric data would imply a $10\times$ larger UV luminosity
($f_{\rm UV}\!\approx\!0.1$). Extrapolating the slab model for the
continuum emission due to accretion of Cruz-S\'aenz de Miera et
al.\ to shorter wavelengths, \'Abrah\'am (2023, private communication)
derive $f_{\rm UV}\!\approx\!0.02$, but this UV excess would not
contain the strong emission lines such as Ly-$\alpha$, due to stellar
activity. A discussion how our modelling results change if larger
values of $f_{\rm UV}$ and $\Mdot$ are used can be found in
Sect.~\ref{sec:depend}.

{\cite{Ansdell2018} observed EX\,Lupi with ALMA in band~6
  continuum and $\rm^{12}CO$ 2-1, measuring a continuum disc radius of
  $62\pm2$\,au and a CO disc radius of $178\pm12$\,au.
  \citet{Hales2018} also imaged the disc of EX\,Lupi with ALMA band~6
  continuum and CO isotopologues, and fitted a disc model with a
  tapering-off radius of $R_{\rm tap}\!\approx\!20\,$au.
  Based on SPHERE-IRDIS polarimetric observations,
  \cite{Rigliaco2020} favoured a scenario where the near-IR light is
  scattered by a circumstellar disk rather than an outflow, and
  suggest that the disc region between about 10 and 30\,au might be depleted
  in $\mu$m-sized grains. \cite{Furlan2009} introduced an index that
  characterises the mid-IR SED slope by}
\begin{equation}
  n_{13-31} = \frac{\log(\nu_{31}F_{\nu_{31}})-\log(\nu_{13} F_{\nu_{13}})}
                  {\log(\lambda_{31})-\log(\lambda_{13})} \ ,
\end{equation}
{where $\lambda_{13}\!=\!13\,\mu$m, $\lambda_{31}\!=\!31\,\mu$m,
  $\nu_{13}$ and $\nu_{31}$ are the corresponding frequencies, and
  $F_{\nu_{13}}$ and $F_{\nu_{31}}$ are the continuum fluxes [Jy] at
  $\lambda_{13}$ and $\lambda_{31}$, respectively. This index has been
  used to identify transitional discs $(n_{13-31}\!>\!1)$, disc
  flaring and dust settling \citep{Furlan2009}. From the Spitzer/IRS
  spectrum of EX\,Lupi shown by \citet{Kospal2023}, with reference to
  \citet{Abraham2009}, we derive a value of
  $n_{13-31}\!\approx\!-0.57$, which suggests a continuous
  disc.}

{\cite{Banzatti2023} defined two classes of discs: (i) compact
  discs with strong low-excitation water lines and (ii) large discs,
  possibly with gaps and rings, with faint low-excitation water lines.
  Banzatti et al.\ associated these classes with different water ice
  transport mechanisms by pebbles, favouring water enrichment in case
  (i), and water depletion in case (ii). Based on the published dust
  radii cited above, and the derived $n_{13-31}$-value, EX~Lupi is a
  relatively large but continuous disc, which does
  not clearly fall into either of these categories. We have not used
  any ALMA data in this paper for the fitting, and have assumed
  standard element abundances for our disc model, see
  Tab.\,\ref{tab:abund}.}

\begin{table}
\begin{center}
\caption{Model parameters for the quiescent state of EX\,Lupi.}
\vspace*{-6mm}
\label{tab:parameter}
\resizebox{90mm}{!}{\begin{tabular}{l|c|c}
\hline
&&\\[-2.2ex]
quantity & symbol$^{\,(1)}$ & value\\
&&\\[-2.2ex]
\hline 
\hline 
&&\\[-2.2ex]
stellar mass                      & $M_{\star}$      & $0.6\,M_\odot$\\
stellar luminosity                & $L_{\star}$      & $0.5\,L_\odot$\\ 
effective temperature             & $T_{\star}$      & $3800\,$K\\
UV excess                         & $f_{\rm UV}$     & $0.011^{\,(2)}$\\
UV powerlaw index                 & $p_{\rm UV}$     & $1.3$\\
X-ray luminosity                  & $L_X$     & $2\times 10^{29}\rm erg/s$\\
X-ray emission temperature        & $T_X$          & $3\times10^7$\,K\\
accretion rate                    & $\Mdot$      &
                                 \!\!$4\times10^{-10}M_\odot$/yr$^{\,(2)}$\!\!\!\\
optical extinction                & $A_V$          & $0^{\,(2)}$\\
\hline
&&\\[-2.2ex]
strength of interstellar UV       & $\chi^{\rm ISM}$ & 1\\
cosmic ray H$_2$ ionisation rate  & $\zeta_{\rm CR}$   
                                          & $1.7\times 10^{-17}$~s$^{-1}$\\
\hline
&&\\[-2.2ex]
disc gas mass$\,^\star$              & $M_{\rm disc}$   & $9.61\times10^{-3}\,M_\odot$\\
disc dust mass$\,^\star$             & $M_{\rm dust}$   & $1.43\times10^{-5}\,M_\odot$\\
inner disc radius$\,^\star$          & $R_{\rm in}$     & 0.0329\,au\\
tapering-off radius$\,^\star$        & $R_{\rm tap}$    & 10.2\,au\\
column density power index$\,^\star$ & $\epsilon$     & -1.19\\
tapering off power index            & $\gamma$       & 1.0\\
extension of inner rim$\,^\star$     & raduc          & 1.19\\
maximum $\Sigma$ reduction          & reduc          & $10^{-4}$\\
dip radius$/R_{\rm in}\,^\star$        & daduc          & 2.95\\
scale height reduction$\,^\star$     & deduc          & 0.290\\
reference gas scale height$\,^\star$ & $H_p(3\,{\rm au})$  & 0.376\,au\\
flaring power index$\,^\star$        & $\beta$        & 0.951\\ 
\hline
&&\\[-2.2ex]
minimum dust particle radius$\,^\star$ & $a_{\rm min}$      & $0.00104\,\mu$m\\
maximum dust particle radius$\,^\star$ & $a_{\rm max}$      & $4.81\,$mm\\
dust size dist.\ power index$\,^\star$ & $a_{\rm pow}$      & 3.87\\
settling method                       & settle\_method   & Riols\,\&\,Lesur\\
settling parameter$\,^\star$           & $\alpha$         & $4.78\times10^{-4}$\\
max.\ hollow volume ratio             & $V_{\rm hollow}^{\rm max}$  & 80\%\\
&&\\[-2.0ex]
dust composition             & $\rm Mg_{0.7}Fe_{0.3}SiO_3$ & 50.1\%\\
(volume fractions)$\,^\star$  & amorph.\,carbon  & 24.9\%\\
                             & porosity         & 25\%\\
\hline
&&\\[-2.2ex]
PAH abundance rel.\ to ISM$\,^\star$ & $f_{\rm PAH}$        & 0.0292\\
ratio of charged/neutral PAHs       & PAH\_charged       & 0.8\\
chemical heating efficiency$\,^\star$ & $\gamma^{\rm chem}$ & 0.619\\
\hline   
&&\\[-2.2ex]
distance                          & $d$              & 157.7\,pc\\
disc inclination                  & $i$              & 32\degr\\
\hline
\end{tabular}}
\end{center}
\vspace*{-2mm}
\small $^{(1)}$: See \citet{Woitke2016} for details and definitions of
parameters.\\
$^{(2)}$: These values are debated in the literature, see text.\\
$^\star$: This parameter was optimised during genetic fitting,
final value given with three digits.
\vspace*{-1mm}
\end{table}

\subsection{Disc shape}

Finding a {\sc ProDiMo} model setup that can roughly reproduce the
observed JWST line and continuum data was a difficult task.  Looking
at Fig.~3 of \citet{Kospal2023}, the water emission lines and the peak
emissions at the Q-branches of \ce{C2H2}, \ce{HCN} and \ce{CO2} in the
13-17\,$\mu$m spectral region all reach similar flux levels of the
order of 30-60\,mJy. This immediately tells us that all these spectral
features must be optically thick, otherwise the certainly very
different concentrations and masses of these molecules in the disc
would result in very different flux levels. This conclusion is
underlined by slab model fits presented by \citet{Kospal2023}, which
resulted in column densities of order $10^{15}-10^{16}\rm\,cm^{-2}$
for \ce{C2H2}, \ce{HCN} and \ce{CO2}, $10^{17}-10^{18}\rm\,cm^{-2}$
for \ce{H2O}, and approximately $10^{19}\rm\,cm^{-2}$ for CO.

In our standard T\,Tauri disc model {(with parameters listed in
  Table~\ref{tab:parameter_standard})}, the lines of \ce{C2H2} and
\ce{HCN} are emitted from radially extended, tenuous layers, high in
the disc (see Fig.\,\ref{fig:molconc}), where UV photons dissociate CO
and \ce{N2}, and the resulting carbon and nitrogen atoms react further
to form some \ce{C2H2} and \ce{HCN} in the presence of
\ce{H2}. {Although we can explain the overall line flux levels of
\ce{H2O} and \ce{CO2} from T\,Tauri stars this way, our standard
models} generally result in too low molecular column densities above
the optically thick dust when compared to the values derived from slab
model fits to observations \citep[e.g.][]{Banzatti2017}.

\begin{figure*}
  \centering
  \begin{tabular}{cc}
    \hspace*{-5mm}
    \includegraphics[width=79mm,height=52mm,trim=0 0 0 0,clip]
                    {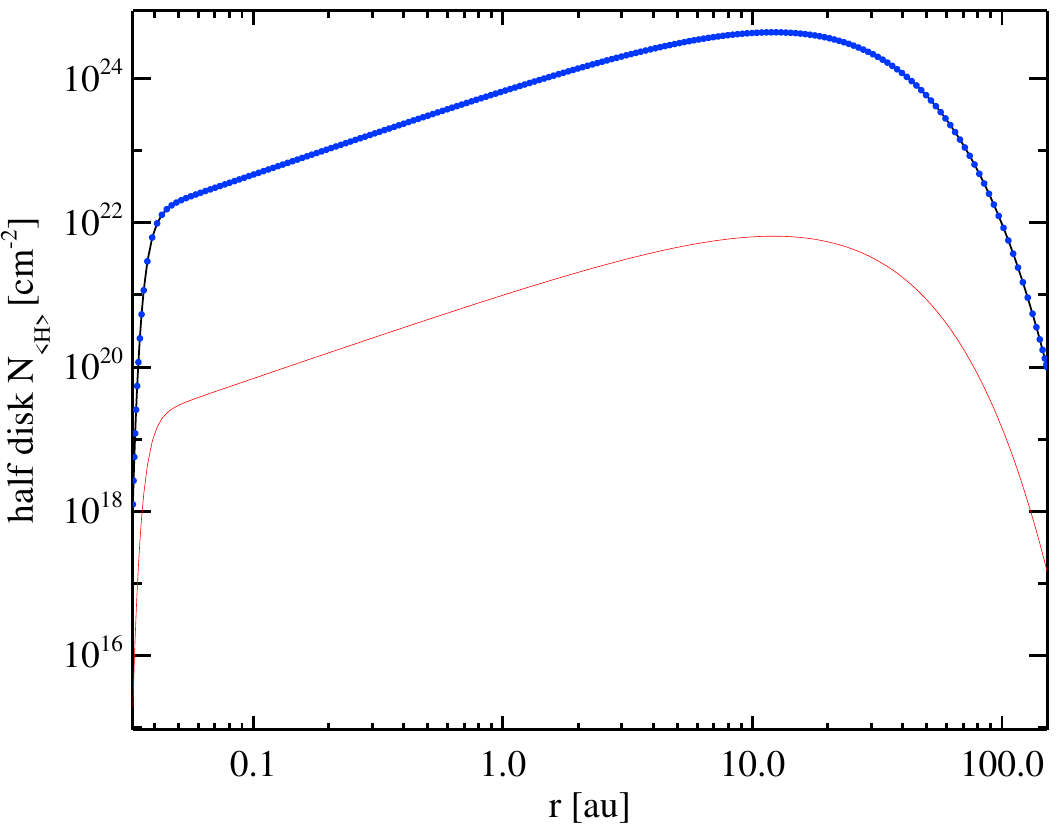} &
    \hspace*{-5mm}
    \includegraphics[width=79mm,height=52mm,trim=0 0 0 0,clip]
                    {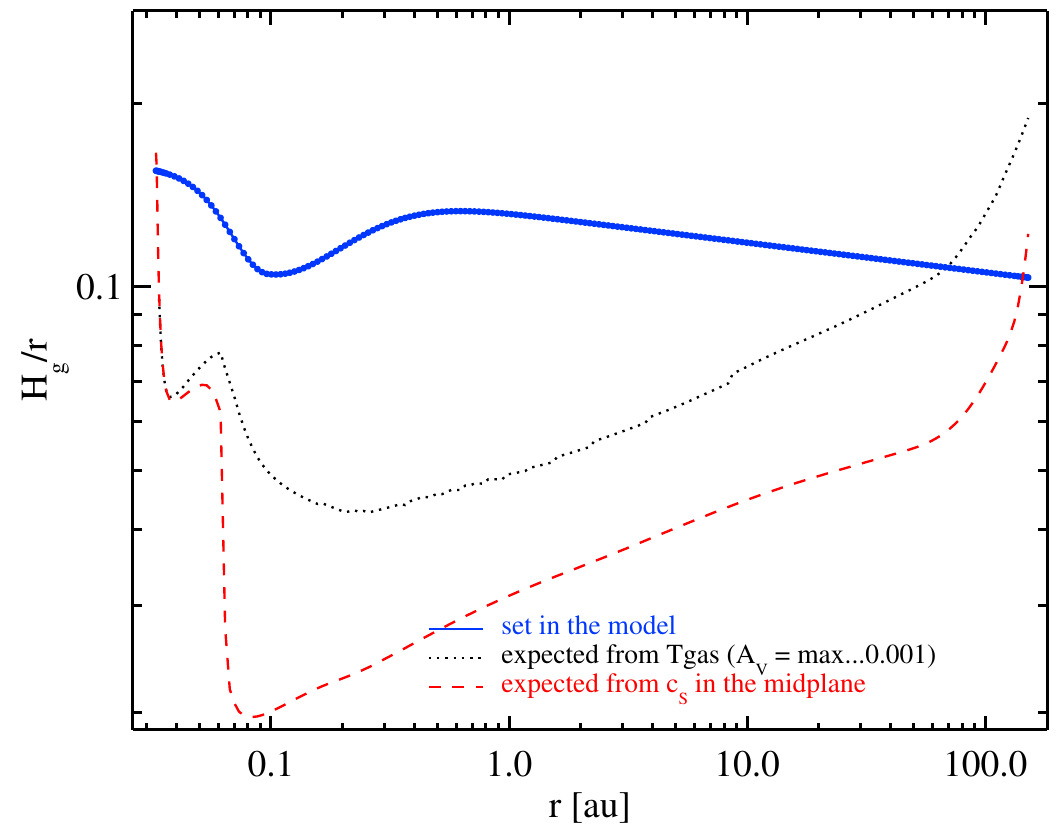}\\[-1mm]
    \hspace*{-5mm}
    \includegraphics[width=79mm,trim=0 0 0 0,clip]
                    {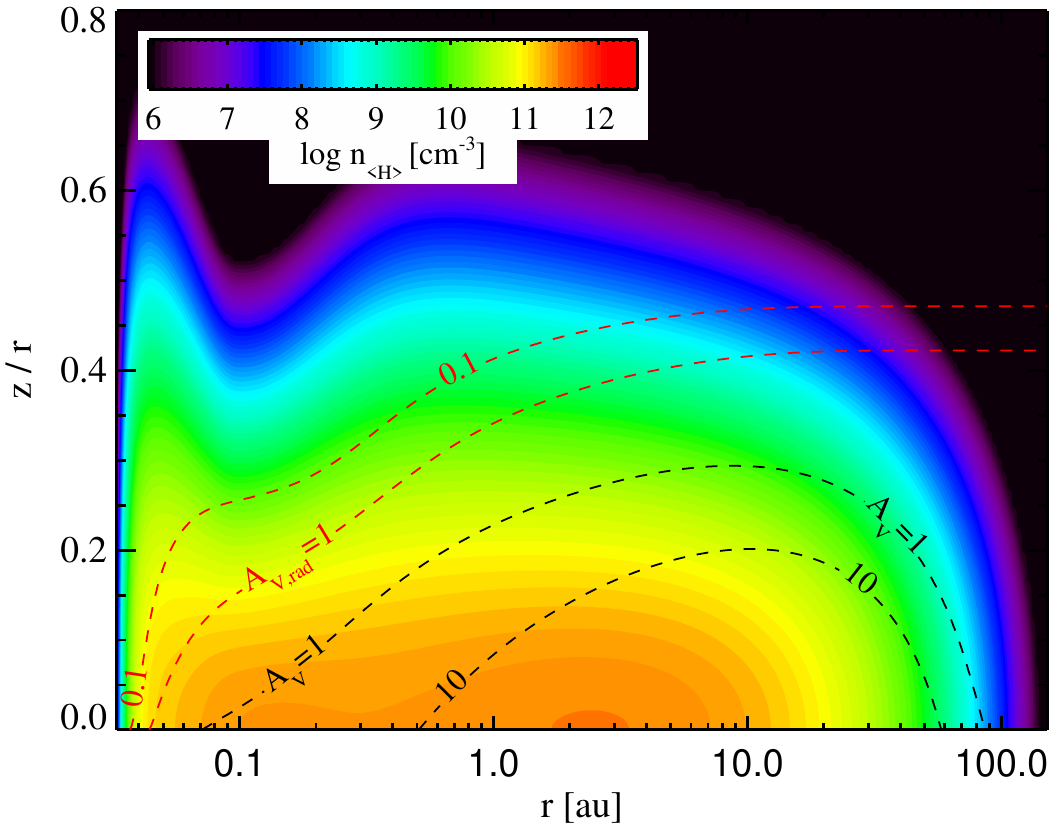} &
    \hspace*{-5mm}
    \includegraphics[width=79mm,trim=0 0 0 0,clip]
                    {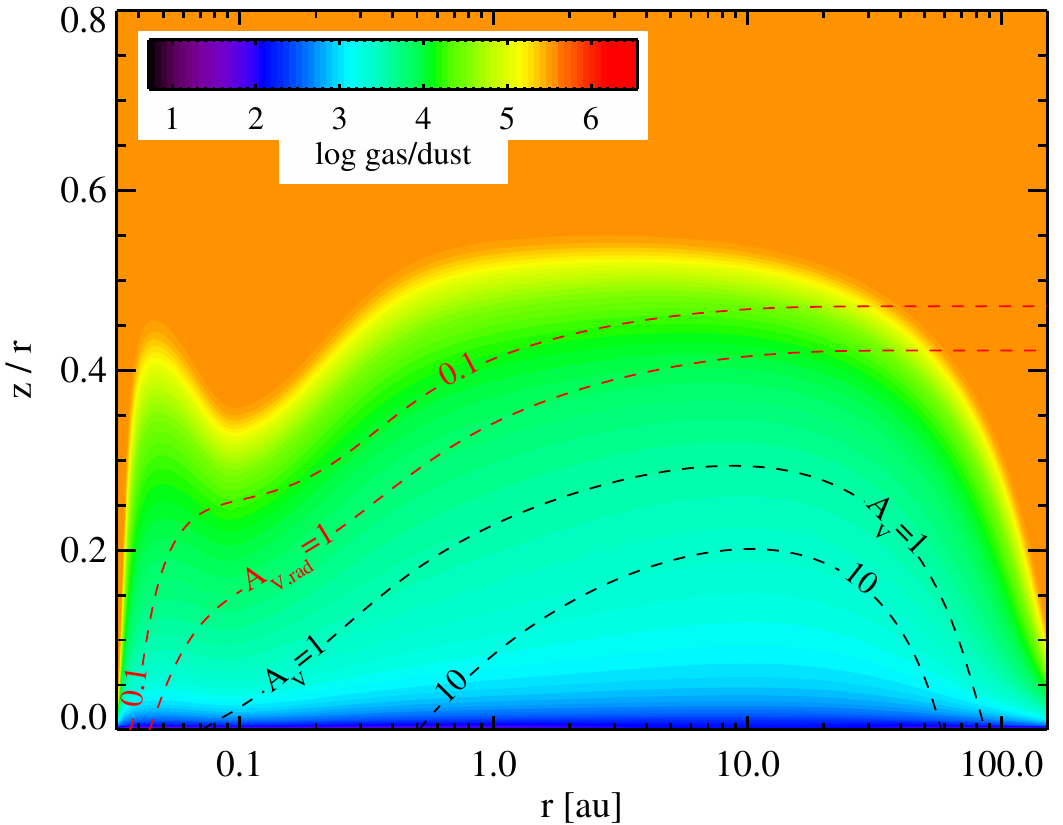}\\[-4mm]
    \hspace*{-5mm}
    \includegraphics[width=79mm,trim=0 0 0 0,clip]
                    {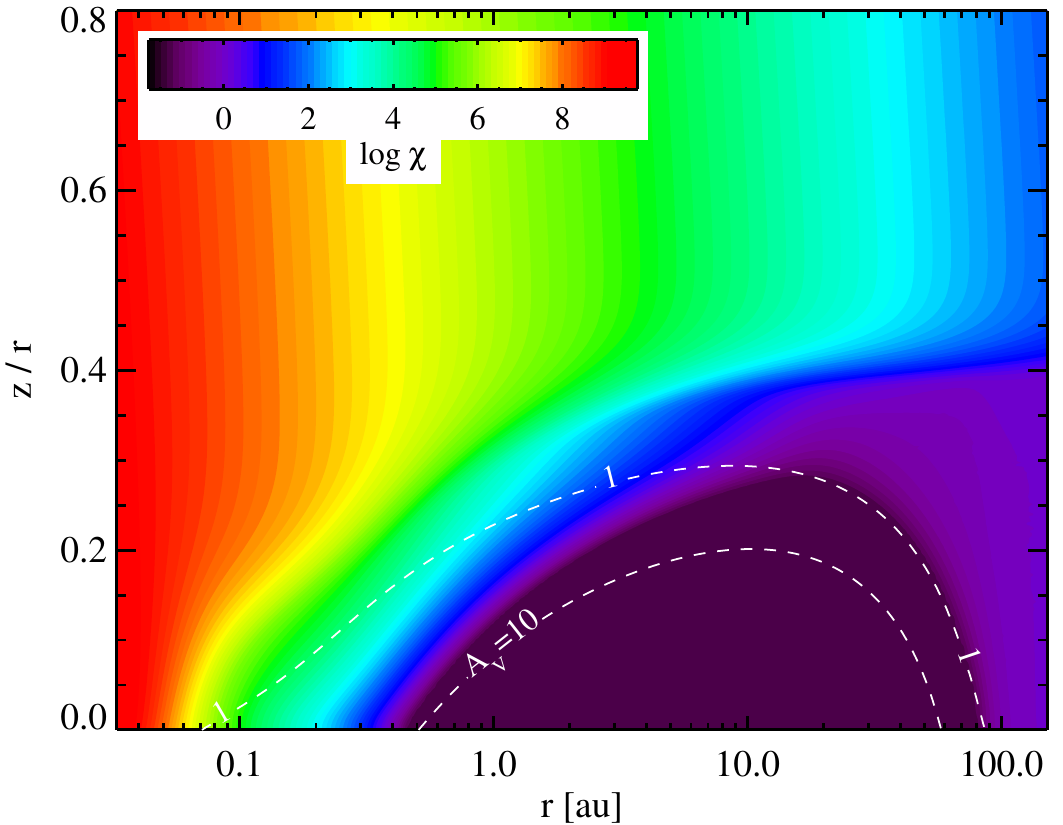} &
    \hspace*{-5mm}
    \includegraphics[width=79mm,trim=0 0 0 0,clip]
                    {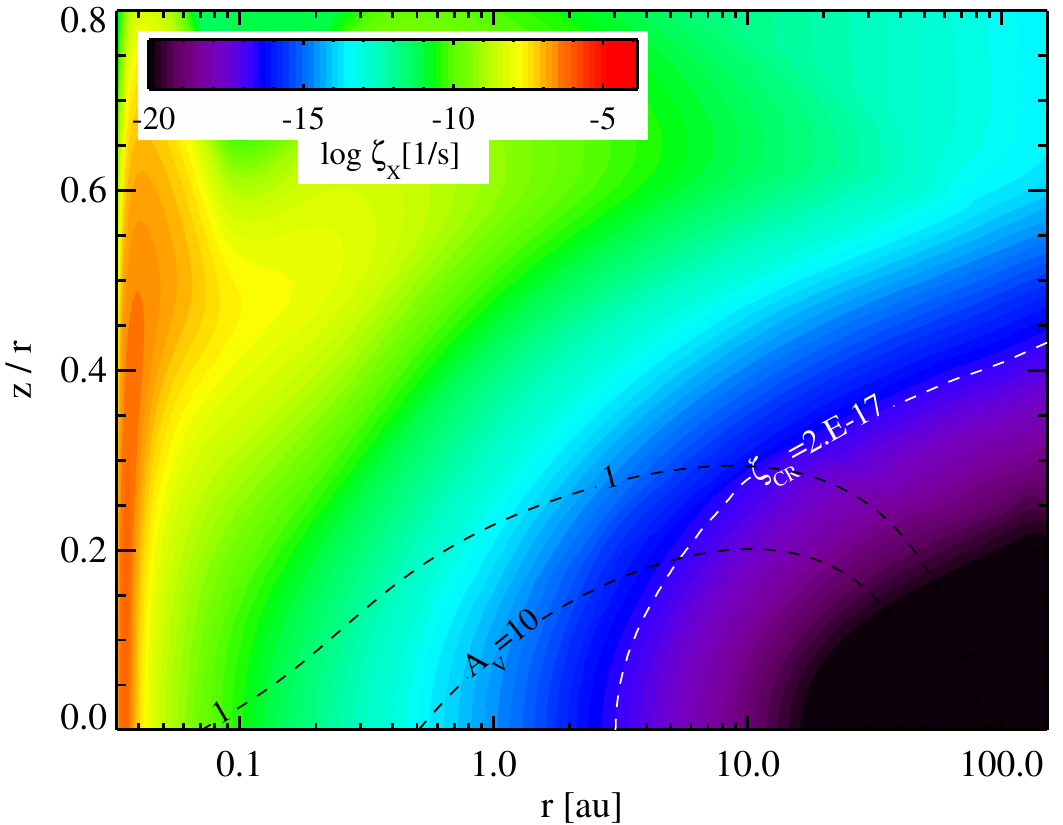}\\[-4mm]
    \hspace*{-5mm}
    \includegraphics[width=79mm,trim=0 0 0 0,clip]
                    {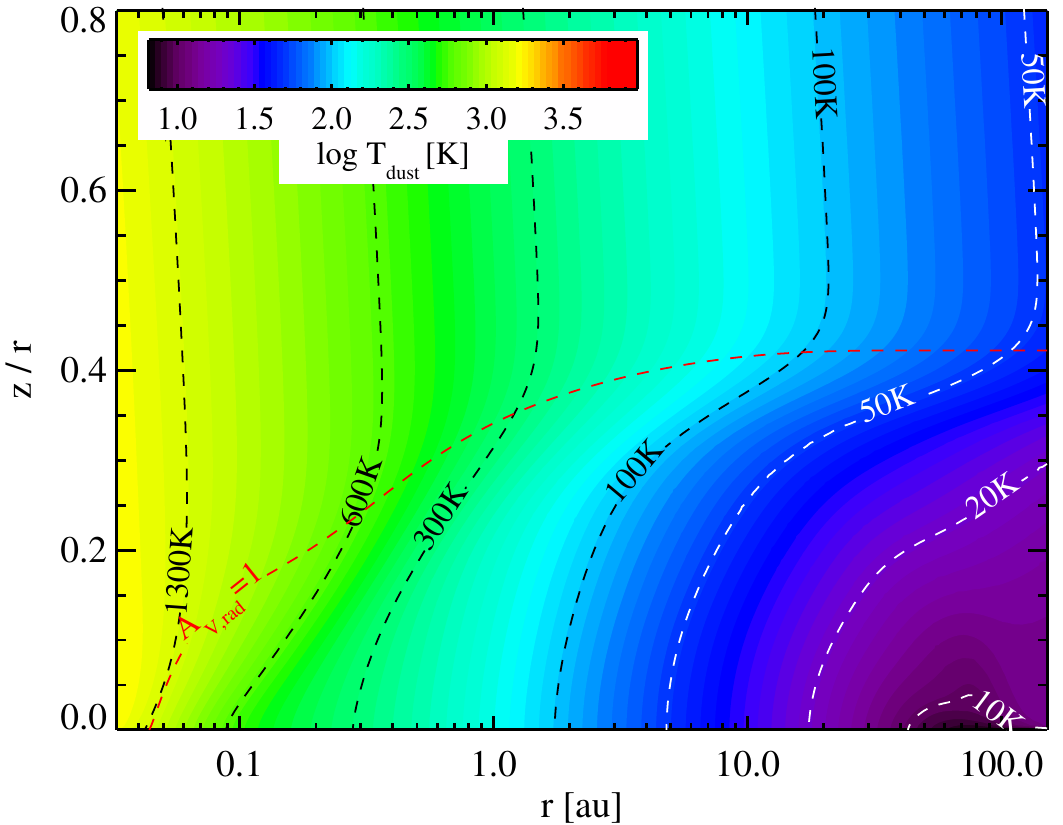} &
    \hspace*{-5mm}
    \includegraphics[width=79mm,trim=0 0 0 0,clip]
                    {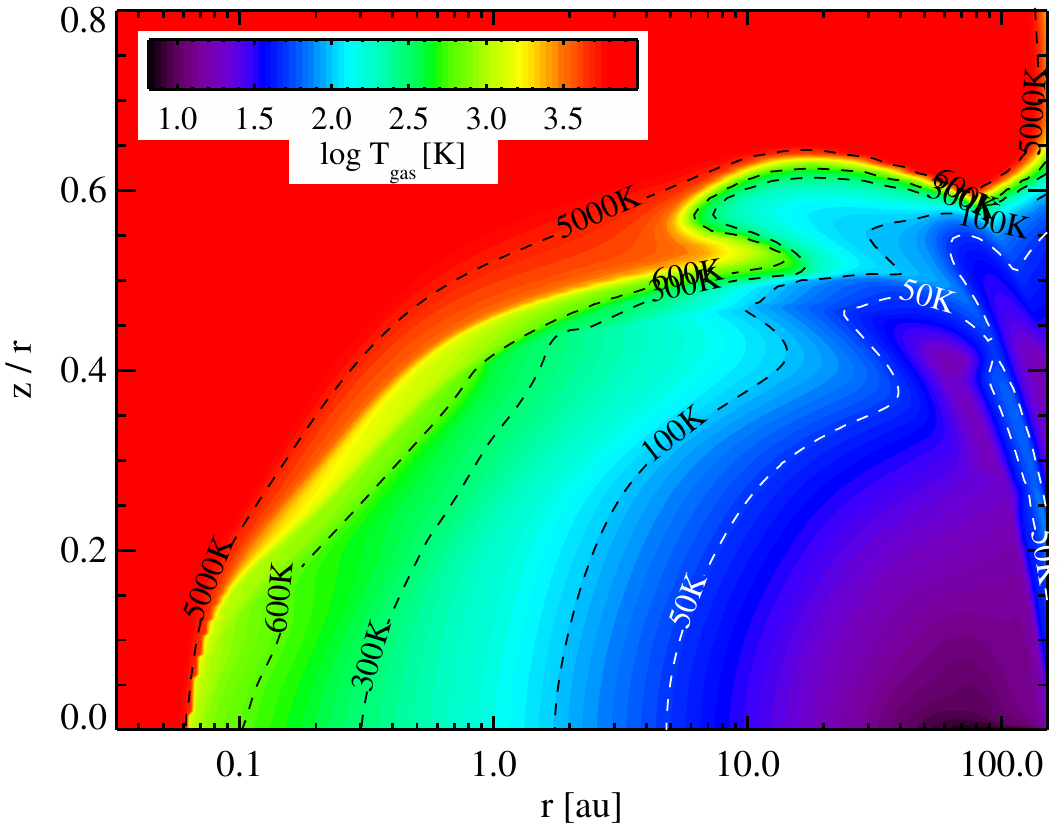}\\[-1mm]
  \end{tabular}
  \caption{{\sc ProDiMo} model for the quiescent state of EX\,Lupi.  The top
    plots show the assumed gas and dust column densities (left) and
    scale heights (right). The second row shows the corresponding gas
    densities (left) and gas to dust mass ratio after settling
    (right). The third row shows UV field strength $\chi$ (left),
    and the X-ray ionisation rate $\zeta_X$ in comparison to the
    cosmic ray ionisation rate (right). The bottom row shows the resulting dust
    (left) and gas (right) temperature structures. Additional contour
    lines include the vertical optical extinction $A_V$ and the radial
    optical extinction $A_{V,\rm rad}$ as indicated.}
  \label{fig:discmodel}
\end{figure*}

The key idea for the EX\,Lupi model presented in this paper was to tap
on the large reservoir of \ce{C2H2} and \ce{HCN} molecules that is
present in our disc models in the midplane regions inside of 1\,au,
which are normally covered by dust opacity. To reveal these regions,
we assume a slowly increasing surface density profile around the inner
rim, combined with strong dust settling, which creates an almost
dust-free gas around the inner dust rim, see
Fig.~\ref{fig:discmodel}. The parameters for this model are listed in
Table~\ref{tab:parameter}. The position of the inner dust rim thereby
becomes wavelength-dependent. Using the radius where the vertical dust
optical depth is one, the results are about 0.1\,au at optical
wavelengths, but about 0.4\,au at $\lambda\!=\!15\,\mu$m in this model,
see Sect.~\ref{sec:means}. For comparison, \cite{Sipos2009} used 
an inner hole of radius 0.2\,au in their SED fits using RADMC to
reproduce the relatively flow continuum fluxes between 3 and 8\,$\mu$m.
This idea was later adopted by \citet{Juhasz2012} and
\citet{Abraham2019}. MIDI interferometry of EX\,Lupi was presented by
\cite{Varga2018}, who measured a half-light radius of $(0.77\pm0.09)$\,au.

\subsection{Model fitting}

Once an initial disc setup was found that could roughly
reproduce the observed SED and JWST molecular line emission features,
within a factor of about 2-5, we utilised the $(12,1)$ genetic
algorithm of \citet{Rechenberg2000}, as explained in
\citet{Woitke2019}, to optimise 17 disc shape, dust material
\& size, and PAH parameters.  The optimised parameters are marked in
Table~\ref{tab:parameter}.  The genetic fitting algorithm seeks to minimise
a total $\chi^2$ 
\begin{equation}
  \chi^2 = \frac{1}{4}\chi_{\rm phot}^2
         + \frac{1}{4}\chi_{\rm JWST}^2
         + \frac{1}{2}\chi_{\rm line}^2 
\end{equation}
to simultaneously fit the photometric data, the absolute JWST
spectrum, and the continuum-subtracted JWST line spectrum.
We used fast {\sc ProDiMo} models with a low resolution spatial grid
$70\times60$ during the fitting phase.  The models calculated the
SED on 500 wavelength points between 0.1\,$\mu$m and 4\,mm.  Based on
these results, we calculated the mean quadratic deviation between
model and photometric fluxes $\chi^2_{\rm phot}$, using the
various filter transmission functions, and the mean quadratic
deviation between model and the absolute JWST spectrum $\chi^2_{\rm JWST}$.
A minimum error of 10\% was assumed for all photometric and spectral
data points.

\begin{figure}
  \vspace*{2mm}
  \centering
  \includegraphics[width=86mm,height=70mm,trim=3 2 7 4,clip]
                  {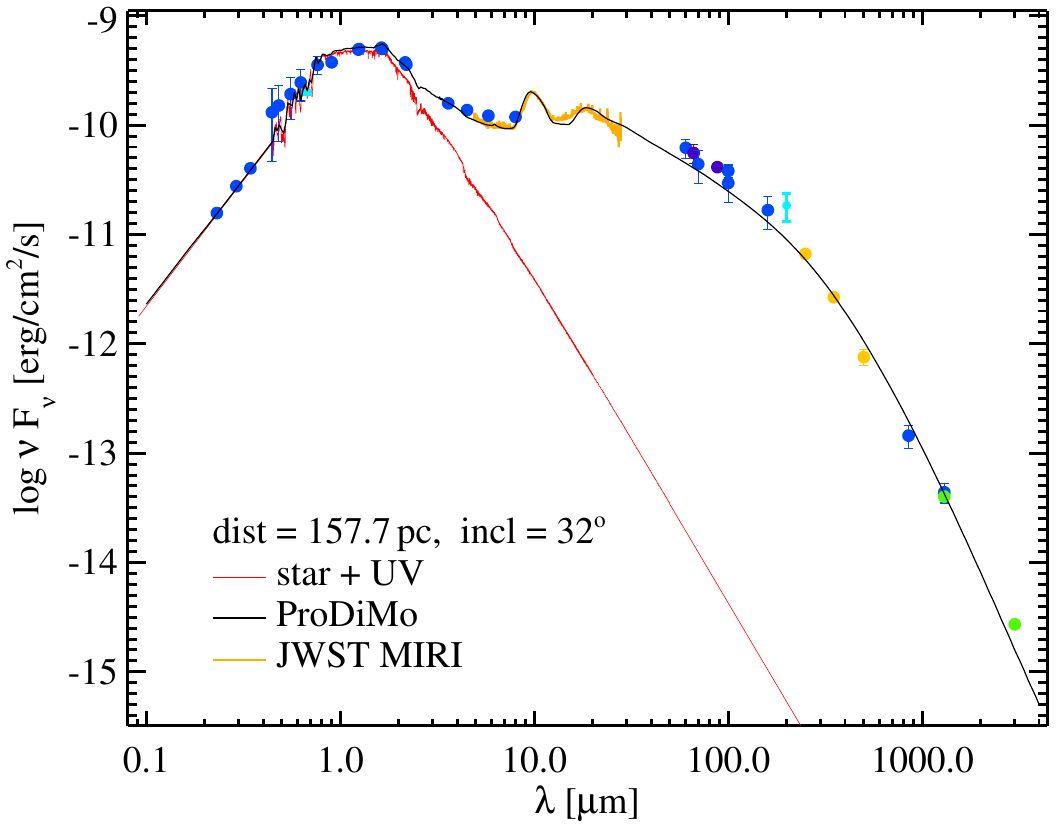}\\[1mm]
  \includegraphics[width=86mm,height=65mm,trim=3 2 7 4,clip]
                  {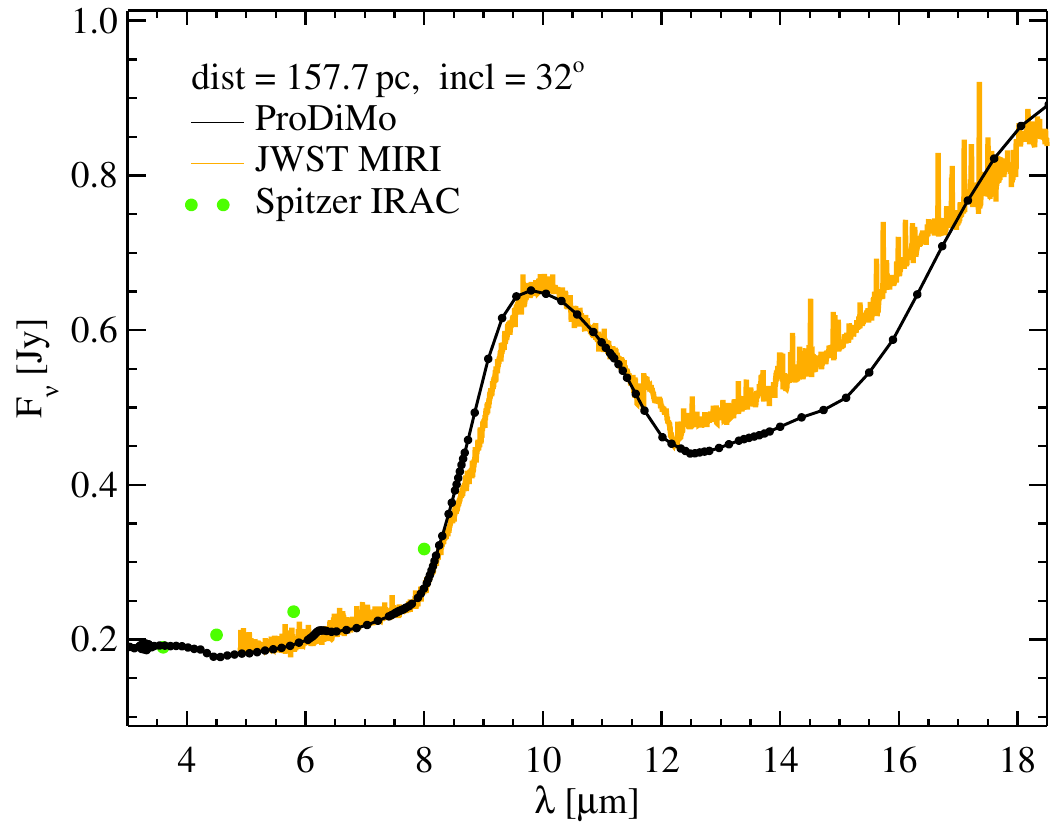}\\[0mm]
  \caption{Fit of the spectral energy distribution (SED) of EX\,Lupi in
    the quiescent state. The orange and green data points in the upper
    plot are Herschel/SPIRE and ALMA measurements. The orange spectrum
    is the JWST/MIRI data. The lower plot shows a magnification of the
    mid-IR spectral region.}
  \label{fig:SED}
\end{figure}

Each {\sc ProDiMo} call was followed by two calls of the Fast Line
Tracer {\sc FLiTs} developed by M.~Min \citep{Woitke2018}, to
calculate high-resolution model spectra between 4.9 and 7.5\,$\mu$m
(10353 lines) and between 13 and 17.7\,$\mu$m (17228 lines),
respectively, both with a velocity resolution of 3\,km/s. Here we
considered the atoms and ions \ce{Ne+}, \ce{Ne++}, \ce{Ar+},
\ce{Ar++}, \ce{Mg+}, \ce{Fe+}, \ce{Si+}, \ce{S}, \ce{S+} and \ce{S++},
and the molecules \ce{H2O}, \ce{OH}, \ce{HCN}, \ce{CO2}, \ce{NH3},
\ce{H2}, \ce{C2H2}, \ce{CH4} and \ce{CO}.  Each FLiTs call calculates
two spectra, a continuum model with only dust opacities and source
functions, and a line\,\&\,continuum model with dust and gas opacities
and source functions.  We then subtracted these two, and convolved the
continuum-subtracted {\sc FLiTs} spectrum to a spectral resolution of
2500\footnote{The spectral resolution of the MIRI instrument depends
  on wavelength, see
  \url{https://jwst-docs.stsci.edu/jwst-mid-infrared-instrument/miri-observing-modes/miri-medium-resolution-spectroscopy},
  varying between about 2000 and 3000, so our choice here is a
  simplification}.  To reduce the observational noise, we applied a
box-filter of size $\lambda^{\rm obs}_{i+1}-\lambda^{\rm obs}_{i-1}$,
where $\lambda^{\rm obs}_i$ are the observational data points, to both
the observational data and the convolved model spectrum. Finally, we
calculated the mean quadratic deviation between the
continuum-subtracted FLiTs model and continuum-subtracted JWST data as
$\chi^2_{\rm line}$, where a $20\times$ higher weight was applied to
the 13.5-14.1\,$\mu$m and 14.8-15\,$\mu$m spectral regions, which
contain the crucial Q-branches of \ce{C2H2}, \ce{HCN} and \ce{CO2}. We
assume an error of $0.5\%\times F_\nu^{\rm obs}$ for the observed
continuum-subtracted flux, where $F_\nu^{\rm obs}$ is the absolute
JWST flux, to reflect the uncertainties in the continuum
subtraction procedure.

Each combined {\sc ProDiMo}-{\sc FLiTs} model takes about 2\,h of
computational time on 16 CPUs.  We run 12 of such models in parallel
to complete one generation. The genetic optimisation was stopped after
about $100-200$ generations, which amounts to a computational cost of about
$150\times12\times2{\rm\,h}\times16{\rm\,CPUs}\!\approx\!60000$\,CPU\,hours
per genetic fitting run.  We experimented with five of such runs,
using slightly different setups (total $\sim\!3\times10^5$\,CPU\,hours), and
selected the best one for publication.

\begin{figure*}
  \centering
  \begin{tabular}{cc}
    \hspace*{-5mm}
    \includegraphics[width=78mm,trim=0 0 0 0,clip]
                    {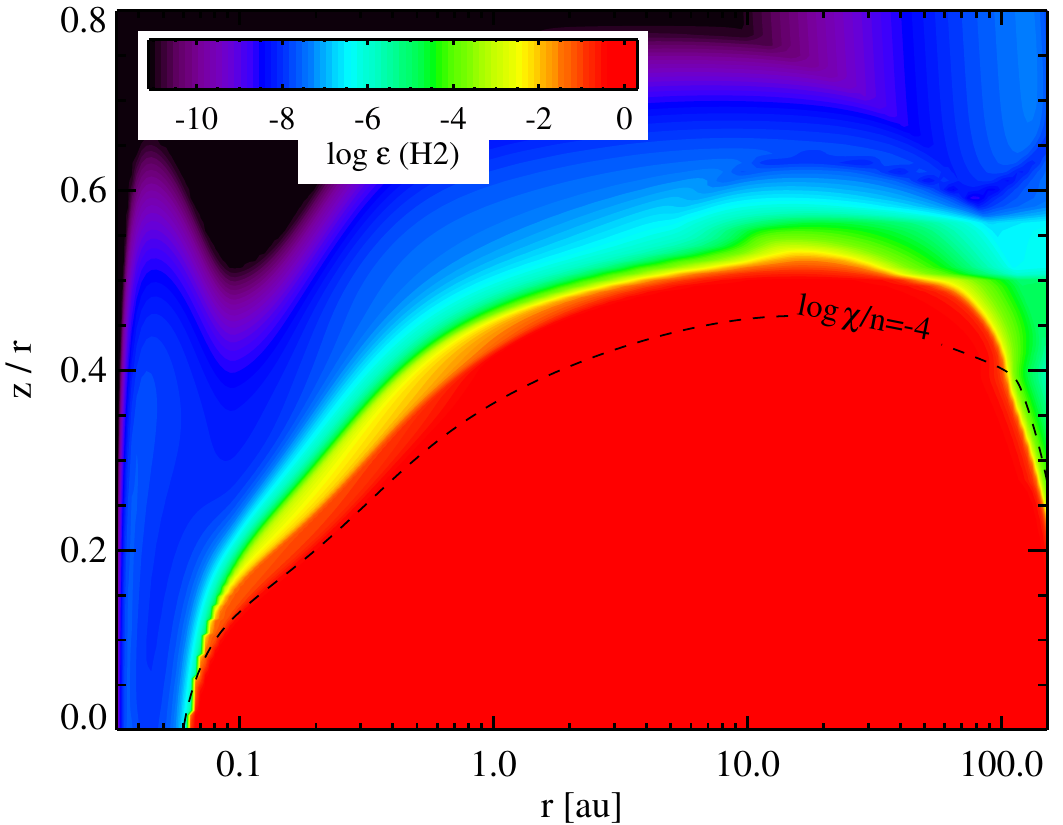} &
    \hspace*{-5mm}
    \includegraphics[width=78mm,trim=0 0 0 0,clip]
                    {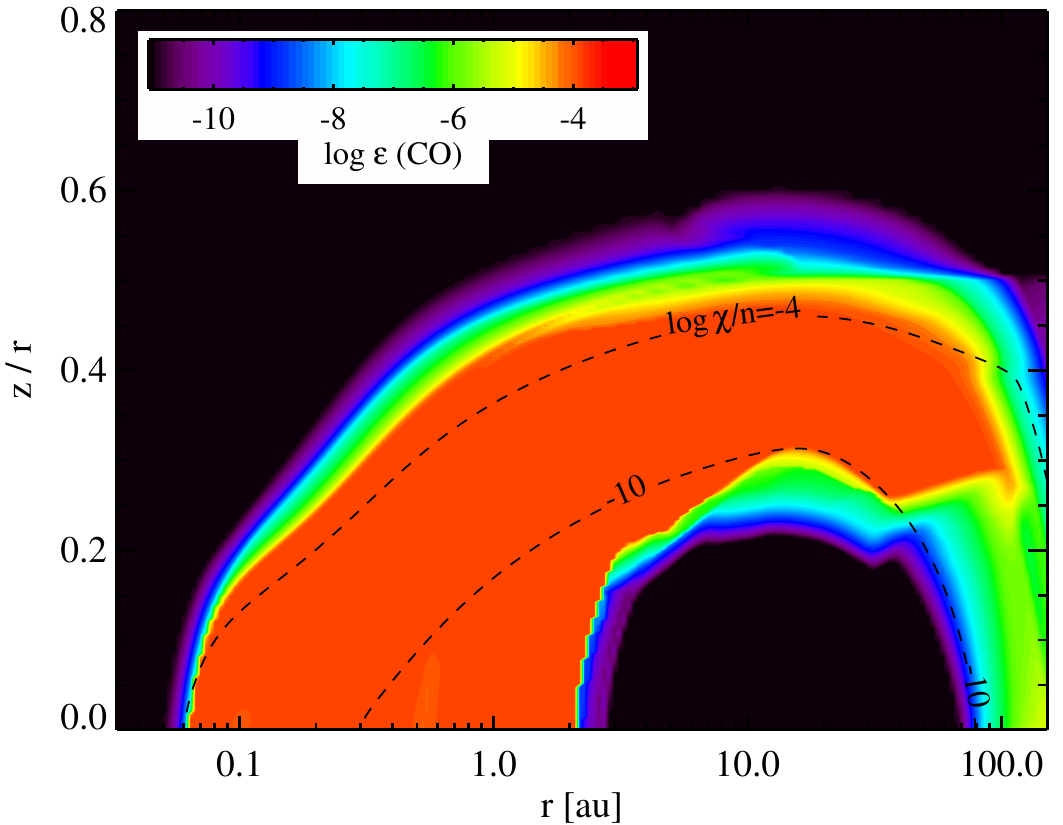}\\[-4.5mm]
    \hspace*{-5mm}
    \includegraphics[width=78mm,trim=0 0 0 0,clip]
                    {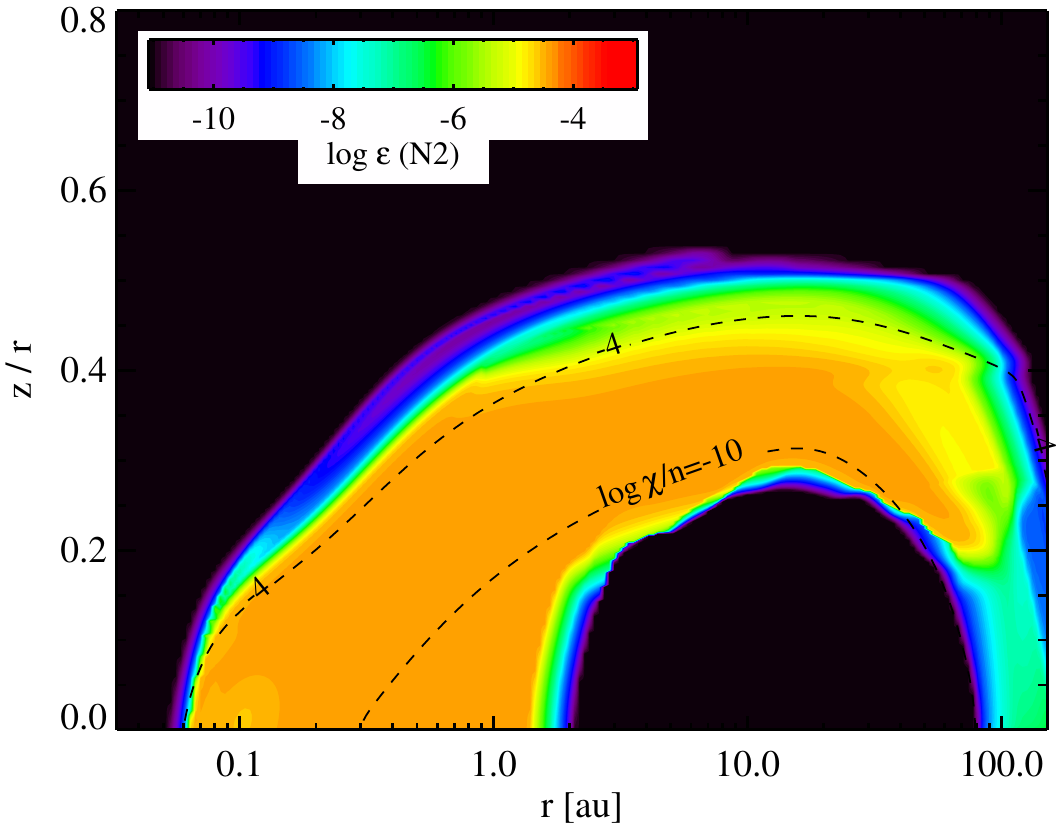} &
    \hspace*{-5mm}
    \includegraphics[width=78mm,trim=0 0 0 0,clip]
                    {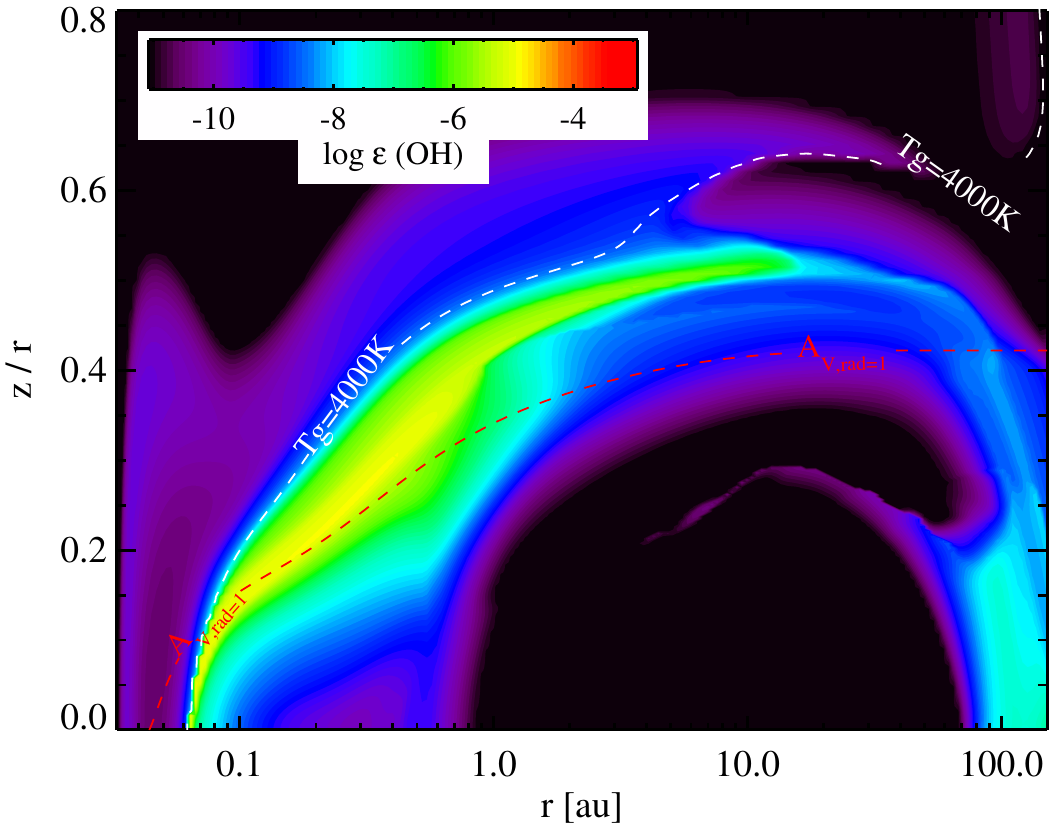}\\[-4.5mm]
    \hspace*{-5mm}
    \includegraphics[width=78mm,trim=0 0 0 0,clip]
                    {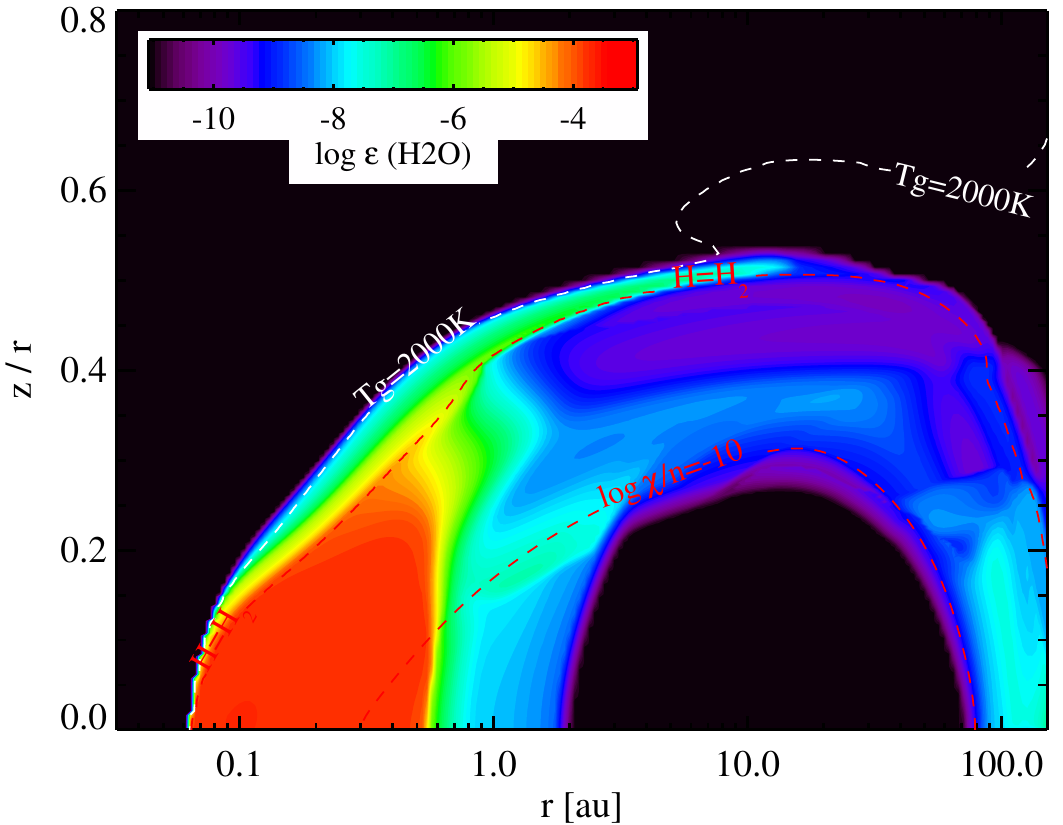} &
    \hspace*{-5mm}
    \includegraphics[width=78mm,trim=0 0 0 0,clip]
                    {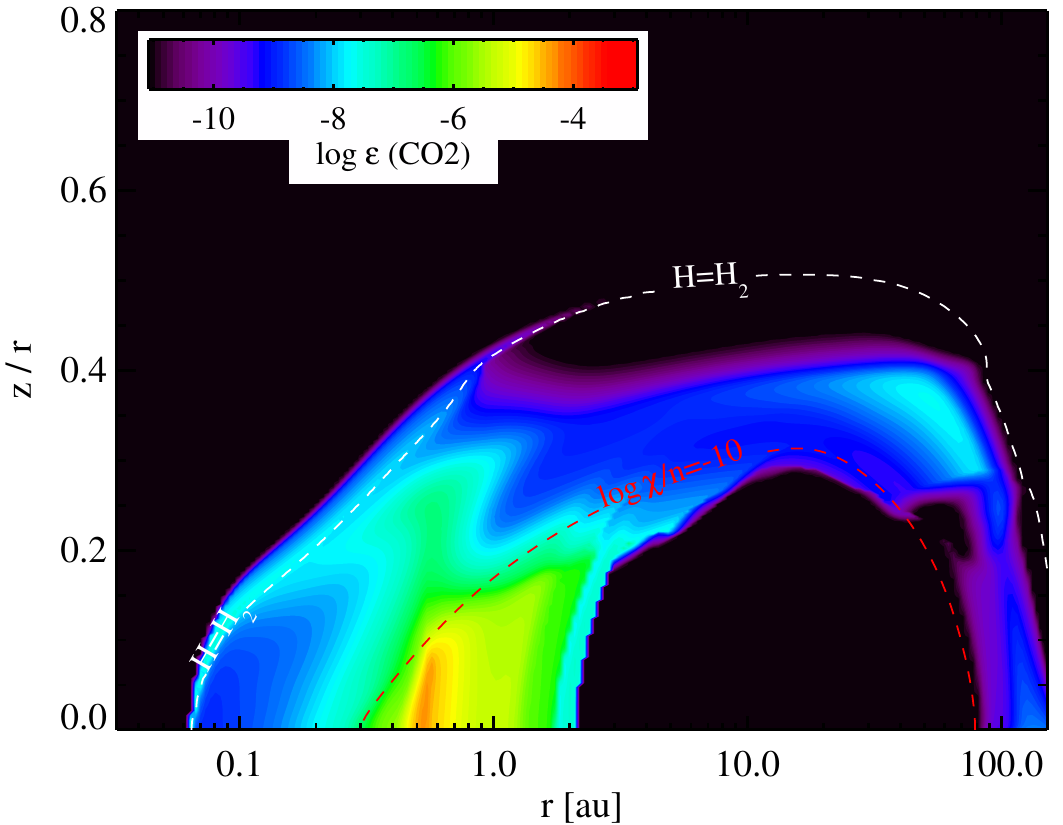}\\[-4.5mm]
    \hspace*{-5mm}
    \includegraphics[width=78mm,trim=0 0 0 0,clip]
                    {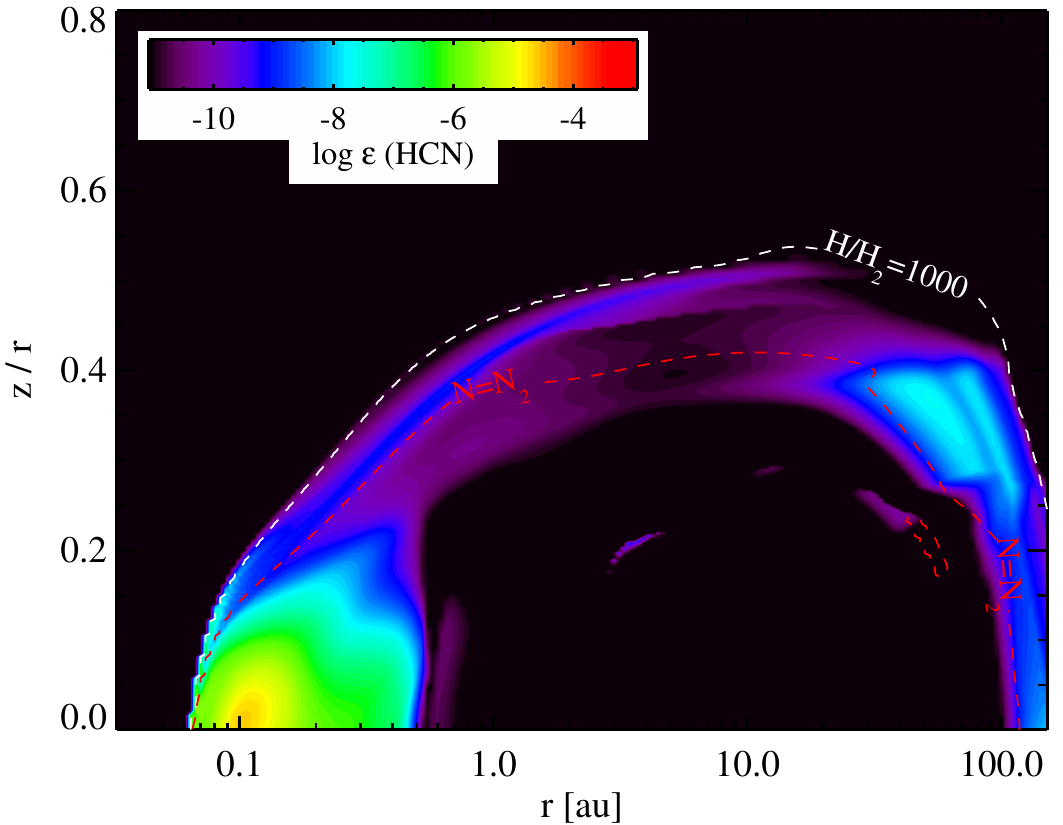} &
    \hspace*{-5mm}
    \includegraphics[width=78mm,trim=0 0 0 0,clip]
                    {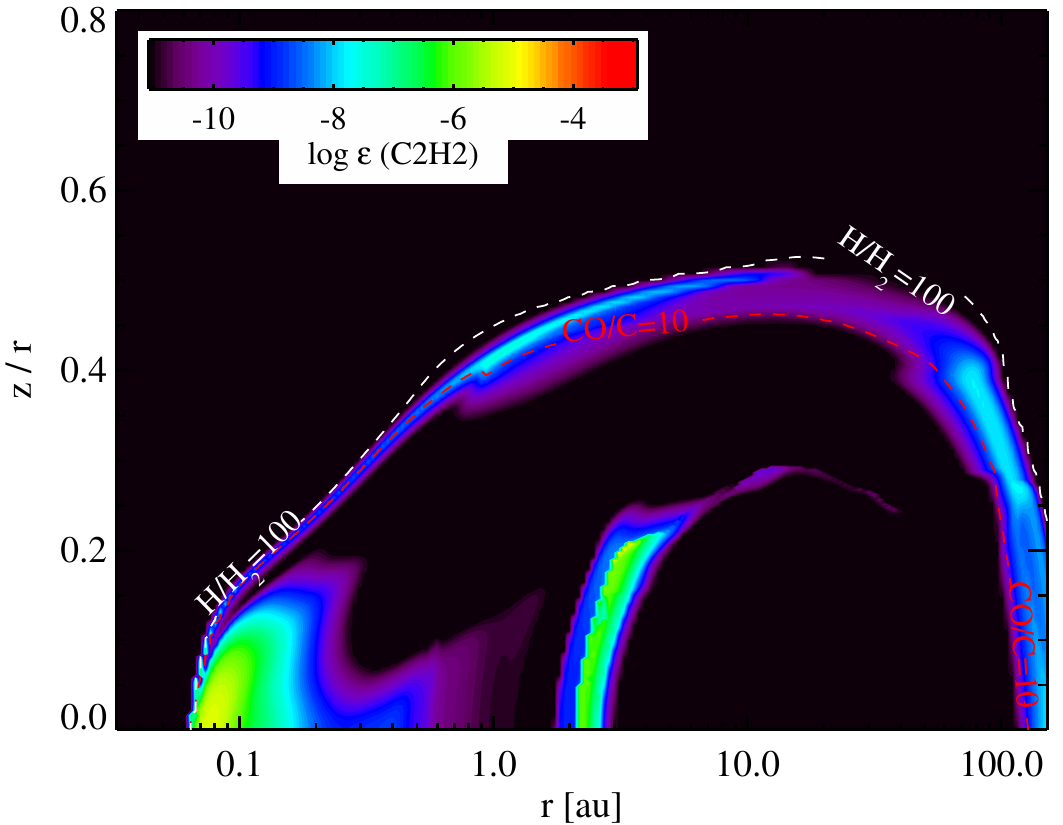}\\[-1mm]
  \end{tabular}
  \caption{Resulting chemical concentrations $\epsilon_i\!=n_i/\nH$ of
    selected molecules in the {\sc ProDiMo} disc model for EX\,Lupi.
    Additional contour lines are plotted as boundaries, including the
    ionisation parameter $\chi/\nH$ (i.e. the UV field strength divided
    by the hydrogen nuclei particle density) and ratios of the
    concentrations of the major molecules. For example, 'H/H2$=$100' means
    that H atoms are 100 times more abundant than H$_2$ molecules.}
  \label{fig:mols}
\end{figure*}

Finally, a high-resolution $200\times200$ {\sc ProDiMo} model was run
for the best-fitting model to produce the results depicted in
Figs.~\ref{fig:discmodel}, \ref{fig:mols} and
\ref{fig:LineOrigin}. All {\sc ProDiMo} models have been performed
with git version e8e68ec6 2023/08/08.

\section{The mid-IR line emitting regions of EX\,Lupi}
\label{sec:analysis}

\subsection{Physico-chemical structure of the best-fitting model}
Figure~\ref{fig:discmodel} shows the assumed gas and dust
distributions in the best-fitting model. As discussed later in this
section, the main line emitting molecules observable with JWST are
located at $r\approx\!0.1-0.5$\,au around the optical disc surface marked
by the $A_{\rm V,rad}\!=\!1$ line, which corresponds to relative
heights $z/r\!\approx\!0.05-0.3$ in this model. The dust settling
according to the applied recipe of \citet{Riols2018} with
$\alpha\!=\!5\times10^{-4}$ is so strong that the local gas/dust ratio
is about $10^4$ along this line, and the calculated UV field strength
$\chi$ about $10^4-10^6$, at densities $10^{10}-10^{11}\rm\,cm^{-3}$,
so we need to discuss the physics and chemistry in a high-density
photon dominated region with X-rays (XDR) with unusually high gas/dust
ratio.

The gas actually results to be cooler than the dust in these regions,
but this does not matter for the line formation in this model, because
the dust is rather transparent at these radii. Therefore, the full
column of gas located here can contribute to the observable line
spectrum in this model. Here, our working definition for molecular column
densities, to compare with the results of slab model fits, is
\begin{equation}
  N_{\rm col}^{\rm line} = \int_{z\big(\tau_{\rm cont}^{\rm ver}(\lambda^{\rm
      line})\,=1\big)}^{\infty} n_{\rm mol}(z)\;dz \ ,
  \label{eq:Ncol}
\end{equation}
i.e.\ the vertical integral over the molecular particle density down
to the point where the dust becomes optically thick at line wavelength.
In contrast, in our standard T\,Tauri disc model (see
Fig.~\ref{fig:molconc}), the high gas column densities in the inner disc
($>\!10^{26}\rm\,cm^{-2}$), combined with a standard gas/dust
ratio of 100 ensure that the disc is optically thick up to about 3
scale heights. In this case, the dust covers most of the molecular gas,
and the molecular column densities defined by Eq.~(\ref{eq:Ncol})
often result to be too low in comparison to observations.

\begin{figure*}
  \centering
  \includegraphics[page=1,width=162mm]{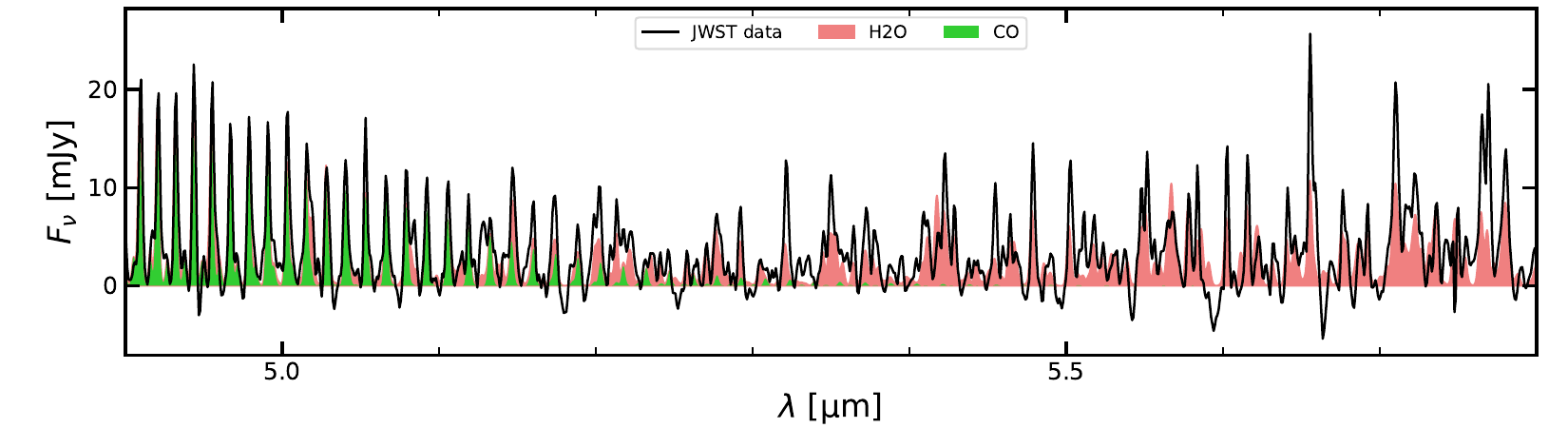}\\[0mm]
  \includegraphics[page=2,width=162mm]{Figures/FLITSmols.pdf}\\[0mm]
  \includegraphics[page=3,width=162mm]{Figures/FLITSmols.pdf}\\[0mm]
  \includegraphics[page=5,width=162mm]{Figures/FLITSmols.pdf}\\[0mm]
  \includegraphics[page=6,width=162mm]{Figures/FLITSmols.pdf}\\[0mm]
  \caption{The continuum-subtracted JWST spectrum (black line) in
    comparison to the {\sc ProDiMo}-{\sc FLiTs} disc model for
    EX\,Lupi.  The coloured areas show the contributions of the
    different molecules to the total model line spectrum.  So they are
    cumulative, i.e. the \ce{CO2}-spectrum is plotted on top of the
    \ce{C2H2}-spectrum, the \ce{HCN}-spectrum on top of the
    (\ce{C2H2}+\ce{CO2})-spectrum, and so on, such that the top of the
    water-spectrum represents the total continuum-subtracted model
    spectrum. The model spectrum has been convolved to $R=2500$. Both
    the observational data and the model spectrum have furthermore
    been slightly box-filtered before plotting, to increase clarity,
    see text. The broad bumps in the observational data, for example around
    6.3\,$\mu$m and in the region $15.5-17.7\,\mu$m, are likely due to
    problems with the early data reduction and continuum subtraction.}
  \label{fig:FLITS}
\end{figure*}

\begin{figure*}
  \centering
  \begin{tabular}{cc}
    \hspace*{-2mm}
    \includegraphics[width=90mm,trim=0 24 0 0,clip]
                    {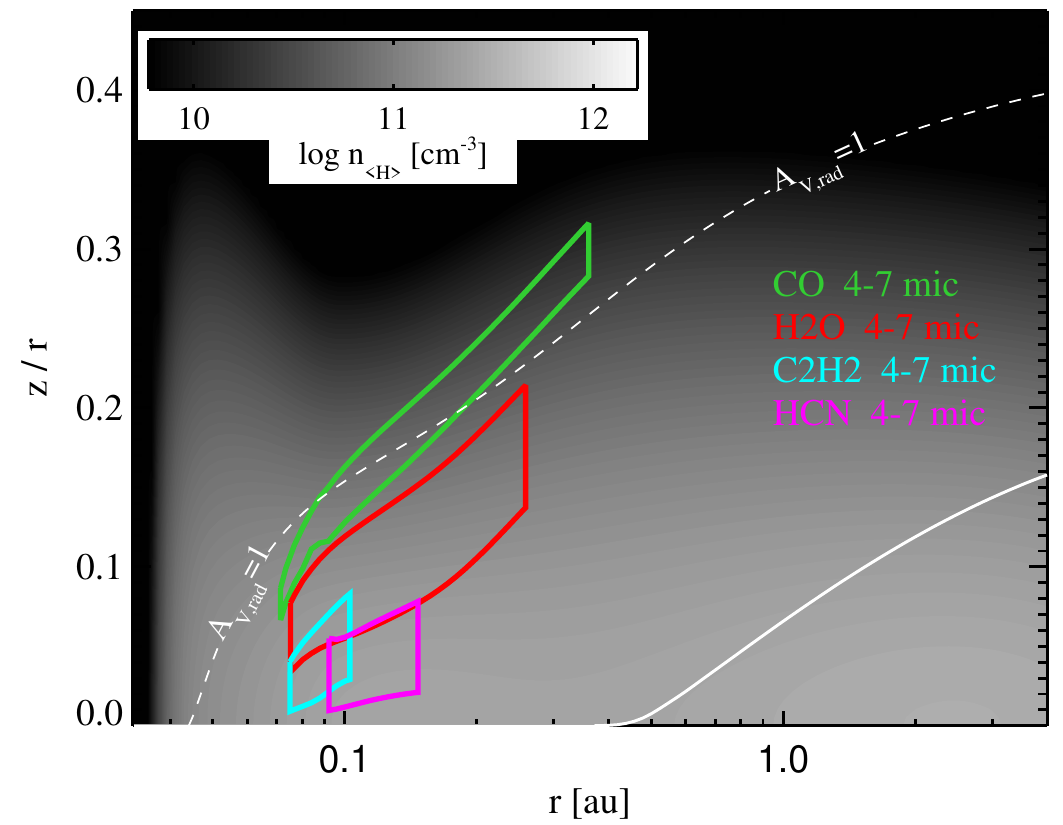} &
    \hspace*{-5mm}
    \includegraphics[width=90mm,trim=0 24 0 0,clip]
                    {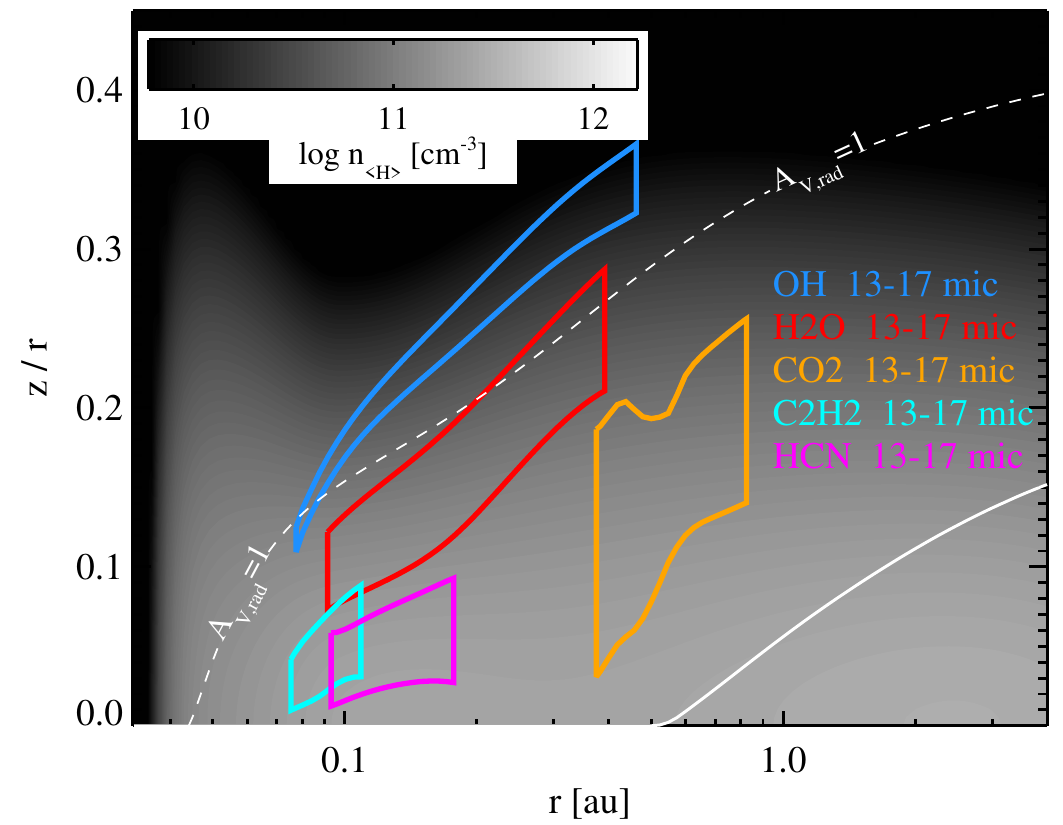}\\
    \hspace*{-1.5mm}
    \includegraphics[width=87.4mm,height=67mm,trim=0 0 6 4,clip]
                    {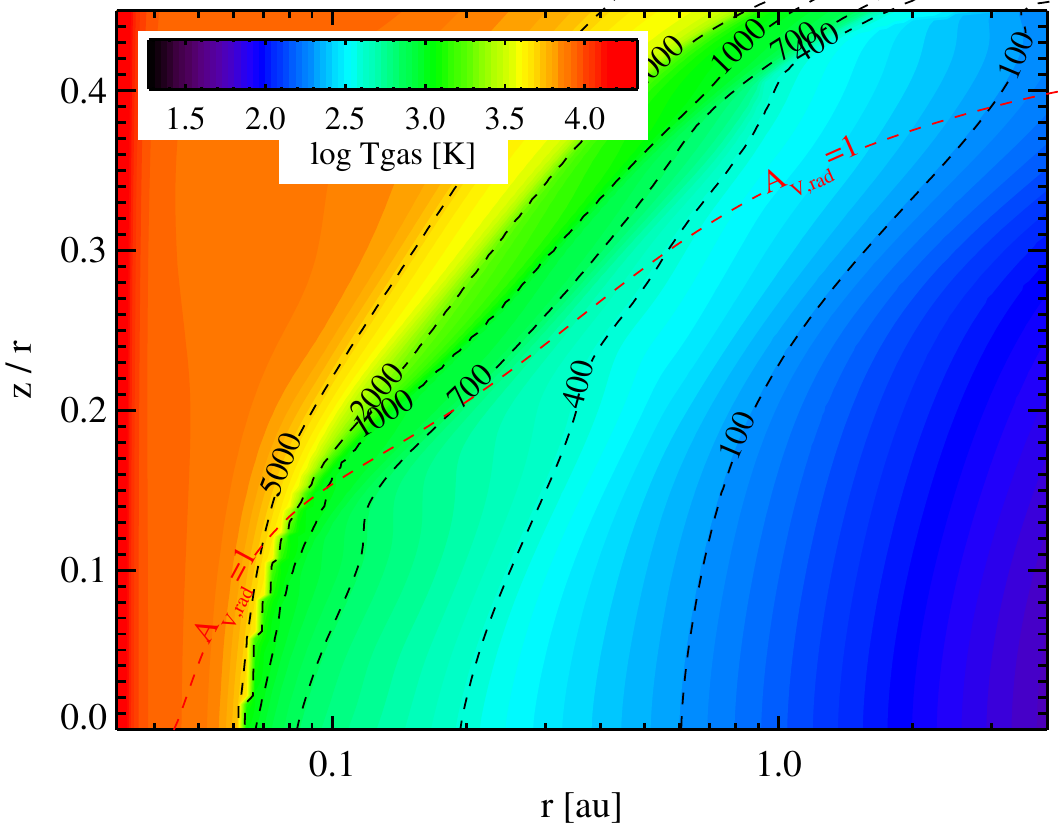} &
    \hspace*{-4.5mm}
    \includegraphics[width=87.4mm,height=67mm,trim=0 0 6 4,clip]
                    {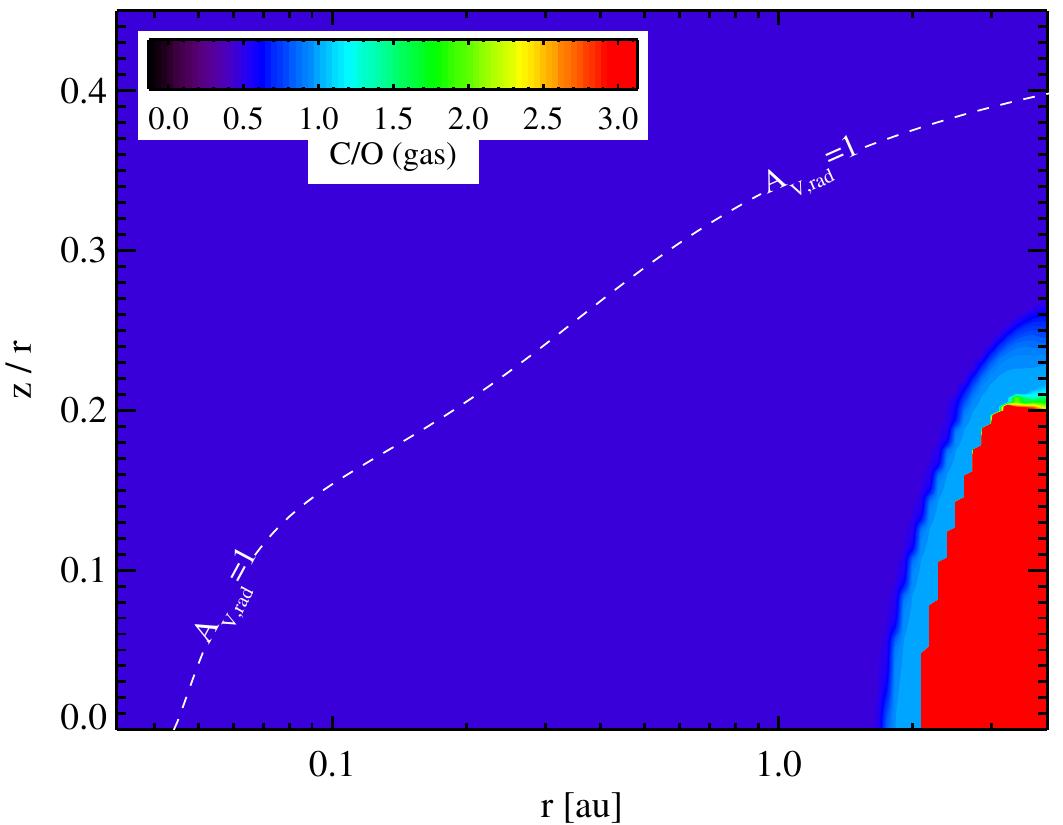} \\[-1mm]
  \end{tabular}
  \caption{{\bf Upper row}: Line emitting regions of different
    molecules in two different spectral bands in the best-fitting
    EX\,Lupi model. The left upper plot considers
    $\lambda\!=\!4-7\,\mu$m, and the right upper plot
    $\lambda\!=\!13-17\,\mu$m. The white lines at the bottom of the
    plots mark the location where the vertical continuum optical depth
    is one, considering an average of the wavelengths of all water
    lines in the considered wavelength region, so this disc would be
    vertically optically thin in the continuum inside of
    $0.4-0.5$\,au, depending on mid-IR wavelength, and inside of about
    0.07\,au in the visual. The {\bf low row} shows the corresponding
    gas temperatures (left) and the carbon-to-oxygen ratio C/O in the
    gas phase (right). {$\rm C/O\!=\!0.46$ is constant by assumption,
    except for the region where water ice is found to be stable, where
    C/O instantly jumps to very large values $>\!1000$.}}
  \label{fig:LineOrigin}
\end{figure*}

Figure~\ref{fig:SED} shows the obtained SED fit, with a magnification
of the mid-IR spectral region in comparison to the JWST spectrum.  We
have not attempted any detailed dust opacity fitting in this model, so
one cannot expect a fit much better than this. We achieved a reduced
$\chi_{\rm phot}$ of 1.45 and $\chi_{\rm JWST}\!=\!0.81$.

Figure~\ref{fig:mols} shows some selected molecular concentrations in
our best-fit model. The JWST line emitting regions are located just
below the $\chi/n\!=\!10^{-4}$ line, where the major phase transitions
$\rm H\to H_2$, $\rm C\to CO$ and $\rm N\to N_2$ occur.  \ce{H2O}
extends a little higher than this, to about $\Tg\!=\!2000\,$K, and OH
even higher, to about $\Tg\!=\!4000\,$K.  The \ce{CO2}, \ce{HCN} and
\ce{C2H2} concentrations are more difficult to understand. These
molecules obtain their maximum concentrations of about
$10^{-6}-10^{-5}$ not far from the midplane, inward of the snowline.
Concerning \ce{HCN} and \ce{C2H2}, there are also some radially
extended, high, tenuous, and spatially thin photodissociation layers
with maximum concentrations $10^{-9}-10^{-8}$, just below $\rm
H/H_2\approx\!1000$ and $100$, respectively, but the major reservoir
of these molecules is the close midplane, between about 0.1\,au
and 0.5\,au, which in this model coincides with the mid-IR line
emitting regions.  Normally, these regions are covered by dust
opacity.

Figure~\ref{fig:FLITS} presents the line spectrum of the best-fitting
{\sc ProDiMo}-{\sc FLiTs} model in comparison to the
continuum-subtracted JWST data \citep{Kospal2023}. The spectral
contributions of \ce{H2O}, \ce{CO}, \ce{C2H2}, \ce{HCN}, \ce{CO2},
\ce{OH} and \ce{H2} are indicated by different colours. Our 2D disc
model can reproduce most of the observed spectral features
surprisingly well. We achieve an overall fit quality with reduced
$\chi_{\rm line}$ values of 3.2 and 2.1 in the spectral regions
$4.9-7.5\,\mu$m and $13-17.7\,\mu$m, respectively. The \ce{HCN} fluxes
are a bit weak, and some of the water line peaks are a bit too
shallow, for example the low-excitation water lines at longer wavelengths.
However, some of the visible deviations are also affected by imperfect data
reduction and continuum subtraction.  We have used the
continuum-subtracted JWST data as published in \citet{Kospal2023}
(private communication A.~Banzatti), reduced with JWST pipeline
version 1.8.2 and continuum-subtracted at that time.

The fit is obviously not as good as what can be achieved with slab
model fits, where one can directly tune the molecular column
densities, the sizes of the emission areas and the emission
temperatures.  But our model is able to simultaneously explain the SED,
the overall spectral shape of the JWST spectrum, and the main
characteristics of the line emissions from CO, from \ce{H2O} in both
considered wavelength bands, and from the Q-branches of \ce{C2H2},
\ce{HCN}, \ce{CO2} between 13.7 and 15\,$\mu$m, even some minor
contributions of OH and \ce{H2}. In our model, all column densities,
emission temperatures and sizes of line emitting areas are consistent
results of our 2D thermo-chemical disc structure, and we are using
this complex structure under an inclination of 32\degr\ to calculate
our spectra.

\begin{table}[!b]
  \vspace*{-1mm}
  \caption{Total line fluxes $\sum\!F_{\rm line}\rm\,[W/m^2]$,
    flux-mean column densities $\langle N_{\rm
      col}\rangle\rm\,[cm^{-2}]$, flux-mean emission temperatures
    $\langle T_{\rm g}\rangle\rm\,[K]$ and flux-mean emission radius
    interval $\langle r_1-r_2\rangle$ of molecules,
    considering all lines emitted in two different spectral regions in
    the 2D disc model, see Eq.\,(\ref{eq:fluxmean}).}
  \label{tab:means}
  \vspace*{-2mm}
  \hspace*{-1mm}
  \resizebox{90mm}{!}{
  \begin{tabular}{cc|cc|ccc}
    \hline
    & \!\!\!\!$\lambda\,[\mu$m] & \!\!\#\,lines\!\!\!\! & $\sum\!F_{\rm line}$ &
      \!\!$\langle\log_{10} N_{\rm col}\rangle$\!\! & $\langle T_{\rm
        g}\rangle$\,[K] & \!\!\!\!\!$\langle r_1-r_2\rangle$\,[au]\!\!\! \\
    \hline
    &&&&&\\[-2.2ex]
     \!\!\ce{CO} & \!\!\!\!4.5-7.5\!\! &  368 & 2.5(-16) & \!\!$19.1\pm0.4$\!\! & \!\!$970\pm290$\!\! & \!0.07-0.36\!\!\\
    \!\!\ce{H2O} & \!\!\!\!4.5-7.5\!\! & 3521 & 3.8(-16) & \!\!$18.9\pm0.4$\!\! & \!\!$700\pm220$\!\! & \!0.08-0.27\!\!\\
    \!\!\ce{CO2} & \!\!\!\!4.5-7.5\!\! &  749 & -6(-21)\\
    \!\!\ce{HCN} & \!\!\!\!4.5-7.5\!\! & 1657 & 3.4(-17) & \!\!$17.6\pm0.4$\!\! & \!\!$610\pm110$\!\! & \!0.09-0.15\!\!\\
   \!\!\ce{C2H2} & \!\!\!\!4.5-7.5\!\! & 2027 & 7.0(-18) & \!\!$17.0\pm0.6$\!\! & \!\!$750\pm90$\!\!  & \!0.08-0.10\!\!\\
    \!\!\ce{NH3} & \!\!\!\!4.5-7.5\!\! & 2430 & 2.2(-19) & \!\!$15.4\pm0.5$\!\! & \!\!$440\pm100$\!\! & \!0.13-0.24\!\!\\
    \!\!\ce{CH4} & \!\!\!\!4.5-7.5\!\! & 1800 & 1.9(-21) & \!\!$13.8\pm0.3$\!\! & \!\!$640\pm120$\!\! & \!0.10-0.14\!\!\\
   \hline
    &&&&&\\[-2.2ex]
    \!\!\ce{H2O} & \!\!\!\!\!\!13-17.7\!\! & 1014        & 1.1(-16) & \!\!$19.1\pm0.4$\!\!& \!\!$620\pm210$\!\! & \!0.11-0.37\!\!\\
    \!\!\ce{CO2} & \!\!\!\!\!\!13-17.7\!\! & 2673        & 2.5(-17) & \!\!$18.0\pm0.9$\!\!& \!\!$270\pm90$\!\!  & \!0.39-0.87\!\!\\
    \!\!\ce{HCN} & \!\!\!\!\!\!13-17.7\!\! & 1763        & 8.5(-17) & \!\!$17.5\pm0.4$\!\!& \!\!$580\pm130$\!\! & \!0.10-0.19\!\!\\
   \!\!\ce{C2H2} & \!\!\!\!\!\!13-17.7\!\! &\!\!10334\!\!& 5.7(-17) & \!\!$16.9\pm0.7$\!\!& \!\!$740\pm110$\!\! & \!0.08-0.13\!\!\\
     \!\!\ce{OH} & \!\!\!\!\!\!13-17.7\!\! &  244        & 6.9(-18) & \!\!$16.8\pm0.6$\!\!& \!\!$1700\pm720$\!\!& \!0.12-0.55\!\!\\
    \!\!\ce{NH3} & \!\!\!\!\!\!13-17.7\!\! & 1178        & 2.5(-19) & \!\!$15.7\pm0.5$\!\!& \!\!$380\pm100$\!\! & \!0.16-0.33\!\!\\
    \hline
  \end{tabular}}
\end{table}

\subsection{Deriving column densities and emission temperatures from
  the 2D disc model}
\label{sec:means}

Table~\ref{tab:means} lists some mean quantities of the molecular
emission features derived from the 2D disc structure, using the line
fluxes and their analysis based on escape probability in
face-on geometry, see Sect.\ref{sec:escprofluxes}.  We consider the
molecular column densities $N_{\rm col}$ as defined in
Eq.\,(\ref{eq:Ncol}) as an integral upwards from the height $z$ in the
disc where the vertical continuum optical depth is one at line
wavelength. This quantity depends on the considered
line, because different lines are emitted from different radii and
from different depths. We also compute mean molecular emission temperatures
$\langle T_{\rm g}\rangle$, where we read off the gas temperatures in
the identified line emitting regions, and the overall radial range of
significant line emission.  All these quantities are calculated by
averaging over all lines of a selected molecule in a given spectral
range, and over all vertical columns in the disc model, weighted by
the line flux that is generated by this column in this line, $\delta
F^{\rm line}_{\rm\,col}$, where $F^{\rm line}=\sum_{\rm cols} \delta
F^{\rm line}_{\rm\,col}$ is the total line flux
\begin{align}
  \langle X\rangle ~=&
  \frac{\sum\limits_{\rm lines} \sum\limits_{\rm cols}
    X^{\rm line}_{\rm\,col} \;\delta F^{\rm line}_{\rm\,col}}
       {\sum\limits_{\rm lines} \sum\limits_{\rm cols}
         \;\delta F^{\rm line}_{\rm\,col}}\ , 
  \label{eq:fluxmean} \\
  \Delta X ~=& \sqrt{\langle X^2\rangle - \langle X\rangle^2}
       \nonumber \ .
\end{align}
The resulting mean values and standard deviations for the
molecular column densities $\langle\log_{10} N_{\rm col}\rangle$,
the emission temperatures $\langle T_{\rm g}\rangle$ and the radial
line emission intervals are listed in Table~\ref{tab:means}. These
data roughly match the characteristics derived from slab model fitting
of the JWST EX\,Lupi observations as published by \citet{Kospal2023},
with CO having the largest column density of order
$10^{19}\rm\,cm^{-2}$, followed by water, and then about
$10^{17}-10^{18}\rm\,cm^{-2}$ for the molecules \ce{C2H2}, \ce{HCN}
and \ce{CO2}. The water emission temperature is found to be about
$600-700\,$K, depending on wavelength. The emission temperatures of
\ce{C2H2} and \ce{HCN} are found to be similar, $580-750\,$K, whereas
\ce{CO2} emits at somewhat lower temperatures $\sim\!300\,$K in the
model. In contrast, the weak OH emission lines are emitted from much
hotter gas $\sim\!1700\,$K. These trends are in general agreement with
slab model fits published for other T\,Tauri stars
\citep{Salyk2011,Grant2023,Tabone2023}.  Interestingly, we also
modelled \ce{NH3} and \ce{CH4} lines, but although the column
densities of these molecules are not small in the model, these lines
do not show up, or even go into absorption at shorter wavelengths.

Figure~\ref{fig:LineOrigin} visualises the line emitting regions of
the various molecules in two different spectral bands. We see a
pattern that re-occurs in basically all {\sc ProDiMo} models. The OH
lines are emitted from the highest disc layers. Here, the OH molecule
is crucial for cooling down the otherwise hot atomic gas, even before
\ce{H2} becomes the dominant molecule. Below that OH-emitting layer,
we find that first CO and then \ce{H2O} are emitting. {The lower
  row of plots shows the corresponding gas temperatures and shows that
  the carbon-to-oxygen ratio C/O in the gas phase is constant $\rm
  C/O\,=\,0.46$ by assumption in the model, except for the region
  where water ice is found to condense, which happens in the midplane
  outside of about 2\,au in this model.}

What is special in this particular disc model for EX\,Lupi is that the
low and increasing surface density profile around the inner rim (only
$\sim\!5\times10^{22}\rm\,cm^{-2}$ at $r\!=\!0.1\,$au) causes very efficient
dust settling, which reveals an almost dust-free gas. Indeed, the
molecular column densities above the $z\,\big(\tau_{\rm cont}^{\rm
  ver}(\lambda^{\rm line})\!\!=\!\!1\big)$\,-level are much larger at
small radii in this model than in our standard T\,Tauri model
($2\times\!10^{26}\rm\,cm^{-2}$ at $r\!=\!0.1\,$au) using a sharp
inner rim.  Thus, line emissions by \ce{C2H2} and \ce{HCN} from deep
layers close to the midplane at small radii becomes visible.  We see
that the line emitting regions of \ce{C2H2} and \ce{HCN} more or less
spatially coincide, whereas the line emitting region of \ce{CO2} is
located further out, which explains the somewhat lower emission
temperature of \ce{CO2}. All these molecules emit from significantly
deeper layers compared to \ce{OH}, CO and \ce{H2O}, which explains why
the emission temperatures of the latter molecules are usually higher.

\begin{figure}
  \centering
  \hspace*{-1mm}
  \includegraphics[width=89mm,trim=5 0 10 0,clip]{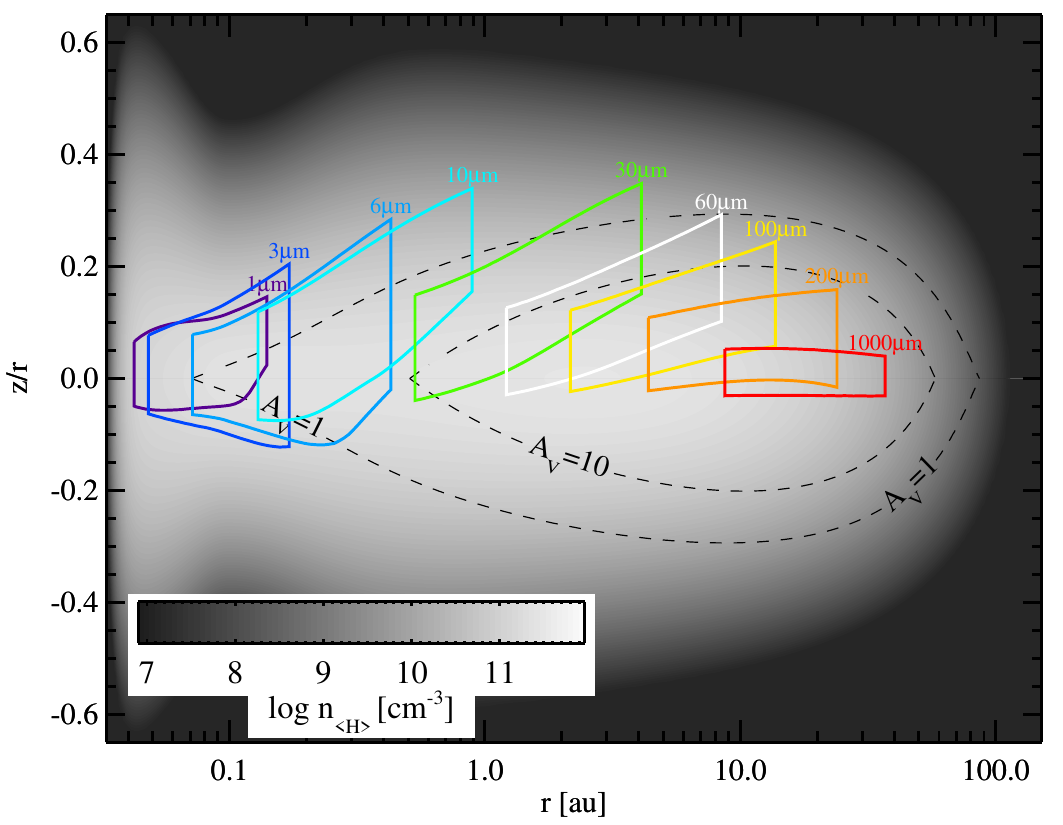} 
  \vspace*{-2mm}
  \caption{Continuum emitting regions responsible for about 50\% of the
    flux at different wavelengths under face-on inclination. Contour
    lines for the vertical optical extinction $A_V$ at 1 and 10 are
    added.}
  \label{fig:ContOrigin}
\end{figure}

Figure~\ref{fig:ContOrigin} shows that at short wavelengths ($\approx
5\,\mu$m), the lines and the continuum are coming from roughly the same
disc regions, so here absorption lines are more likely to occur.
In contrast, at longer wavelengths ($\approx 20\,\mu$m), the continuum is
produced by more distant disc regions than the lines, so a simple
addition of continuum and line emission is possible, as is
implicitly done in slab model analysis.

\subsection{Chemical pathways}
\label{sec:chempaths}

For each molecule, we have analysed the chemistry in the centre of the
respective line forming regions as depicted in
Fig.~\ref{fig:LineOrigin}.  For OH, \ce{H2O} and \ce{CO2}, we do not
need a very sophisticated chemistry to understand their formation, a
few basic photoreactions in combination with neutral-neutral
reactions, and the \ce{H_2}-formation on grains, are sufficient.
Previous work on such chemical pathways can be found e.g.\ in
\cite{Thi2005}, \cite{Glassgold2009}, \cite{Woitke2009} and \cite{Najita2011}.

At the top of the photodissociation layer at $r\!=\!0.2\,$au, where
$A_{\rm V,rad}\!\approx\!0.07$ (optically thin), the gas temperature is
of the order of 3000\,K, carbon is mostly ionised, oxygen atomic, and
there is almost no molecular hydrogen yet, $n_{\rm H}/n_{\rm
  H_2}\!\approx\!10000$. In these circumstances, the first IR active
molecule to form is OH with concentrations $\sim\!10^{-7}$ via
\begin{equation}
  \rm H ~+~ H ~\stackrel{dust}{\longrightarrow}~ H_2
              ~\stackrel{O}{\longrightarrow}~ OH ~+~ H \ .
\end{equation}  
The amount of OH that can form this way is limited by the
photodissociation of OH, and by the back reaction $\rm
OH+H\to O+H_2$.

In only slightly lower layers, where $A_{\rm V,rad}\!\approx\!0.3$, the
gas temperature is $\sim\!1500\,$K, carbon is still mostly ionised
and oxygen atomic, and $n_{\rm H}/n_{\rm H_2}\!\approx\!200$. Here,
OH reacts further to form water with concentrations $\sim\!10^{-7}$ 
\begin{equation}
  \rm OH ~+~ H_2 ~\longrightarrow~\rm H_2O ~+~ H \ ,
\end{equation}  
limited by \ce{H2O} photodissociation.

\ce{CO2} forms in similar ways as water, but only in deeper regions
where \ce{H2} and CO are already abundant, see \cite{Thi2005}, for
example. At $r\!=\!0.5\,$au, at a height where $A_{\rm
  V,rad}\!\approx\!3$ and $A_{\rm V}\!\approx\!0.2$, we find
$\Tg\!\approx\!\Td\!\approx\!360\,$K. Carbon is mostly present in form
of CO, oxygen in \ce{H2O}, and hydrogen in \ce{H2}. Under these
circumstances, \ce{CO2} reaches concentrations of $\sim\!10^{-7}$ via
\begin{align}
  \rm OH ~+~ C  &~\longrightarrow~\rm CO ~+~ H \\
  \rm OH ~+~ CO &~\longrightarrow~\rm CO_2 ~+~ H \ ,
\end{align}  
limited by \ce{CO2} photodissociation.

\begin{figure}[!b]
  \centering
  \includegraphics[width=91mm]{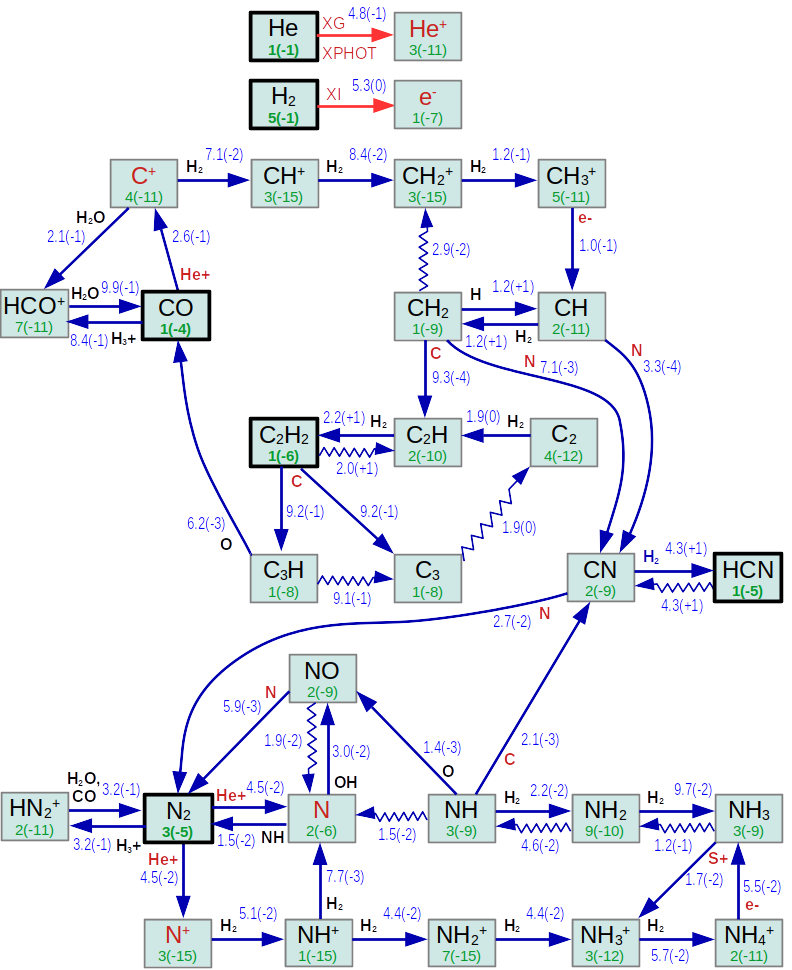}
  \caption{Reaction pathways to form \ce{HCN} and \ce{C2H2} in the
    close midplane, where $\nH\!=\!2\times10^{11}\rm\,cm^{-3}$,
    $\Tg\!=\!640\,$K, $\Td\!=\!630\,$K, $A_{\rm V,rad}\!=\!7$ and
    $A_{\rm V}\!=\!0.7$. The major species are marked with thick
    boxes. Red species are rare and created by X-ray processes. The
    green numbers in the boxes are the particle concentrations $n_{\rm
      mol}/\nH$, the blue numbers are the rates
    $\rm[cm^{-3}s^{-1}]$. Wiggly arrows indicate photodissociation,
    and red arrows indicate X-ray processes: XPHOT is the X-ray
    primary ionisation of He, and XG and XI are the X-ray secondary
    ionisations by fast electrons.}
  \label{fig:pathway}
  \vspace*{-1mm}
\end{figure}

The formation of \ce{HCN} and \ce{C2H2} in the close midplane is more
complicated, see Fig.~\ref{fig:LineOrigin}. Previous works on the
chemical pathways of these species have been published, for example,
by \cite{Walsh2015} and \cite{Bosman2022a,Bosman2022b}. Similar to
\ce{CO2}, UV-shielded conditions are required, but X-rays also play an
important role here, and this is why these line emitting regions need
to be so close to the star, before the X-rays are absorbed by the
gas. At $r\!=\!0.1\,$au, at a height where $A_{\rm V,rad}\!\approx\!7$
and $A_{\rm V}\!\approx\!0.7$, we find $\Tg\!\approx\!640\,$K and
$\Td\!\approx\!630\,$K.  As before, carbon is in CO, oxygen in
\ce{H2O}, nitrogen in \ce{N2} and hydrogen in \ce{H2}. The chemical
destruction timescales due to X-ray ionisations are about
$\zeta_X\!\approx\!10^{-10}\rm\,s^{-1}$ in this region, i.e.\ typical
molecular lifetimes are a few 100\,yrs.

\begin{figure*}
  \centering
  \begin{tabular}{cc}
    \hspace*{-2mm}
    \includegraphics[width=88mm,height=84mm,trim=0 0 0 0,clip]
                    {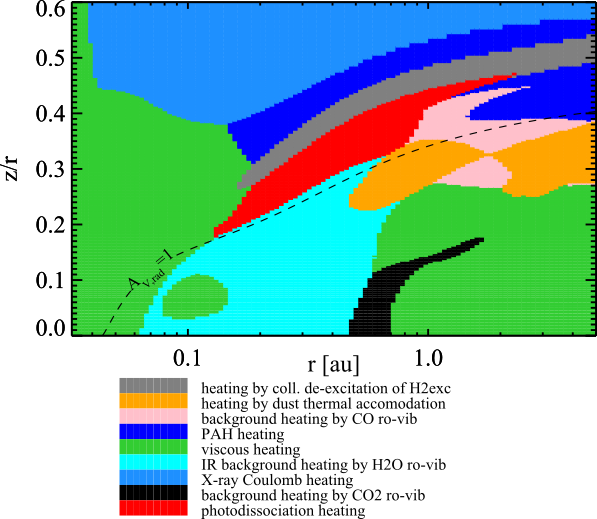} &
    \hspace*{-2mm}
    \includegraphics[width=88mm,height=84mm,trim=0 0 0 0,clip]
                    {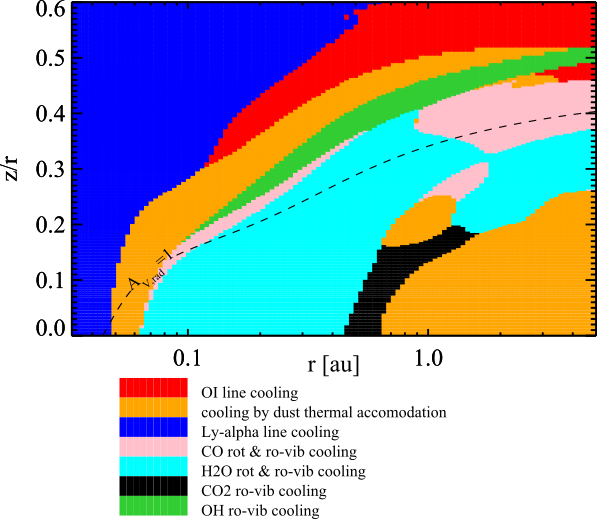} 
  \end{tabular}
  \caption{Leading heating (left) and cooling process
    (right) in the best-fitting EX Lupi model. The plots show a
    magnification of the inner disc relevant for the IR molecular line
    emissions. The dashed contour line marks the disc surface, where
    the radial optical extinction $A_{\rm V,rad}\!=\!1$.}
  \label{fig:heatcoolEX}
\end{figure*}

Figure~\ref{fig:pathway} shows the chemical pathways in this region
leading to \ce{HCN} and \ce{C2H2}.
The starting point is the X-ray driven dissociation of \ce{CO},
\ce{N2} and \ce{H2}, which creates small amounts of \ce{C+}, \ce{N},
\ce{N+}, and \ce{e-}.
By fast \ce{H2}-addition reactions, \ce{C+} forms \ce{CH3+}, which
recombines dissociatively to CH, which is in kinetic equilibrium with
\ce{CH2}. These two neutral molecules react with N and C to eventually
form \ce{HCN} and \ce{C2H2}. Another reaction chain starts with
\ce{N+}.  By fast \ce{H2}-addition reactions, \ce{N+} forms \ce{NH4+},
which recombines dissociatively to \ce{NH3}.  \ce{NH3}, \ce{NH2} and
NH are in kinetic equilibrium with each other (neutral-neutral
reactions vs.\ photodissociation), and the reaction $\rm NH + C \to CN
+ H$ contributes to the production of CN, in addition to the reactions
$\rm CH_2 + N \to CN + H_2$ and $\rm CH + N \to CN + H$ from the
reaction chain starting with \ce{C+}.  CN is in kinetic equilibrium
with HCN. We find large concentrations of \ce{HCN} and \ce{C2H2} this
way, $\sim\!10^{-5}$ and $10^{-6}$, respectively.

Neutral carbon primarily forms via the $\rm C_3 +h\nu \to C_2 + C$
reaction, and is mainly destroyed by the $\rm C_2H_2 + C \to C_3H + H$
and $\rm C_2H_2 + C \to C_3 + H_2$ reactions as indicated in the
diagram.  Concerning the $\rm C + H_2O \to HCO + H$ reaction, if we
had used reaction parameters from the KIDA database (see
Sect.~\ref{sec:network}) the concentration of C atoms would be
strongly reduced, and the formation of \ce{HCN} and \ce{C2H2} strongly
suppressed.  In the situation shown in Fig.~\ref{fig:pathway}, the
$\rm C + H_2O \to HCO + H$ reaction only contributes with a rate of
$7\times10^{-2}\rm\,cm^{-3}s^{-1}$ to the destruction of neutral
carbon, i.e. with less than 4\%.

\subsection{Chemical and cooling timescales}

The chemical relaxation timescales \citep[see Eq.\,117
  in][]{Woitke2009} in the line emitting regions are of order
$0.1-10\,$\,yrs. Although this is short compared to overall disc
evolutionary timescales, this could mean that the chemistry is not yet
fully relaxed, because EX\,Lupi went through a phase of $7\times$
stronger accretion about 5 months before the JWST spectrum was taken
\citep{Kospal2023}.  The cooling timescales are shorter, about
$10^{-3}-0.1\,$yrs, so we do not expect any significant memory effects
concerning the gas temperature in the line emitting regions.

The most important heating and cooling processes in the inner disc of
the EX\,Lupi model are shown in Fig.~\ref{fig:heatcoolEX}. In the IR
line emission regions around and below the $A_{\rm V,rad}\!=\!1$ line,
inside of about 0.5\,au, there are three significant heating
mechanisms: (1) viscous heating, (2) absorption of IR continuum
photons emitted from the star and from the disc by water molecules and
(3) photodissociation heating. This heating is mostly balanced by
water, CO and OH line cooling. Below the $A_{\rm V,rad}\!=\!1$ line,
relevant for the HCN and \ce{C2H2} line emissions, there is simply a
balance by water line heating and cooling. Therefore, this region has
a rather specific temperature, namely the radiative equilibrium
temperature in consideration of the water line opacity.  Since the
water line opacity has a redder characteristic than the dust opacity,
this region is generally found to be somewhat cooler than the local
dust temperature $\Tg\!<\!\Td$. However, for larger mass accretion
rates, the viscous heating will eventually cause $\Tg\!>\!\Td$.

\subsection{Dependencies on $L_{\rm UV}$, $L_X$, and $\Mdot$}
\label{sec:depend}

The analysis of the chemical pathways in Sect.~\ref{sec:chempaths}
suggests that the line fluxes of OH and \ce{H2O} should depend on the
UV-luminosity of the star, whereas the \ce{HCN} and \ce{C2H2} lines
should be excited by X-rays.  Table~\ref{tab:UV-Xray} shows that in
our models, indeed, the OH and \ce{H2O} lines become stronger with
increasing stellar UV luminosity.  Both the sizes of the line emitting
areas and the emission temperatures become larger when $L_{\rm UV}$ is
increased.  This trend is confirmed by observations
\citep[e.g.\ Fig.~3 in][]{Banzatti2023}, which show a positive
correlation of the \ce{H2O} line luminosity with accretion luminosity,
which physically implies larger $L_{\rm UV}$.  The high-UV model can
be considered to represent a case, in which a larger optical
extinction $A_V$ was assumed for EX\,Lupi, see discussion in
Sect.~\ref{sec:star}.  From a theoretical point of view, the positive
correlation with $L_{\rm UV}$ is expected from the global energy
budget of the gas in the inner disc, which is heated by stellar UV
photons, and mainly cools through water emission lines.  As long as
the UV is mostly absorbed by the gas and not by the dust, a higher
stellar UV luminosity should therefore result in stronger water
emission lines.  Both the OH and water line luminosities are not much
affected by the X-ray luminosity. Concerning an increase of the mass
accretion rate, OH is not affected, but the water lines become
slightly brighter.  This is because the water emissions come from
slightly deeper regions than OH, where viscous heating is more
relevant, see Fig.~\ref{fig:heatcoolEX}.

\begin{table*}
  \caption{Total line fluxes $\sum_{\rm line}F_{\rm mol}^{\rm line}\rm\,[W\,m^{-2}]$,
    column densities $\langle N_{\rm
      col}\rangle\rm\,[cm^{-2}]$, emission temperatures
    $\langle\Tg\rangle\,$[K] and emitting radii $\langle
    r_1-r_2\rangle$\,[au] of selected molecules, dependent on stellar
    UV luminosity, X-ray luminosity, and mass accretion rate,
    considering all lines between 13 and 17.7\,$\mu$m. }
  \label{tab:UV-Xray}
  \vspace*{-8mm}
  \hspace*{-1.5mm}
  \begin{center}
  \resizebox{135mm}{!}{
  \begin{tabular}{cr||c|c|c|c||c}
    \hline
    &&&&&&\\[-2.0ex]
    && {\bf fiducial model} & high$^{(1)}$ $L_{\rm UV}$ & high$^{(1,3)}$ $L_X$
    & high$^{(1)}$ $\Mdot$ & old ProDiMo$^{(2)}$ \\
    \hline
    &&&&&&\\[-2.2ex]
    & $\sum_{\rm line}F_{\rm mol}^{\rm line}$ &
    \bf 6.9(-18) & 1.9(-17) & 6.9(-18) & 6.9(-18) & 3.6(-18) \\
    \ce{OH}  & \!\!\!\!$\langle\log_{10} N_{\rm col}\rangle$ &
    16.8     & 16.8     & 16.8     & 16.8     & 16.8     \\
    & $\langle\Tg\rangle$ & 
    1700     & 1760     & 1700     & 1710     & 1460     \\
    & $r_1-r_2$ &
    $0.12-0.55$ & $0.15-1.13$ & $0.12-0.56$ & $0.14-0.57$ & $0.10-0.54$ \\
    \hline
    &&&&&&\\[-2.2ex]
    & $\sum_{\rm line}F_{\rm mol}^{\rm line}$ &
    \bf 1.1(-16) & 1.5(-16) & 1.1(-16) & 1.5(-16) & 1.2(-16) \\
    \ce{H2O} & \!\!\!\!$\langle\log_{10} N_{\rm col}\rangle$ &
    19.1     & 19.0     & 19.0     & 19.1     & 19.0     \\
    & $\langle\Tg\rangle$ &
    620      & 640      & 620      & 650      & 600      \\
    & $r_1-r_2$ &
    $0.11-0.37$ & $0.12-0.43$ & $0.11-0.38$ & $0.11-0.36$ & $0.11-0.31$ \\
    \hline
    &&&&&&\\[-2.2ex]
    & $\sum_{\rm line}F_{\rm mol}^{\rm line}\rm$ &
    \bf 2.5(-17) & 2.1(-17) & 3.0(-17) & 2.6(-17) & 9.3(-17) \\
    \ce{CO2} & \!\!\!\!$\langle\log_{10} N_{\rm col}\rangle$ &
    18.0     & 17.9     & 18.0     & 18.0     & 18.1     \\
    & $\langle\Tg\rangle$ &
    270      & 270      & 270      & 270      & 310      \\
    & $r_1-r_2$ &
    $0.39-0.87$ & $0.38-0.89$ & $0.36-0.86$ & $0.40-0.87$ & $0.26-0.65$ \\
    \hline
    &&&&&&\\[-2.2ex]
    & $\sum_{\rm line}F_{\rm mol}^{\rm line}$ &
    \bf 8.5(-17) & 4.7(-17) & 8.1(-17) & 2.5(-16) & 1.0(-18) \\
    \ce{HCN} & \!\!\!\!$\langle\log_{10} N_{\rm col}\rangle$ &
    17.5     & 17.1     & 17.5     & 18.1     & 14.9     \\
    & $\langle\Tg\rangle$ &
    580      & 570      & 550      & 660      & 590      \\
    & $r_1-r_2$ &
    $0.10-0.19$ & $0.10-0.21$ & $0.10-0.20$ & $0.09-0.18$ & $0.11-0.31$ \\
    \hline
    &&&&&&\\[-2.2ex]
    & $\sum_{\rm line}F_{\rm mol}^{\rm line}$ &
    \bf 5.7(-17) & 3.1(-17) & 1.0(-16) & 1.5(-16) & 1.3(-18) \\
    \ce{C2H2} & \!\!\!\!$\langle\log_{10} N_{\rm col}\rangle$ &
     16.9     & 16.3     & 17.2     & 17.2     & 13.8     \\
    & $\langle\Tg\rangle$ &
     740      & 740      & 690      & 770      & 410      \\
    & $r_1-r_2$ & 
    $0.08-0.13$ & $0.08-0.72$ & $0.08-0.12$ & $0.08-0.13$ & $1.1-6.2$ \\
    \hline
  \end{tabular}}
  \end{center}
  \vspace*{-2mm}
  \small
  \hspace*{28mm}$^{(1)}$: Increased by a factor
  of three with respect to the parameter value given in
  Table~\ref{tab:parameter}.\\
  \hspace*{28mm}$^{(2)}$: Same model parameters as the fiducial model,
  but before the code changes as described in
  Sect.~\ref{sec:CodeChanges}.\\
  \hspace*{28mm}$^{(3)}$: A higher value of the parameter $\Mdot$
  does not change $L_{\rm UV}$ in the model, whereas in nature it should.
  \vspace*{-1mm}
\end{table*}

The \ce{HCN} and \ce{C2H2} lines behave in a different way. They
actually become weaker with increasing UV, because photodissociation
limits the concentrations that can build up at given X-ray induced
production rates.  Table~\ref{tab:UV-Xray} shows how the column
densities of \ce{HCN} and \ce{C2H2} decrease when $L_{\rm UV}$ is
increased.  However, only the \ce{C2H2} lines become clearly stronger,
and the \ce{C2H2} column density clearly larger, when $L_X$ is
increased. The \ce{HCN} line fluxes and column densities remain on a
similar level as $L_X$ is varied.  The line emitting areas and
emission temperatures of all considered molecules do not change much
when $L_X$ is changed -- it is the column densities that are affected,
a chemical effect.

Table~\ref{tab:UV-Xray} also shows the results for a model with three
times larger mass accretion rate, a value that is close to the $\Mdot$
reported by \cite{Wang2023} for EX\,Lupi.  While the
\ce{H2O}, OH and \ce{CO2} line fluxes and their characteristics remain
rather similar, the \ce{HCN} and \ce{C2H2} lines show a strong
response.  Their line fluxes increase by factors 3.0 and 2.6,
respectively, while also increasing their column densities and
emission temperatures. This effect is likely related to the fact that
these emission lines come from relatively deep disc layers, where the
viscous heating is more relevant, see Fig.~\ref{fig:heatcoolEX}. In
addition, the line photons have to outshine the continuum photons,
which come from even deeper layers, to create an emission line. The
observable line fluxes hence depend on the contrast between line and
continuum emission temperatures. Although the viscous heating due to
$\Mdot$ only slightly increases $\Tg$ locally in the deep line emitting
regions, it has a profound impact on the temperature contrast $\Tg-\Td$,
thus creating an almost linear response to increasing $\Mdot$.

\section{Summary and conclusions}
\label{sec:conclusions}

We have implemented a number of new theories and improvements into our
thermo-chemical disc modelling code {\sc ProDiMo}.  These improvements
include a new escape probability method for spectral lines, relevant
for the calculation of line heating/cooling rates, a new treatment of
dust settling, and a new concept to apply UV molecular shielding
factors to photorates in 2D disc geometry. We have shown that these
measures altogether result in significant changes to our predicted
mid-IR molecular line spectra. The last column in
Table~\ref{tab:UV-Xray} shows that the same disc model with the
previous code version concerning escape probability, molecular
shielding and dust settling results in molecular emission features
that are too strong for \ce{CO2} by a factor of 4, and too weak for
HCN and \ce{C2H2} by factors of 80 and 40, respectively.  The code
improvements allow us to get closer to the observed molecular
properties derived from slab model fits of Spitzer/IRS and JWST/MIRI
observations, such as column densities, molecular emission
temperatures, and line emitting areas.

Based on the new {\sc ProDiMo} code, we have developed a disc model
for the quiescent state of the T\,Tauri star EX\,Lupi that can
simultaneously fit the spectral energy distribution, the overall
spectral shape of the JWST spectrum, and all important molecular
emission features in the continuum-subtracted JWST spectrum between
4.9\,$\mu$m and 7.5\,$\mu$m, and between 13\,$\mu$m and 17.7\,$\mu$m,
including \ce{CO}, \ce{H2O}, \ce{OH}, \ce{C2H2}, \ce{HCN}, \ce{CO2}
and \ce{H2}.  According to our knowledge, this is the first time that
a full Spitzer or JWST line spectrum could be fitted using the complex
2D disc structure of a thermo-chemical model, with consistent dust
opacities, molecular concentrations and calculated dust and gas
temperatures.

{Our model uses standard element abundances with solar
  carbon-to-oxygen ratio ($\rm C/O\!=\!0.46$) throughout the disc. We
  do not claim, however, that this is a unique solution. If the
  transport by pebble drift and subsequent sublimation of water
  \citep[e.g.][]{Booth2019, Banzatti2023} or carbonaceous grains and
  PAHs \citep[e.g.][]{Nazari2023, Tabone2023} can be taken into
  account in a quantative way, it is very likely that other disc
  shapes can fit the JWST data, too.}

Our model explains the observed molecular emission features by a
slowly increasing surface density profile around the inner rim in
combination with strong dust settling, which creates an almost
dust-free gas with large molecular column densities above the settled
dust. According to this model, the observed line emissions are
produced mostly by the inner disc $0.1-0.5\,$au, with a certain
layered structure, where hot OH emits from the top, warm CO and
\ce{H2O} from the middle, and \ce{C2H2}, \ce{HCN} and \ce{CO2} from
deeper layers.  The \ce{CO2} emitting region is located slightly
further away from the star as compared to \ce{C2H2} and \ce{HCN},
leading to lower \ce{CO2} emission temperatures.  This pattern is in
agreement with the general trends published for a number of T\,Tauri stars
observed with Spitzer \citep[e.g.][]{Salyk2011, Najita2013,
  Banzatti2020} and JWST \citep{Grant2023, Tabone2023, Banzatti2023}.

At the top of the photo-dissociation layer, the OH molecule plays a
key role to form \ce{CO} and \ce{H2O}, and to cool down the
predominantly atomic gas from about 4000\,K to 2000\,K, before the
water cooling takes over, which establishes temperature conditions
wherein, subsequently, other molecules can form as well.  This happens
at densities of about $\sim\!10^9-10^{10}\rm\,cm^{-3}$ in the inner
disc, where it seems critical to have a non-LTE treatment of
the statistical populations of the excited ro-vibrational states of the
OH molecule.  The required collisional rates are available in the
literature, see for example \cite{Rahmann1999,Tabone2021}, but are not
yet implemented in {\sc ProDiMo}.
Concerning the line emissions from \ce{C2H2}, \ce{HCN} and \ce{CO2},
this seems less critical, because these lines are emitted from a
denser gas $>\!10^{11}\rm\,cm^{-3}$.


The existence of large amounts of HCN and \ce{C2H2} molecules in the
oxygen-rich gas inside of the snowline is a consequence of an X-ray
driven chemistry in our model, where the most abundant CO and \ce{N2}
molecules are partly dissociated by \ce{He+} created by the X-ray
irradiation. Similar conclusions have been presented by
\cite{Walsh2015}. These free atoms and ions quickly form \ce{C2H2} and
HCN in large concentrations of about $10^{-6}$ to $10^{-5}$ in our
model. But this effect is limited to the regions close to the inner rim, before
the X-rays get absorbed by the gas. The chemical and cooling
timescales are short in the line emitting regions, smaller than
about 10\,yrs and 0.1\,yrs, respectively.

Our models show that the water and OH line luminosities increase with
stellar UV luminosity, whereas the HCN and \ce{C2H2} lines show the opposite
trend. We predict a positive correlation between \ce{C2H2} line
luminosity and stellar X-ray luminosity.  Concerning the mass
accretion rate $\Mdot$, our models show an almost linear response
of both the HCN and \ce{C2H2} line luminosities, whereas other
molecular emissions are less affected.

\begin{acknowledgements}
   P.\,W.\ acknowledges funding from the European Union
   H2020-MSCA-ITN-2019 under Grant Agreement no.\,860470 (CHAMELEON).
   I.\,K. and A.\,M.\,A. acknowledge support from grant
   TOP-1 614.001.751 from the Dutch Research Council (NWO).
\end{acknowledgements}

\bibliographystyle{aa}
\bibliography{references}

\appendix
\section{Additional figures and tables}

{Table~\ref{tab:abund} shows the element abundances assumed throughout
this paper, $n_{\langle\rm el\rangle}/n_{\langle\rm H\rangle} =
10^{\,\epsilon_{\rm\,el}-12}$.  The assumed carbon-to-oxygen ratio is 0.46.
This table is a copy of Table~5 in \citet{Kamp2017}.  In case of the
best-fitting EX\,Lupi model, the PAH abundance is larger by a factor
of 2.9, so $\epsilon_{\rm\,PAH}\!=\!5.94$.}

\begin{table}
  \centering
  \caption{Element abundances assumed throughout this paper.}
  \vspace*{-2mm}
  \begin{tabular}{cr|cr}
    \hline
    element & $\epsilon_{\rm\,el}$\ \ \ & element & $\epsilon_{\rm\,el}$\ \ \ \\
    \hline
    &\\[-2.2ex]
    H   & 12.00 & Mg  &  4.03 \\
    He  & 10.98 & Si  &  4.24 \\
    C   &  8.14 & S   &  5.27 \\
    N   &  7.90 & Ar  &  6.08 \\
    O   &  8.48 & Fe  &  3.24 \\
    Ne  &  7.95 & PAH &  5.48 \\
    Na  &  3.36 &     &       \\
    \hline
  \end{tabular}
  \label{tab:abund}
\end{table}

\begin{table}
\begin{center}
\caption{Parameters for the {\sc ProDiMo} standard T\,Tauri disc model.}
\vspace*{-6mm}
\label{tab:parameter_standard}
\resizebox{90mm}{!}{\begin{tabular}{l|c|c}
\hline
&&\\[-2.2ex]
quantity & symbol$^{\,(1)}$ & value\\
&&\\[-2.2ex]
\hline 
\hline 
&&\\[-2.2ex]
stellar mass                      & $M_{\star}$      & $0.7\,M_\odot$\\
stellar luminosity                & $L_{\star}$      & $1.0\,L_\odot$\\ 
effective temperature             & $T_{\star}$      & $4000\,$K\\
UV excess                         & $f_{\rm UV}$     & $0.01$\\
UV powerlaw index                 & $p_{\rm UV}$     & $1.3$\\
X-ray luminosity                  & $L_X$          & $10^{30}\rm erg/s$\\
X-ray emission temperature        & $T_X$          & $2\times10^7$\,K\\
accretion rate                    & $\Mdot$        & 0 \\
\hline
&&\\[-2.2ex]
strength of interstellar UV       & $\chi^{\rm ISM}$ & 1\\
cosmic ray H$_2$ ionisation rate  & $\zeta_{\rm CR}$   
                                          & $1.7\times 10^{-17}$~s$^{-1}$\\
\hline
&&\\[-2.2ex]
disc gas mass                       & $M_{\rm disc}$   & $1\times10^{-2}\,M_\odot$\\
disc dust mass                      & $M_{\rm dust}$   & $1\times10^{-4}\,M_\odot$\\
inner disc radius                   & $R_{\rm in}$     & 0.07\,au\\
tapering-off radius                 & $R_{\rm tap}$    & 100\,au\\
column density power index          & $\epsilon$     & 1.0\\
tapering off power index            & $\gamma$       & 1.0\\
reference gas scale height          & $H_p(100\,{\rm au})$  & 10\,au\\
flaring power index                 & $\beta$        & 1.15\\ 
\hline
&&\\[-2.2ex]
minimum dust particle radius        & $a_{\rm min}$      & $0.05\,\mu$m\\
maximum dust particle radius        & $a_{\rm max}$      & $3\,$mm\\
dust size dist.\ power index        & $a_{\rm pow}$      & 3.5\\
settling method                     & settle\_method   & Riols\,\&\,Lesur\\
settling parameter                  & $\alpha$         & $1\times10^{-3}$\\
max.\ hollow volume ratio           & $V_{\rm hollow}^{\rm max}$  & 80\%\\
&&\\[-2.0ex]
dust composition             & $\rm Mg_{0.7}Fe_{0.3}SiO_3$ & 60\%\\
(volume fractions)           & amorph.\,carbon  & 15\%\\
                             & porosity         & 25\%\\
\hline
&&\\[-2.2ex]
PAH abundance rel.\ to ISM          & $f_{\rm PAH}$        & 0.01\\
PAH opacities not considered in RT  & --                 & -- \\
chemical heating efficiency         & $\gamma^{\rm chem}$  & 0.2\\
\hline   
\end{tabular}}
\end{center}
\vspace*{-2mm}
\small $^{(1)}$: See \citet{Woitke2016} for details and definitions of
parameters.
\vspace*{-1mm}
\end{table}

{Figure~\ref{fig:Qheatcool} shows vertical cuts through the disc at
$r\!=\!0.1$\,au, 0.3\,au, 1\,au, 3\,au, 10\,au and 30\,au through our
  standard T\,Tauri disc model after the {\sc ProDiMo} changes
  discussed in Sect.~\ref{sec:newProDiMo}. This figure
shows more details as compared to Figure~\ref{fig:heatcool}, but both
figures show the same model.}

\begin{figure*}
  \begin{tabular}{ccc}
    \hspace*{-5mm}
    \includegraphics[page=1,width=58mm,trim=30 -76 80 360,clip]
                    {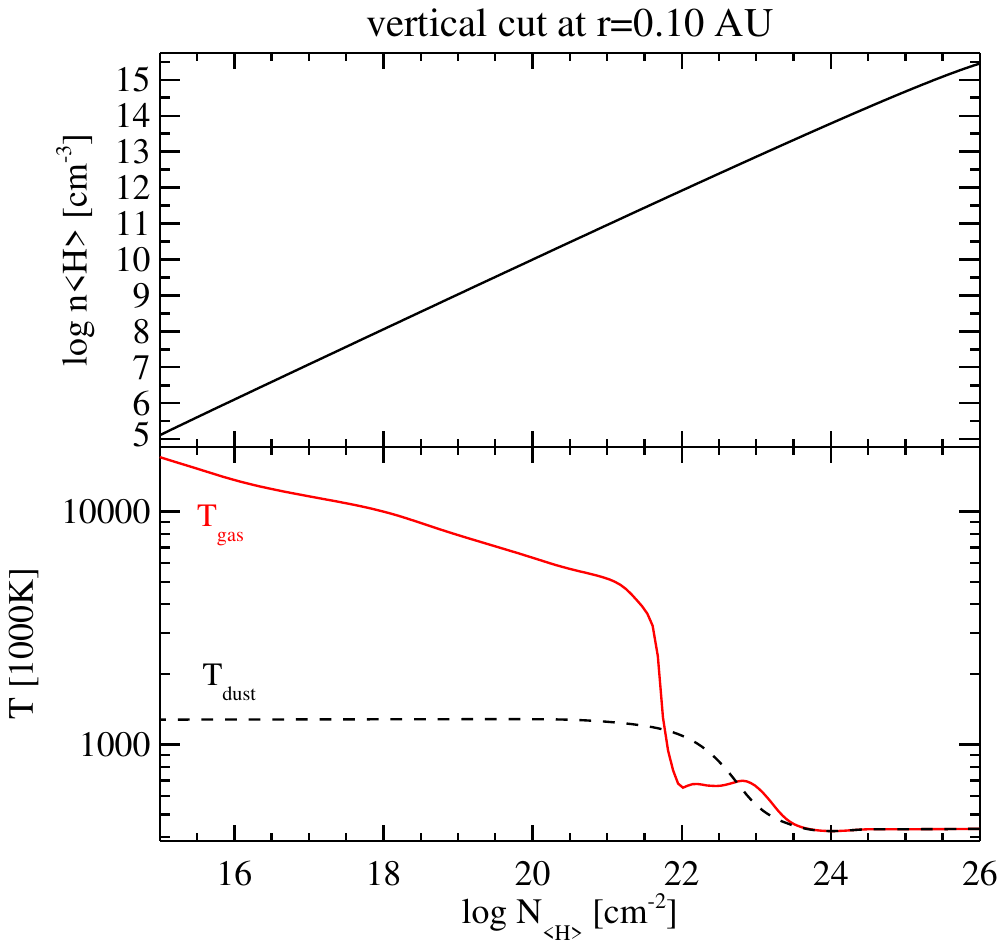} &
    \hspace*{-5mm}
    \includegraphics[page=2,width=64mm,trim=50 20 80 330,clip]
                    {Figures/Cut0.1.pdf} &
    \hspace*{-5mm}
    \includegraphics[page=3,width=64mm,trim=50 20 80 330,clip]
                    {Figures/Cut0.1.pdf} \\
    \hspace*{-5mm}
    \includegraphics[page=1,width=58mm,trim=30 -76 80 360,clip]
                    {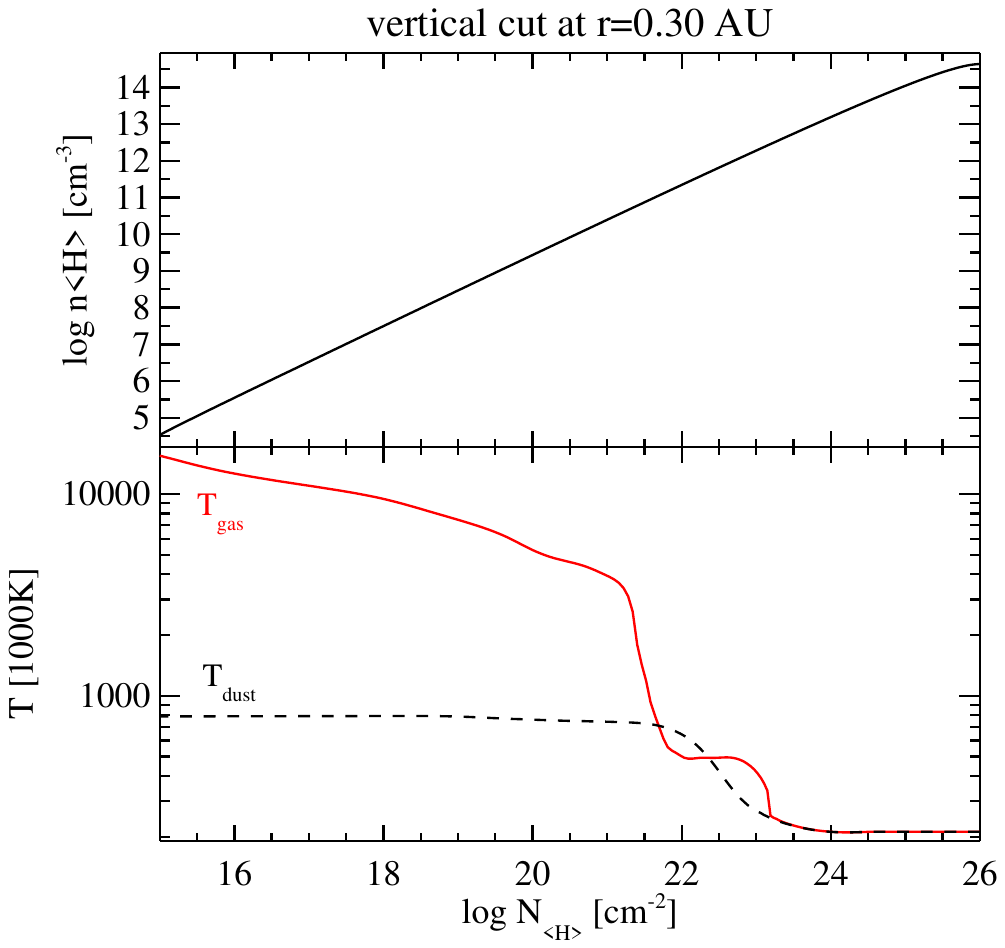} &
    \hspace*{-5mm}
    \includegraphics[page=2,width=64mm,trim=50 20 80 330,clip]
                    {Figures/Cut0.3.pdf} &
    \hspace*{-5mm}
    \includegraphics[page=3,width=64mm,trim=50 20 80 330,clip]
                    {Figures/Cut0.3.pdf} \\
    \hspace*{-5mm}
    \includegraphics[page=1,width=58mm,trim=30 -76 80 360,clip]
                    {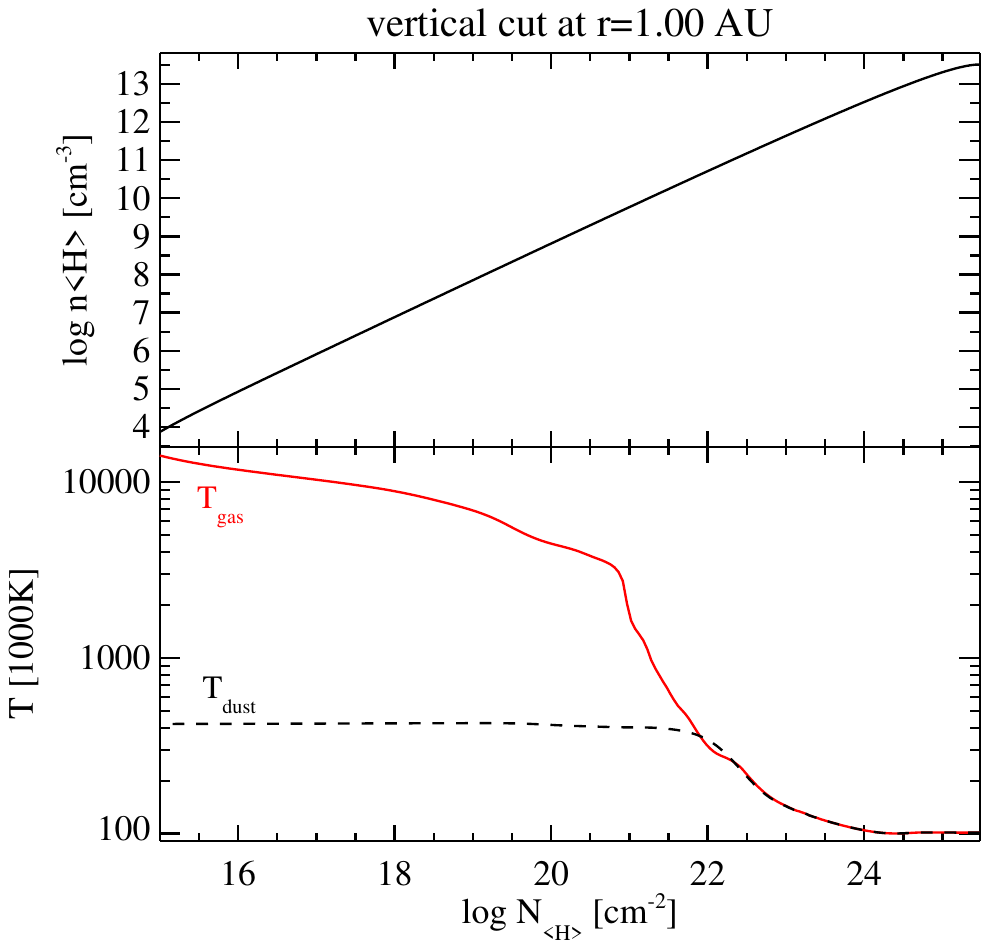} &
    \hspace*{-5mm}
    \includegraphics[page=2,width=64mm,trim=50 20 80 330,clip]
                    {Figures/Cut1.0.pdf} &
    \hspace*{-5mm}
    \includegraphics[page=3,width=64mm,trim=50 20 80 330,clip]
                    {Figures/Cut1.0.pdf} 
  \end{tabular}
  \caption{Vertical cuts at $r\!=\!0.1$\,au, 0.3\,au, and 1\,au through
    the standard {\sc ProDiMo} T\,Tauri disc model after the code
    modifcations described in Sect.\,~ref{sec:CodeChanges}. The
    $x$-axis is the hydrogen nuclei column density measured from the
    top, relevant line formation typically happens between about
    $10^{21}\rm\,cm^{-2}$ and $10^{24}\rm\,cm^{-2}$. The left plots
    show the assumed hydrogen nuclei particle density $\nH$ and the
    calculated gas and dust temperatures. The middle and left plots
    show the relevant heating and cooling rates, annoted below, where
    the thick gray line is the total heating/cooling rate. We note
    that $\sum Q_{\rm heat}\!=\!\sum Q_{\rm cool}$. An individual
    heating or cooling rate is plotted when it reaches at least 15\% of the
    total heating/cooling rate anywhere in the depicted vertical
    range.}
  \label{fig:Qheatcool}
\end{figure*}
\addtocounter{figure}{-1}
\begin{figure*}
  \begin{tabular}{ccc}
    \hspace*{-5mm}
    \includegraphics[page=1,width=58mm,trim=30 -76 80 360,clip]
                    {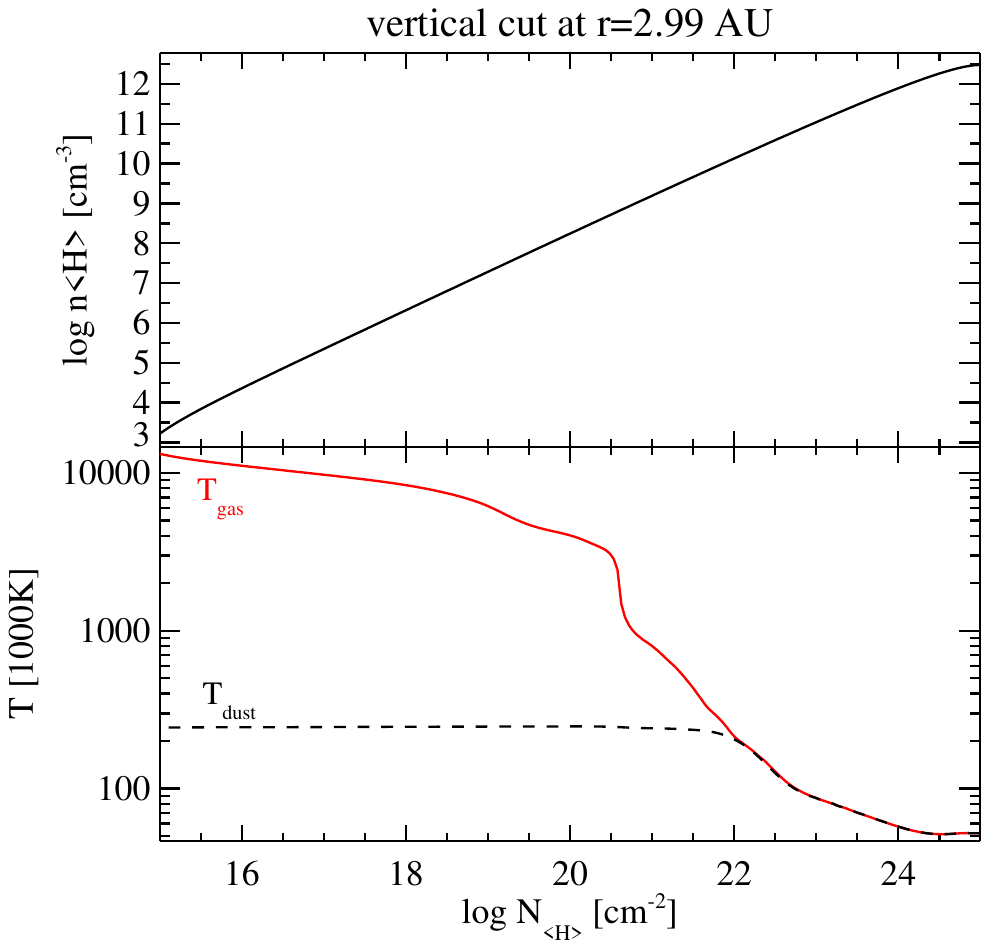} &
    \hspace*{-5mm}
    \includegraphics[page=2,width=64mm,trim=50 20 80 330,clip]
                    {Figures/Cut3.0.pdf} &
    \hspace*{-5mm}
    \includegraphics[page=3,width=64mm,trim=50 20 80 330,clip]
                    {Figures/Cut3.0.pdf} \\
    \hspace*{-5mm}
    \includegraphics[page=1,width=58mm,trim=30 -76 80 360,clip]
                    {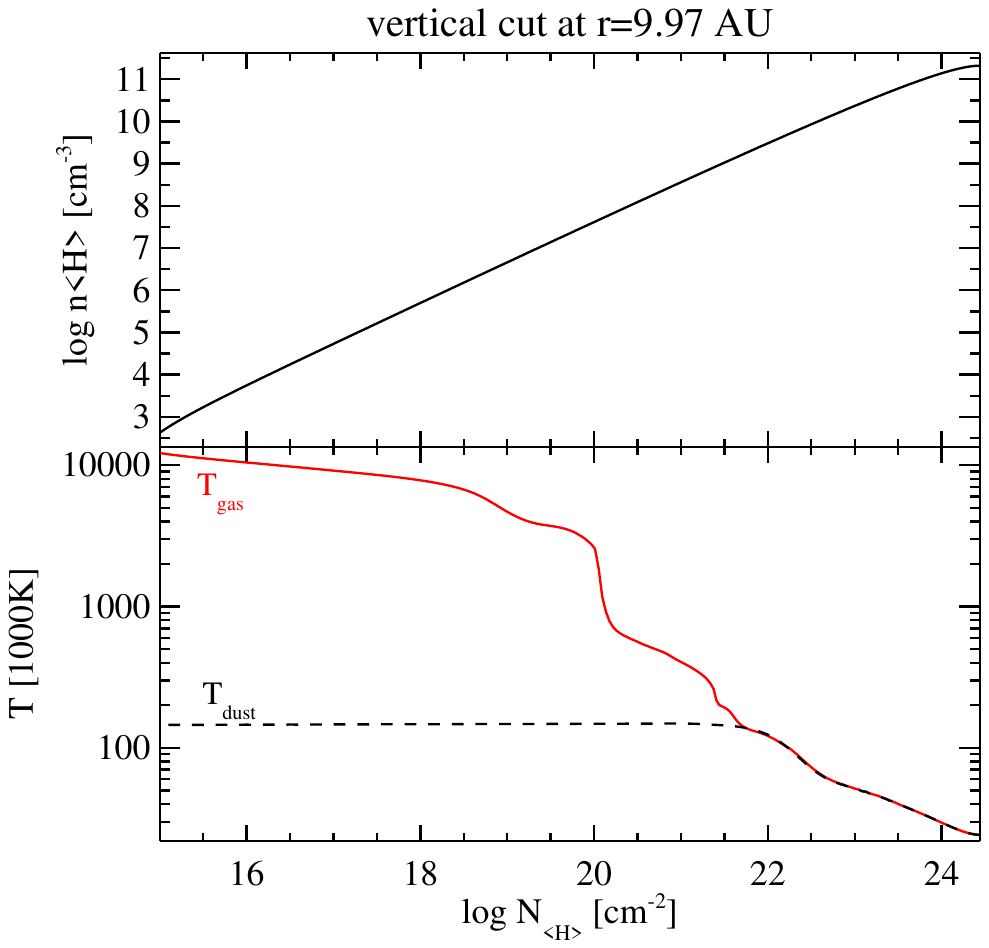} &
    \hspace*{-5mm}
    \includegraphics[page=2,width=64mm,trim=50 20 80 330,clip]
                    {Figures/Cut10.0.pdf} &
    \hspace*{-5mm}
    \includegraphics[page=3,width=64mm,trim=50 20 80 330,clip]
                    {Figures/Cut10.0.pdf} \\
    \hspace*{-5mm}
    \includegraphics[page=1,width=58mm,trim=30 -76 80 360,clip]
                    {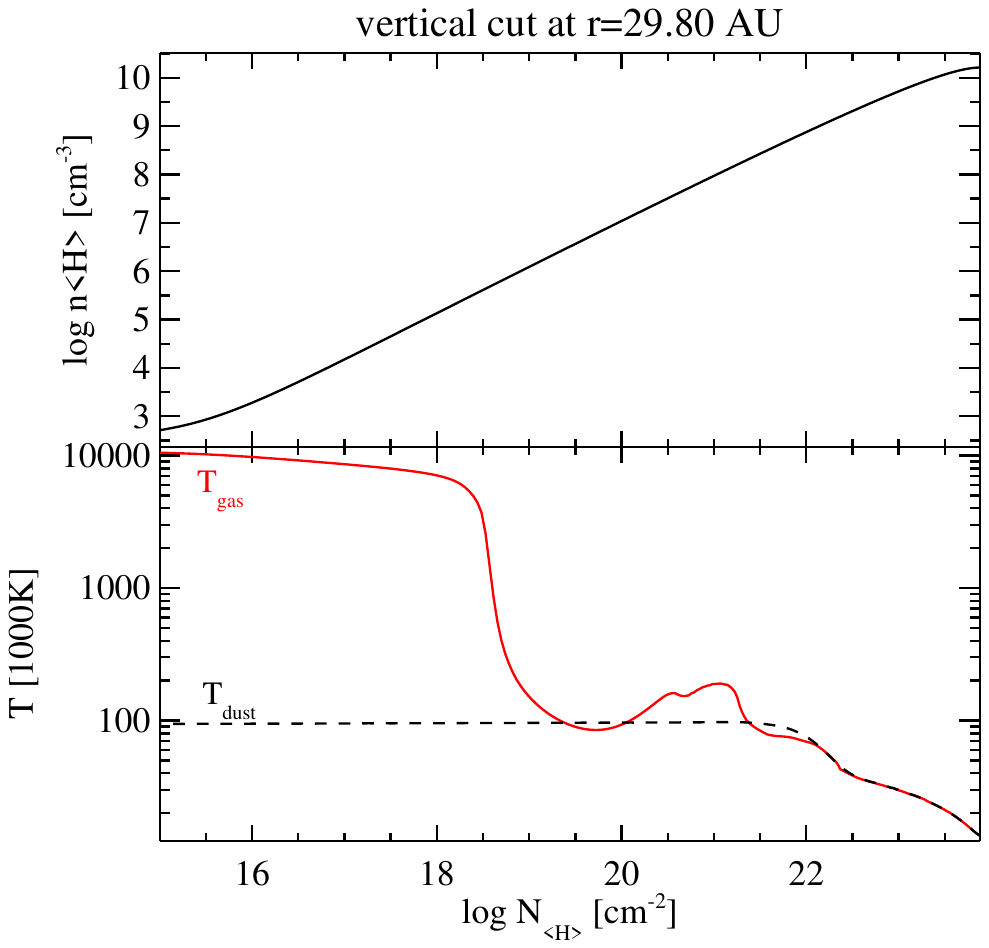} &
    \hspace*{-5mm}
    \includegraphics[page=2,width=64mm,trim=50 20 80 330,clip]
                    {Figures/Cut30.0.pdf} &
    \hspace*{-5mm}
    \includegraphics[page=3,width=64mm,trim=50 20 80 330,clip]
                    {Figures/Cut30.0.pdf} 
  \end{tabular}
  \caption{continued. Vertical cuts at $r\!=\!3$\,au, 10\,au, and 30\,au}
\end{figure*}

\end{document}